\documentclass[11pt]{article}

\usepackage[T1]{fontenc}
\usepackage{lmodern}
\usepackage{microtype}
\usepackage[dvipsnames]{xcolor}

\usepackage{booktabs}
\usepackage{amssymb}
\usepackage{tabularx}
\usepackage{array}
\usepackage{makecell}
\usepackage{longtable}

\usepackage{graphicx}

\usepackage[
  backend=biber,
  style=apa,
  natbib=true,
  url=true
]{biblatex}
\usepackage[hidelinks]{hyperref}

\title{Algorithm Validation as a Policy Audit: \\
Evidence from Race-blind Charging}
\author{
    Muskan Walia \\
    New York University
    \and
    Joe Nudell \\
    Harvard University
    \and
    Alex Chohlas-Wood \\
    New York University
}
\date{\today}

\begin{document}

\newcommand{\joe}[1]{\textbf{\textcolor{magenta}{joe: #1}}}
\newcommand{\acw}[1]{\textbf{\textcolor{orange}{#1~--Alex}}}
\newcommand{\todo}[2]{\textbf{\textcolor{red}{TO-DO (#1): #2}}}

\maketitle

\begin{abstract}
\noindent
California recently required that 
all prosecutors in the state
conduct a ``race-blind charging'' decision 
by reviewing case documents
where selected race-related proxies
have been redacted.
We validate an open-source LLM-based algorithm 
called \texttt{bc2}
that we developed
to conduct this redaction automatically,
which was used 
to facilitate race-blind review
on over 119,000 real-world cases in 2025.
We evaluate two distinct questions:
whether \texttt{bc2} faithfully implements 
the state's requirements,
and whether the state's requirements---even when faithfully implemented---advance the goal of race-blind decision-making.
To do so, 
we draw on a corpus of 
nearly 5,000 real-world police reports 
we assembled
from jurisdictions across the United States.
We find that the latest version of \texttt{bc2}
faithfully implements the legal mandate 
on 96.7\% of narratives in our sample 
under a stringent document-level measure.
This performance is 
a substantial improvement 
over earlier versions of \texttt{bc2},
and better
than leading open-source redaction methods.
Our validation also shows 
that California's mandate
misses key proxies for race,
including location information.
Redacting these additional proxies on top of the state mandate
(as \texttt{bc2} does)
eliminates 
43.1\%
of the predictive signal 
that remains after compliance with the state mandate. 
These findings show that validation can do more than
assess technical compliance.
Validation 
can also improve algorithms
as well as
help policymakers achieve
underlying policy goals.

\end{abstract}

\section{Introduction}

In 2022, California unanimously passed a law 
that mandates all prosecutors in the state 
implement a new charging procedure 
called ``race-blind charging.'' 
Under this mandate, 
prosecutors must make initial charging decisions 
by reading crime report narratives 
where 
explicit mentions of race or ethnicity, 
person names, and 
the skin color or complexion
of those involved
have been redacted.
This legislation targets longstanding concerns 
that racial bias may influence prosecutorial decision-making
and seeks to align prosecutorial decision-making 
with the longstanding ideal of blind justice. 

Under the state's mandate,
prosecutors' offices must institute a process
to redact race-related information
from narratives
before the charging prosecutor's review.
Since asking humans to conduct this redaction
would be costly and difficult to scale,
state regulators encouraged offices
to use artificial intelligence (AI) 
to automate redaction 
and make implementation feasible~\citep{doj2024guidelines}.
Yet automating the redaction of race-related information
from free-text narratives is far from straightforward. 
For example, any such algorithm 
needs to distinguish between the word ``Will'' 
used as a person's name---which should be redacted---and the verb, which should not. 
Similarly, the word ``black'' 
may refer to a person's race 
(and thus require redaction)
or may describe the color of an object 
(and thus require preservation,
in case it is relevant
to the charging decision).
Challenges like these make 
accurate automated redaction a technically demanding task.

Recognizing this difficulty, 
state regulators required 
that any automated redaction algorithm 
be validated to perform as expected. 
This requirement reflects 
legitimate and widespread concerns 
about the reliability of AI systems, 
particularly when used in
high-stakes settings
like the prosecutor's charging decision. 
Yet when describing what validation should entail, 
the state merely suggested that prosecutors 
confirm the accuracy and reliability of an application 
on ``a representative set of documents'' 
and monitor ``the performance over time 
to ensure that it remains accurate and reliable''~\citep{doj2024guidelines}. 
Implicit in this suggestion 
is the notion that validation only serves 
a single, narrow function: 
confirming that an existing algorithm 
performs adequately.

In this paper, 
we develop and apply a series of tests 
to validate one such redaction algorithm we developed,
called \texttt{bc2}. 
\texttt{bc2} is an open-source application 
that combines large language models (LLMs) 
and computer vision models 
to redact race-related information 
from police report narratives. 
\texttt{bc2} is now used in at least 
50\% 
of California prosecutor’s offices, 
covering a constituent population of nearly 18 million people,
where it was used to facilitate 
a race-blind charging review 
on at least 119,030 cases in 2025. 
To assess algorithmic performance, 
we constructed a benchmark dataset 
by drawing on a corpus of 
4,879 police reports spanning 
239 jurisdictions across 
42 states
that we assembled 
through public records requests. 

We find that \texttt{bc2} 
largely accomplishes the redaction task 
mandated by the state: 
only 3.3\% 
of narratives 
retain any explicit mention of 
race, a person name, or a description of skin color 
after redaction. 
Moreover, validation itself drove much of this performance,
as errors we identified during validation 
reduced \texttt{bc2}'s failure rate by nearly 80\%.
We also show that \texttt{bc2} 
outperforms leading open-source redaction algorithms 
on our benchmark, 
including a state-of-the-art system 
released by OpenAI in April 2026. 
At the same time, our validation suggests
that compliance with the state's mandate alone 
leaves substantial racial signal intact. 
By targeting additional proxies on top of the state mandate
(i.e., location information 
and a broader range of physical descriptions),
\texttt{bc2} removes 
43.1\% of the signal remaining
after state-required redaction, 
as measured by how accurately the arrested person's race 
can be predicted from a redacted narrative.

In doing so, we show that algorithmic validation 
has a greater role to play in the responsible deployment of AI 
than merely verifying that an algorithm works as intended. 
Validation can also improve the performance of an algorithm 
by identifying errors and revealing opportunities for refinement. 
At the same time, 
validation can also reveal 
that the rules such applications 
are designed to implement 
may not adequately implement 
policymakers' underlying goals. 

\section{Related work}

California's race-blind charging mandate
was motivated, in part, by a concern that prosecutors 
may be subject to ``unconscious'' bias in charging decisions~\citep{ab2778}.
This concern is reflected in a broader academic literature 
investigating whether the race of an arrestee 
affects prosecutorial decision-making, 
though empirical evidence on bias in charging decisions 
remains mixed~\citep{rehavi2014racial, berdejo2018criminalizing,sloan2024racial,wu2016racial, macdonald2020analysis}. 

To address this concern,
both scholars and policymakers have proposed blinding a prosecutor to the race or ethnicity of those involved in a criminal incident~\citep{chohlas2021blind,sah2015blinding,robertson2019raceclass}.
This mirrors a broader class of interventions that seek to reduce discrimination by limiting decision-makers' access to sensitive demographic attributes, like race or gender, 
e.g., blind auditions in orchestras~\citep{goldin2000orchestrating}.
Relatedly, audit and correspondence studies take a complementary approach, experimentally manipulating specific race-related cues, such as a person's name, while holding other features constant~\citep{bertrand2004emily}. 
These approaches reflect a view of race not as a single observable attribute, but as a collection of signals that may be communicated through names, language, location, appearance, and other contextual markers---what \citet{sen2016bundle} describe as a ``bundle of sticks.''
For race-blind charging, this means that redaction must obscure not only explicit racial identifiers, but also the surrounding cues through which race may still be inferred.

Because redacting this information can be labor-intensive and difficult to scale, California explicitly endorsed the use of AI to automate race redaction and make race-blind charging feasible in practice~\citep{doj2024guidelines}.
This endorsement reflects the growing use of algorithmic and AI systems across the justice system, a development that has long been the subject of vigorous debate~\citep{angwin2016machine,kleinberg2018human,dressel2018accuracy,lin2020limits}.
Critics have warned that algorithms used in high-stakes settings may reproduce or exacerbate racial disparities, underscoring the importance of careful system design and evaluation~\citep{chohlaswood2023designing,corbett2023measure}.
These concerns are especially salient for generative AI systems, which may produce inaccurate or fabricated outputs, commonly referred to as ``hallucinations''~\citep{dahl2024large,magesh2025hallucination,charlotin2025hallucinations}.

At the same time, recent high-profile audits of algorithmic systems suggest that careful validation can improve performance in high-stakes settings~\citep{marquand2026auditing}.
For example, audits of commercial facial-recognition and speech-recognition systems documented large disparities for darker-skinned women and Black speakers, respectively, and 
follow-up work suggests that these findings helped pressure vendors to improve subgroup performance, including reported improvements in commercial facial-analysis systems and increased attention to Black speech patterns as a benchmark for speech recognition~\citep{buolamwini2018gendershades,koenecke2020racial,raji2019actionable,radford2023robust,speechmatics2021breakthrough}.

These examples suggest that validation can play an important role in diagnosing failure points and guiding concrete improvements to deployed systems.
Thus, advocates have called on algorithms to be validated before they are deployed~\citep{ccj2025principles,raji2020closing}, mirroring California's guidance for race-blind charging~\citep{doj2024guidelines}.
Yet traditional programmatic validation 
is difficult to achieve at scale 
for applications that use generative AI,
in part because unstructured inputs and outputs limit the use of traditional methods~\citep{shankar2025docetl}.
To address this challenge, researchers have begun using LLMs as evaluators of unstructured outputs. 
This ``LLM-as-a-judge'' approach 
employs LLMs as critics, 
identifying instances 
where outputs fail to align with specified policy goals~\citep{zheng2023judging}.
This approach is part of a broader turn 
toward using LLMs as research aides, 
including the use of LLMs to simulate human responses
in a new set of approaches called ``silicon sampling''~\citep{barrie2026ai,hullman2026human,egami2023using}.

\section{Data}
\label{sec:data}

To make validation possible, we assembled a benchmark dataset
of police reports via 819 public records requests filed across all 50 American states.
For each jurisdiction, we initially requested all crime reports describing assault incidents that occurred on a single calendar day. 
If this request yielded few records, we expanded the request scope to include felony offenses and, if necessary, all crime types occurring over a one-week period until at least a few dozen reports were obtained.
This process ultimately resulted in 4,879 police reports 
from 239 jurisdictions across 42 states; 
the release of such records was prohibited in the remaining 8 states.
Details regarding our cataloguing of these reports are provided in Appendix~\ref{apx:prr_cataloguing}.

The documents in our full benchmark dataset reflect the wide variety of layouts, document quality, and writing styles typical of police reports referred to prosecutor’s offices. 
Reports range in length from 1 to 168 pages and include everything from computer-native PDFs to low-quality scanned pages of handwritten witness statements. Narrative sections vary substantially in length, ranging from 3 to 15,516 words. 
While the publicly released records in our dataset may differ in some respects from the confidential 
versions provided to prosecutors---for example, many jurisdictions 
released records 
in which certain identifying details 
had already been obscured 
to protect personally identifying information (PII)
as part of the public records process---they nevertheless provide a diverse and realistic corpus for evaluating redaction for
race-blind charging. 
Figure~\ref{fig:reports} illustrates this diversity by presenting example pages from several redacted reports. 

For our validation,
we filter to two subsets of documents designed to test specific aspects of \texttt{bc2}’s performance. We call the first subset the \textit{coverage sample}.
We assembled this sample to include documents with a high likelihood of containing at least one type of race-related information, mirroring the prevalence of such information in the confidential police reports prosecutors receive. 
We therefore exclude reports with narratives under 250 words, which frequently lack names, physical descriptions, or other race-related information. 
We also exclude records containing pre-existing masking of identifying information (e.g., a black bar over apparent identifying information) from the public records process because their inclusion would downwardly bias 
the prevalence of race-related information 
in unredacted documents.
To identify such records, we developed a page-level classifier that examined each PDF page for black-box masking, marker-like obscuration, placeholder text, or unnatural gaps within lines of text, marking a document as having been altered for public release if any page showed clear evidence that information had been masked.
To assess the performance of this approach, we had a human reviewer label a subsample of documents for evidence of pre-existing masking. We then used those labels to verify that the automated procedure performs reasonably well. Additional details are provided in Appendix~\ref{apx:preexisting_redactions}.
After following these steps, the coverage sample contains 1067 documents 
from 88 jurisdictions, 
totaling 24,947 pages.

The second subset, which we refer to as the \textit{prediction sample}, is designed to test whether \texttt{bc2} makes it difficult to guess the race of the arrestee.
Because this analysis requires a race label for the arrestee,
we automatically extract this information from structured report metadata and unstructured narrative text 
and restrict the sample to records in which the arrestee's race is explicitly identified.
These labels reflect how race was recorded in the report and may not correspond to the arrestee's self-identified race or ethnicity.
Details on the evaluation of this automated extraction are provided in Appendix~\ref{apx:arrestee_race}.
To avoid ambiguity regarding which individual’s race is being predicted, we further restrict the sample to reports involving a single arrestee; this exclusion removes relatively few records while simplifying the prediction task. 
Applying these criteria resulted in a prediction sample of
1,445 documents
from across 36 states,
totaling 10,420 pages.

Separately, we drew on a different set of public records requests to verify use and deployment of \texttt{bc2} across California. 
In early 2025, shortly after the mandate took effect, we requested records regarding algorithm adoption and policies. 
Of California's 58 counties, 
29 reported using \texttt{bc2}, 
17 reported using alternative solutions,
5 reported not complying with the mandate,
and 7 did not respond.
In early 2026, we conducted a follow-up round of public records requests to all prosecutors' offices seeking information about \texttt{bc2} usage during 2025, including case volumes and reported error rates.
Responses indicated that \texttt{bc2} was used to enable race-blind review 
on at least 119,030 cases during 2025, with participating offices reporting incomplete redaction on 0.698\% of reports successfully processed by the application. 
Appendix~\ref{apx:usage_stats} reports office-level case volumes and reported error rates.

\begin{figure}[!t]
\centering
\includegraphics[width=0.6\textwidth]{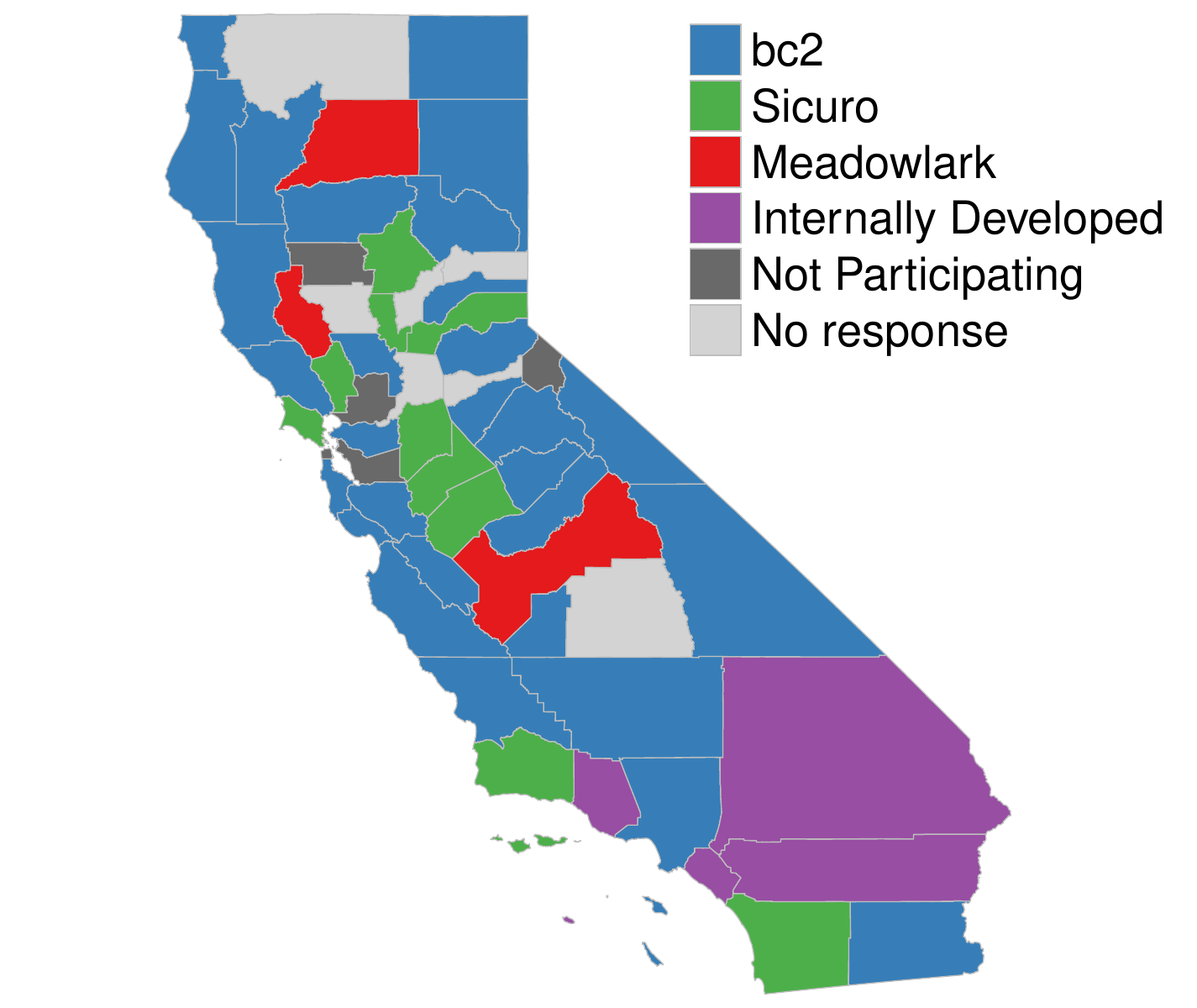}
\caption{
\textit{
California prosecutors' offices using \texttt{bc2} (in blue)
as determined by a series of public records requests.
Sicuro and Meadowlark are closed-source
vendor solutions.
Offices using \texttt{bc2}
later indicated that the algorithm 
had been used to enable race-blind review on  119,030 cases in 2025.
}
}
\label{fig:map}
\end{figure}

\section{Methods and results}
\subsection{Algorithm overview}
The algorithm we developed, \texttt{bc2}, was designed to address several key challenges in the requirements set forth by California’s mandate.
One primary difficulty is that race-related information can appear anywhere within a police report narrative,
and there is rarely a reliable indication of where such information is located within a narrative. 
This challenge is compounded by the fact that police reports 
are typically provided in digital formats
which do not expose their underlying structure. 
Because different police departments use a wide variety of document templates, 
there is also substantial variation in report structure across jurisdictions. 
As a result, an automated algorithm 
required to work at scale 
must reliably infer the structure 
of a wide variety of crime report layouts
to identify the location of the narrative. 
Finally, because the intended deployers of this redaction software 
are budget-constrained government agencies, 
it is important to design approaches 
that are accessible and transparent to a government IT employee
while remaining affordable to run. 
These requirements rule out more complex solutions---like fine-tuning relatively expensive computer vision models---in favor of a generalizable approach 
that can accommodate diverse document structures 
while remaining practical to deploy and maintain. 

\begin{figure}[t]
\centering
\includegraphics[width=\linewidth]{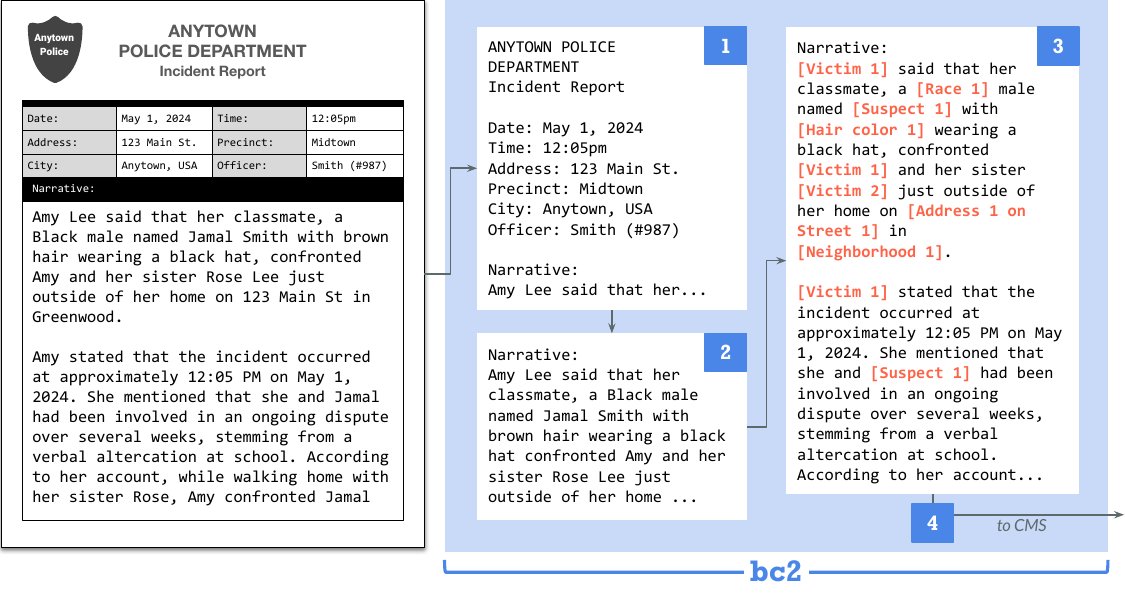}
\caption{
\textit{
\texttt{bc2} pipeline.
In step (1), \texttt{bc2} accepts documents in many formats
(including scanned and native PDFs)
and uses computer vision models
to extract any text in the report.
In step (2),
\texttt{bc2} uses an LLM
to isolate any free-text narratives 
in the extracted text.
In step (3),
\texttt{bc2} uses a new LLM call
to redact race-related information 
in the provided narrative(s).
After quality checks in step (4),
the redacted narrative
is sent to a case management system (CMS)
so a prosecutor can review the redacted narrative
and file a race-blind charging decision.
}
}
\label{fig:pipeline}
\end{figure}

\texttt{bc2} addresses these challenges
by applying three operations 
to transform police reports 
into redacted narratives 
(Figure~\ref{fig:pipeline}).
First, the algorithm converts 
the entirety of each document 
into raw text using 
Azure Document Intelligence’s 
\texttt{prebuilt-read} model.
Second, it isolates free-text narratives
from this raw text using 
\texttt{GPT-4.1}. 
Third, it redacts race-related information 
from the extracted narratives 
by removing explicit mentions of 
race, 
physical descriptions, 
names, and 
location information,
again using 
\texttt{GPT-4.1} to do so. 

\texttt{bc2} also includes specialized tooling 
to handle several less common scenarios. 
Some crime reports include multiple documents 
associated with the same incident, 
requiring \texttt{bc2} to use consistent placeholders
across documents 
(e.g., so that Jane Doe 
is always replaced with the same unique placeholder, 
like \texttt{[Victim 3]}). 
Other reports exceed the context window 
of the LLMs we use
and are thus too long to be processed 
in a single language model call. 
To address both challenges, \texttt{bc2} 
maintains a roster of previously redacted entities 
and supplies this information as context to subsequent model calls 
when processing additional documents or continuing the redaction of a long report. 
Finally, 
the algorithm includes quality checks to detect hallucinations 
or other modifications to the underlying narrative.
Outputs containing substantial changes\footnote{
We define ``substantial'' 
as the character-level error rate of the redacted narrative, 
computed from the diff of the input and output texts 
after accounting for the bracketed redactions that we expect. 
By default, the maximum error rate is 0.1, 
but it is configurable. 
We found that setting the threshold too low 
leads to redactions being rejected for benign reasons, 
such as altering the epenthetic ``n'' 
in the article ``a(n)'' 
preceding a placeholder.
} 
beyond the intended bracketed redactions
are automatically rejected and retried.
We also designed a cloud-based software application  
to coordinate these transformations in high-volume environments.
The prompts used for the narrative extraction and redaction stages are provided in Appendices~\ref{apx:bc2_extract_prompt} and \ref{apx:bc2_redact_prompt},
and information about the software application 
that handles requests to \texttt{bc2} 
is available in Appendix~\ref{apx:app}.

Of the two LLM-based transformations 
in the \texttt{bc2} pipeline, 
redaction is the most challenging 
because race-related information 
can appear in many forms and contexts 
throughout a crime report narrative. 
We therefore focus this paper on our validation 
of the redaction step.
Validation results for the narrative extraction step 
are reported in Appendix~\ref{apx:narr_extraction}.

\subsection{Redaction coverage}
We first assess whether \texttt{bc2} 
reliably redacts race-related proxies.
Note that California only requires the redaction of three categories of information: explicit mentions of race or ethnicity, 
physical descriptions (in particular, skin color or complexion), 
and person names. 
\texttt{bc2} also targets additional information 
above and beyond the state mandate
that we thought would be likely
to provide strong cues about the race of the arrestee---in particular, location information
like addresses, neighborhoods, commercial establishments, and landmarks;
we also target a wider range of physical descriptions,
including hair color, eye color, 
and hairstyles.

Accordingly, we assess whether \texttt{bc2} 
successfully redacted \textit{all}
explicit mentions of race, 
select physical descriptions, 
and person names from each crime report narrative
(as required by \citet{doj2024guidelines})
as well as whether \texttt{bc2} 
successfully redacted all location information
(as we required above and beyond the state mandate).
In this setup, a single missed redaction---even just one in a 10,000-word narrative---renders the entire record unsuccessful, 
making this a particularly stringent assessment.
One challenge with this assessment 
is that it requires a qualitative review 
of redacted narratives to determine if \texttt{bc2} 
missed any race-related information. 
While a manual qualitative review is possible in principle, 
it would be prohibitively time-consuming to conduct at scale,
and would need to be repeated whenever \texttt{bc2} is modified. 
These constraints make comprehensive manual evaluation difficult 
and potentially infeasible 
for both developers and government agencies 
seeking to validate \texttt{bc2} and other such algorithms. 

We therefore employ a validation approach 
that can be conducted quickly, automatically, and at low cost
by using an LLM critic to review narratives for missed redactions.
(This approach is typically referred to as \textit{LLM-as-a-judge},
but we avoid this term here 
to prevent confusion 
with human judges in the legal system.)
Because the LLM critic is used only for evaluation 
rather than large-scale deployment, 
we use a more expensive and capable model,
\texttt{GPT-5.5},
compared to what \texttt{bc2} uses.
We also use a set of prompts distinct from 
\texttt{bc2}'s redaction pipeline
(Appendix~\ref{apx:critic_prompts}).
However, because the redaction model 
and the LLM critic may share systematic blind spots, 
we use design-based supervised learning (DSL) 
to correct for potential statistical bias 
in the critic's assessments~\citep{egami2023using}. 
Specifically, we combine LLM critic-generated labels 
with a small stratified random sample of human labels 
to produce doubly robust bias-corrected estimates of redaction performance. Additional details on this estimation procedure are provided in Appendix~\ref{apx:dsl}. 

We find that \texttt{bc2} largely achieves its redaction goals 
(Table~\ref{tab:redaction-coverage}).
Only 3.3\% of narratives 
contain an explicit reference to race, physical descriptions, or person names---the three legally required targets of redaction---and only 6.7\% 
contain one of the race proxies targeted by \texttt{bc2}, 
including location information. 
For reference,
we compare these results 
to the original, unredacted narratives.
98.2\% of unredacted narratives 
contain at least one of the three legally required categories,
and 99.7\% 
contain at least one of the four categories targeted by \texttt{bc2}. 

\begin{table}[t]
\centering
\begin{tabularx}{\textwidth}{@{}l*{4}{>{\raggedright\arraybackslash}X}@{}}
\toprule
& \multicolumn{4}{c}{Narratives with at least one mention} \\
\cmidrule(lr){2-5}
&
Name
&
\makecell[l]{Explicit \\race/eth.}
&
\makecell[l]{Physical\\description}
&
Location \\[-1.2ex]
\midrule
Unredacted narratives
    & 97.8\%
    & 25.5\%
    & 23.1\%
    & 97.8\% \\
\midrule
Microsoft Presidio
    & 71.3\%
    & (20.7\%)
    & (20.2\%)
    & (94.8\%) \\
OpenAI Privacy Filter
    & 66.6\%
    & (25.3\%)
    & (22.6\%)
    & (88.0\%) \\
OpenAI GABRIEL (5.4)
    & 22.8\%
    & (25.4\%)
    & (23.0\%)
    & (53.3\%) \\
\midrule
\texttt{bc1} (2019)
    & 95.5\%
    & 7.0\%
    & 8.7\%
    & 85.1\% \\
\texttt{bc2} (Pre-validation)
    & 11.1\%
    & 0.5\%
    & 1.5\%
    & 21.6\% \\
\textbf{\texttt{bc2}}
    & \textbf{2.1\%}
    & \textbf{0.2\%}
    & \textbf{1.1\%}
    & \textbf{3.8\%} \\
\bottomrule
\end{tabularx}
\caption{\textit{
Proportion of narratives containing one or more instances of race-related
information after applying varying redaction methods. These statistics were
calculated on the ``coverage sample,'' i.e., among narratives containing at
least 250 words and no public-record pre-redactions
($N=1067$). Lower values indicate more complete
redaction. Values in parentheses indicate an information category that is not
fully targeted by that redaction method.
}}
\label{tab:redaction-coverage}
\end{table}

We note that many of the examples 
characterized as leaks 
are borderline cases for redaction, 
highlighting the skeptical disposition of our LLM critic. 
For example, 
the LLM critic occasionally flags physical characteristics 
like a person's height or hair length,
even though these attributes 
are not the types of physical descriptions 
targeted by \texttt{bc2}.
In other cases, the LLM critic classifies 
interstate highways, county names, or county government entities as leaks. 
However, \texttt{bc2} is only designed to redact
precise location information---like addresses, neighborhoods, commercial establishments, and landmarks---and avoid redaction of less precise geographic context 
that is 
both likely known to the prosecutor 
(i.e., they know what county and state they work in)
and potentially useful for understanding the incident.

We extend these results by comparing \texttt{bc2} 
to leading open-source solutions for PII redaction
(Table~\ref{tab:redaction-coverage}).
PII redaction algorithms do not target 
the exact same categories of information as \texttt{bc2},
but both redaction tasks target person names.
As a result, we focus exclusively 
on the incidence of person names
as the comparison metric
against open-source PII redaction algorithms.

We observe that \texttt{bc2} outperforms leading
open-source PII redaction algorithms 
at the specific task of name redaction.
For example, 
OpenAI's Privacy Filter 
(one state-of-the-art solution, as of April 2026)
missed at least one name 
on 66.6\% 
of cases in our coverage sample.
OpenAI's GABRIEL missed at least one name on
22.8\% of cases
when using GPT-5.4 as a backend.\footnote{
\texttt{GABRIEL}'s \texttt{deidentify()} function 
technically implements \textit{pseudonymization} instead of redaction. 
To evaluate coverage, we replaced pseudonyms 
with opaque placeholders like those we use with \texttt{bc2},
since our LLM critic 
would otherwise interpret \texttt{GABRIEL}'s pseudonyms as leaked names. 
In the process, we identified and patched a bug in the deidentification procedure that causes failures when the model proposes multiple pseudonyms for the same entity.} 
One older open-source solution, 
Microsoft Presidio,
missed at least one name on 
71.3\%
of cases.

We also find that our own redaction solutions 
have improved over time---both as the result of moving to LLM-based approaches
and as the result of this validation.
Our 2019 solution,
\texttt{bc1},
used regex and named entity recognition to identify race-related information~\citep{chohlas2021blind}.
However, it required substantial customization per jurisdiction,
an effort that would not be practical across the 600 law enforcement agencies
that exist in California.
As a result, 
we did not attempt this customization
on our 88 jurisdiction coverage sample,
and the results in Table~\ref{tab:redaction-coverage}
thus reflect its likely performance at scale without customization.

Validation also improved redaction coverage
for \texttt{bc2} over time.
The first version of \texttt{bc2}
used sensible prompts to target redaction of race-related information,
but these prompts missed important edge cases that appear in real-world narratives,
resulting in at least one leak of
race-related information
on 30.0\% of cases.
For example, the prompts did not consistently recognize combined race-and-gender abbreviations, 
such as W/F or B/M, or account for foreign-language text that could convey race or ethnicity. 
The revised prompts instructed the model to redact these abbreviations and replace foreign-language text with a marked English translation.
Targeting these issues
substantially improved our performance on our coverage sample,
though we note the exact estimate we report
may be biased downward 
due to the coverage sample's
use as both a learning and evaluation sample.
Still,
our relative improvement 
in performance for \texttt{bc2}---from errors on 30.0\% to only 6.7\% of cases---was driven by issues we discovered and fixed during validation.

\subsection{Race predictability}
\label{sec:race_predictability}

Next, we assess whether redacting race-related information 
makes it difficult to guess 
the reported race or ethnicity
of the arrested person.
Even if race-related information is perfectly removed 
as specified by California's law
or as intended by \texttt{bc2}, 
prosecutors
may still be able to infer the race or ethnicity 
of the arrested person
with reasonable confidence
from information remaining in the redacted narrative. 
Indeed,
racial cues are unavoidable to some degree,
since the alleged crime type 
itself may be correlated with race.
And in practice,
there are good reasons to preserve other information
that could correlate with race.
Many potential informative cues---e.g., quotes from a victim in African-American Vernacular English---should likely be preserved in their original form
for the prosecutor to make a properly informed decision 
based on an unmodified set of facts for the case.
As a result,
redaction for race-blind charging
must strike a balance in deciding which information to remove,
seeking to maximize available information 
while minimizing available \textit{race-related} information. 

This balance underscores
why it is important to carefully measure race predictability
in redacted reports.
Yet a traditional evaluation of race predictability 
would require large numbers of human reviewers---ideally legal professionals with charging decision experience---to repeatedly read police reports and guess the race of the arrestee. 
This approach would be costly in practice and difficult to replicate,
preventing open, transparent, and repeated benchmarking 
of \texttt{bc2} and other such race-redaction algorithms.  

As an alternative, 
we design a silicon sampling approach, 
using large language models 
to assess whether narratives
contain any clues about whether the arrestee
is non-White.
We note at least two inherent differences
between human raters and silicon samples:
on the one hand,
language models may lack the advanced mental model 
that a senior prosecutor acquires after years 
of reviewing police reports in their county;
on the other hand, 
language models are trained on enormous volumes of text,
and thus may have superhuman abilities 
to infer one's race or ethnicity~\citep{kozlowski2025simulating}. 
Yet their use in our validation 
makes repeated assessment cheap, scalable, and replicable---important attributes if 
developers, governments, and researchers 
are to adopt or extend 
this validation.

The silicon sampling approach we designed
was inspired by the ``bundle of sticks'' 
conception of race~\citep{sen2016bundle}, 
which emphasizes that racial identity 
may be communicated through many channels
aside from physical appearance---including the way one speaks,
the institutions one engages with,
and the social world one inhabits. 
We thus ask an LLM to assess latent racial signal
across a broad set of channels,
including 
explicit mentions of one's race or ethnicity;
person names;
physical appearance, including
hair, complexion, facial features, and eye color;
geographic context,
including addresses, neighborhoods, 
commercial establishments, and 
any listed jurisdictions from the city to state level;
nationality, ancestry, language, accent, and communication style;
clothing and accessories;
religious and cultural practices;
employment, education, and justice-system history;
social affiliations and group memberships;
vehicles, weapons, pets, consumer brands;
behavior, and substance use;
and housing or built-environment context.

We generate these assessments in two steps.
In the first step, 
we use \texttt{grok-4.3 (Medium)}
to assess whether a given channel
(e.g., built-environment context)
is present in the narrative.\footnote{
We experimented with asking other LLMs to produce these covariates,
but most models were reluctant to assign strong probabilities
in favor or against any specific race or ethnic group,
possibly due to post-training alignment.
} Then, if \texttt{grok-4.3} determines that 
the narrative contains information in this channel,
we next ask \texttt{grok-4.3}
to return a score between
0--100,
with low values representing the LLM's assessment 
that the given channel
contains evidence \textit{against} a non-White arrested person,
high values representing evidence in the given channel
\textit{in favor} of a non-White arrested person,
and 50
representing the neutral midpoint between these extremes.
Alongside these covariates,
we also ask the model to assess and return
the number of times race or ethnicity is mentioned in the narrative,
the number of children mentioned in connection with the arrestee,
and indicators for non-nuclear household structure,
public-transit use or dependence,
and the timing of the incident.
We additionally ask \texttt{grok-4.3}
to score, on a scale from
0--100,
how consistent the facts of the incident are
with each of 13
broad offense categories
(e.g., robbery or theft).
The prompts we used to create these covariates 
are provided in Appendix~\ref{apx:interpretable_feature_prompts}.
Altogether, we call these covariates the
``interpretable features,''
since they each represent an easy-to-interpret 
model-derived rating
on a single potential channel 
for race-related information.

We then use these interpretable features 
to fit a series of logistic regressions
on the prediction sample (Section~\ref{sec:data}).
Our models are trained to predict 
whether the arrestee is non-White,
since this prediction task 
correlates with the prosecutor's ability 
to privilege White arrestees over non-White arrestees.
In practice, 
most of the prediction sample
consists of reports involving either White or Black arrestees,
with 
54.7\% of our sample
listing a White arrestee,
36.1\% of our sample
listing a Black arrestee,
and 9.2\% of our sample
listing an arrestee of Asian, Hispanic, or other descent.
So the prediction task 
for our model is largely a question 
of whether the arrestee is White or Black.
However, 
our findings are qualitatively similar 
when specifically predicting 
whether the arrestee is Black, 
or is from another race or ethnic group. 

Each logistic regression
is the result of an ablation
intended to measure 
the latent racial signal in a narrative
under different redaction information regimes.
Our first model has access 
to all interpretable features described above,
representing the full set of channels
available in an unredacted narrative.
For our second model,
we ablate the interpretable features
for 
explicit mentions of race or ethnicity,
skin tone,
and person names---the channels that must be redacted 
for race-blind charging 
under California regulations.
For our third model,
we ablate all channels
targeted by \texttt{bc2},
including those required by the state
as well as hairstyle and eye color,
and location channels
like addresses and neighborhood names.
Our final model 
is fit on
a minimal ``baseline'' set of channels,
comprised only of 
county demographics,
alleged offenses,
and structural features of the narratives
(like the number of times any race/ethnicity is mentioned).
Each model is fit
via leave-one-out cross-validation (LOOCV),
and we retain the predicted probabilities 
from the holdout report
to calculate reported metrics.

\begin{figure}[!t]
\centering
\includegraphics[width=0.75\linewidth]{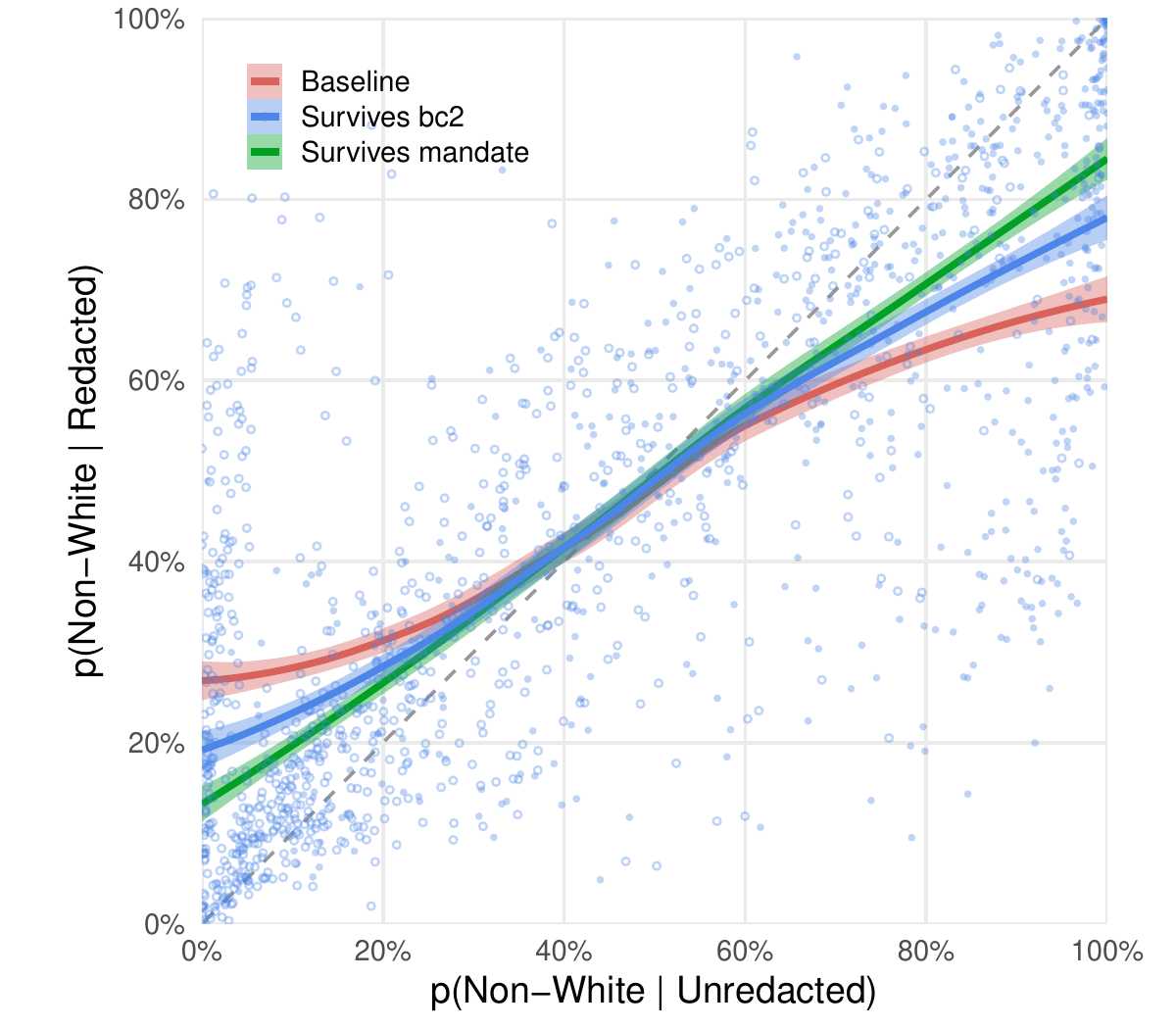}
\caption{
\textit{
Predicted probabilities that the arrested person is non-White
under both unredacted and redacted information regimes.
Probabilities were produced 
with logistic regression
for the held-out sample 
in leave-one-out cross-validation.
X-axis values are generated from all modeled features,
representing all information available in an unredacted narrative.
Y-axis values are generated from different subsets of LLM-derived features,
reflecting the information remaining after a successful redaction
by different methods 
(Table~\ref{tab:race-prediction-auc}),
with LOESS smoothing applied to each method.
Plotted points represent predicted probabilities
under the information that survives \texttt{bc2} only.
Solid points represent a non-White arrestee;
hollow points represent a White arrestee.
The diagonal reference line 
represents no change in predictability
from unredacted narratives.
}
}
\label{fig:scatterplot}
\end{figure}

\begin{table}[t]
    \centering
    \begingroup
    \setlength{\tabcolsep}{4pt}

    \makebox[\textwidth][c]{        \begin{tabular}{@{}l*{5}{c}@{}}
        \toprule
        & \multicolumn{4}{c}{\textbf{Interpretable features (logit)}}
        & \shortstack{\textbf{Embedding}\\\textbf{(GBT)}} \\
        \cmidrule(lr){2-5}
        \cmidrule(l){6-6}

        & \textbf{(1)}
        & \textbf{(2)}
        & \textbf{(3)}
        & \textbf{(4)}
        & \textbf{(5)} \\
        \midrule

        Within-county ROC AUC
        & 0.821
        & 0.702
        & 0.627
        & 0.529
        & 0.673 \\

        95\% CI
        & {\footnotesize 0.722--0.866}
        & {\footnotesize 0.659--0.726}
        & {\footnotesize 0.579--0.664}
        & {\footnotesize 0.501--0.587}
        & {\footnotesize 0.626--0.760} \\

        \midrule
        \multicolumn{6}{@{}l}{\textit{Information included}} \\
        \addlinespace[2pt]

        Baseline
        & \(\checkmark\)
        & \(\checkmark\)
        & \(\checkmark\)
        & \(\checkmark\)
        & \(\checkmark\) \\

        Survives \texttt{bc2}
        & \(\checkmark\)
        & \(\checkmark\)
        & \(\checkmark\)
        &
        & \\

        Survives mandate, not \texttt{bc2}
        & \(\checkmark\)
        & \(\checkmark\)
        &
        &
        & \\

        Redacted by mandate
        & \(\checkmark\)
        &
        &
        &
        & \\

        Unredacted embedding
        &
        &
        &
        &
        & \(\checkmark\) \\

        \midrule

        Narratives
        & 1,445
        & 1,445
        & 1,445
        & 1,445
        & 1,445 \\

        \bottomrule
        \end{tabular}    }

    \caption{
    \textit{
    Race-predictability under different redaction methods.
    Scores are produced
    by calculating the within-county ROC AUC
    with predictions from logistic regression under leave-one-out cross-validation.
    Model (1) is trained on
    all extracted features, 
    corresponding to the information in an
    unredacted report.
    Models (2) and (3) are trained on 
    information remaining after a successful redaction
    under differing standards for race-blind charging.
    Model (4) is trained on
    only basic information 
    that must be known to a prosecutor
    when making a charging decision.
    For robustness, we fit model (5),
    which is trained on
    semantic embeddings generated from unredacted narratives
    as well as the baseline feature set.
    }
    }
    \label{tab:race-prediction-auc}
    \endgroup
\end{table}

We find that \texttt{bc2} reduces the ability
to predict one's race from a redacted narrative
relative to the unredacted version---especially among reports that previously contained a strong signal (Figure~\ref{fig:scatterplot}).
For example,
among the 532 unredacted narratives
with a predicted probability
greater than 90\%
or less than 10\%
of being non-White, 
\texttt{bc2}
moved 360
(67.7\%) of these reports
to predicted probabilities
between 10\%--90\%
after redaction.
We also observe that these probabilities
lie close to the baseline information regime,
where a prosecutor can already 
make a weakly informed guess
about the race of an arrestee.
These results suggest that \texttt{bc2}
removes substantial race-related information---especially among reports where there was previously a clear racial signal.

Examining predicted probabilities 
shows one of the limits 
of the race-blinding approach
in the prosecutorial context.
The county and state in which a case originates
are already known to a prosecutor,
who is typically employed by (and reviews cases from)
a single county.
So county-level information 
cannot be hidden from the prosecutor 
in any meaningful way.
In their own county,
prosecutors may form a strong prior
about the race of an arrestee
based on the composition of historical arrestees.
For example,
if most arrestees are from
disadvantaged groups,
prosecutors may have reasonably high confidence
that a new case involves an arrestee 
from a disadvantaged group---even if they have yet to see
\textit{any} information on the specific case 
under consideration.
These location cues may be amplified 
in our national validation sample,
which covers a much wider range of jurisdictions 
than county prosecutors would typically see in practice. 
Concretely, knowing that a report originated in Idaho
rather than Mississippi 
may inform reasonably strong priors about the race of the arrestee 
aside from any details about the case itself.

To address these concerns,
we next examine whether race-blinding
reduces a prosecutor's ability
to rank cases within their own county
on the probability that a case
involves a non-White arrestee.
Specifically,
we examine the within-county
ROC AUC:
we first calculate ROC AUC separately
for each county in our sample
and then calculate a pair-weighted average
of this statistic
across all counties in our sample.
This statistic represents 
a county prosecutor's ability
to correctly rank
non-White narratives
in their jurisdiction
as more likely to contain a non-White arrestee
compared to White narratives.

We find that \texttt{bc2}
measurably reduces the ability 
to rank cases within a given county
above and beyond the redaction targets
required by California's guidelines (Table~\ref{tab:race-prediction-auc}).
First, we find that unredacted narratives
provide plenty of information to rank cases,
with a within-county AUC of
0.821.
Next, we see that the information regime
corresponding to California's guidelines
measurably reduces a prosecutor's ability
to rank cases,
with a within-county AUC of
0.702.
Yet this result suggests that there is still
some information available to prosecutors
to distinguish between non-White and White arrestees.
Next, we examine predictability
under the information available 
after a \texttt{bc2} redaction,
and find that \texttt{bc2} removes
43.1\%
of the remaining information 
on top of the state mandate
by redacting location information
and a wider range of physical descriptions.
These results highlight 
how validation itself
can uncover shortcomings 
in the well-intended guidelines 
that policymakers devise
to achieve underlying policy goals.

As a robustness check,
we also evaluate race-predictability 
by representing report narratives with semantic embeddings.
Although embeddings 
may capture residual race-related information 
that our interpretable features miss, 
making them an important check,
they are difficult to interpret 
and cannot be ablated in the same way as the interpretable features.
As a result,
we follow a similar modeling procedure 
as with our interpretable attributes, 
using LOOCV on the prediction sample,
but now using gradient-boosted trees (GBTs)
to take advantage of possible non-linearity 
in semantic embeddings.
Using 256-dimension embeddings 
generated by OpenAI's text-embedding-3-large
from unredacted narratives,
our model has an AUC of 0.673.
This relatively low AUC 
suggests that our sample size is too small
for a model to adequately learn much signal
from semantic embeddings.

\section{Discussion}

In this paper,
we show that \texttt{bc2}
both faithfully implements 
California's race-blind charging mandate
and improves on this mandate
by targeting additional proxies
for redaction.
On our coverage sample,
\texttt{bc2} 
misses at least one 
explicit mention of race,
person name,
or physical description
on 3.3\% of documents,
and misses any category of information 
targeted by \texttt{bc2}
(e.g., including location information)
on 6.7\%
of documents.
Validation helped our team reduce the error rate
compared to our original prompt
by almost 80\%,
resulting in PII redaction quality
better than leading open-source solutions.
We find that California's mandate 
allows a substantial amount
of race signal to remain in redacted documents.
We further show that the full set of proxies 
targeted by \texttt{bc2}
reduces the amount of race signal by 
43.1\%
compared to the standard allowed by the state.

These findings collectively suggest
California regulators should consider
updating their mandate to include
the redaction of location information---and a wider set of physical descriptors than skin tone---if they want implementations of ``race-blind charging'' 
to better reflect the name of the practice.
These findings will become increasingly relevant 
if other states consider 
passing race-blind charging mandates
and must implement standards of their own.
And to some extent,
the findings we share here 
may be complemented or complicated
by a randomized controlled trial 
we recently concluded
to measure race-blind charging's 
causal impact
on charging decisions.

Although this study
was focused on the race-blind charging context,
we note several wider implications
beyond the scope of race-blind charging.
First,
our study raises important questions
about the human experience of race-blindness.
In settings where decision-makers 
may form reasonably strong priors
about the identity of redacted individuals
(e.g., because a disadvantaged group
dominates the population under consideration),
it may not be reasonable to assume that 
decision-makers avoid mental representations
of the disadvantaged group.
Instead, in a blinded context,
they may assume that every person under consideration
is a member of the disadvantaged group,
penalizing everyone equally 
but resulting in substandard decisions overall.
Second,
although our validation makes heavy use of LLMs
as a proxy for understanding race information,
humans 
(and experienced prosecutors in particular) 
may be better or worse
than LLMs at inferring race from redacted crime report narratives.
More research 
is required to understand how humans conceptualize and comprehend
blinded materials.

One final implication follows from subjecting \texttt{bc2} 
and California’s mandate 
to the same race-predictability test. 
Throughout the paper,
we described one as an algorithm and the other as a policy. 
Yet both ultimately perform the same function: 
they reflect normative objectives
through a set of rules
(in this case governing what race proxies 
a prosecutor may see). 
One rule is expressed in law 
and the other in code, 
but both can be evaluated 
by asking whether they produce 
the information environment 
that race-blind charging 
aims to create. 
This common test suggests that algorithms 
should not be treated as a special class of objects 
separate from public policies. 
Instead, they should be understood as 
computational instantiations of policy choices---explicit attempts to translate abstract social goals 
into institutional practice. 
Their validation should therefore extend 
beyond technical fidelity 
to the harder question that has long applied to policy itself: 
whether these rules, 
once implemented, 
bring us closer to the society 
they were designed to create.

\section*{Acknowledgments}

This work was generously supported by 
the Abdul Latif Jameel Poverty Action Lab,
Arnold Ventures,
Microsoft's Justice Reform Initiative,
Stanford Human-Centered Artificial Intelligence (HAI),
and
Stanford Impact Labs.

\printbibliography

@inproceedings{chohlas2021blind,
    author = {Chohlas-Wood, Alex and Nudell, Joe and Yao, Keniel and Lin, Zhiyuan Jerry and Nyarko, Julian and Goel, Sharad},
    title = {Blind {J}ustice: {A}lgorithmically {M}asking {R}ace in {C}harging {D}ecisions},
    year = {2021},
    isbn = {9781450384735},
    publisher = {Association for Computing Machinery},
    address = {New York, NY, USA},
    url = {https://doi.org/10.1145/3461702.3462524},
    doi = {10.1145/3461702.3462524},
    booktitle = {Proceedings of the 2021 AAAI/ACM Conference on AI, Ethics, and Society},
    pages = {35–45},
    numpages = {11},
    location = {Virtual Event, USA},
    series = {AIES '21}
}

@legislation{ab2778,
  title        = {Assembly {B}ill {N}o. 2778: {C}rimes: {R}ace-{B}lind {C}harging},
  shorttitle   = {AB-2778},
  jurisdiction = {California},
  date         = {2022-09-29},
  number       = {AB-2778},
  note         = {Chapter 806, Statutes of 2022},
  url          = {https://leginfo.legislature.ca.gov/faces/billNavClient.xhtml?bill_id=202120220AB2778}
}

@article{wu2016racial,
    author = {Jawjeong Wu},
    title ={{R}acial/{E}thnic {D}iscrimination and {P}rosecution: {A} {M}eta-{A}nalysis},
    journal = {Criminal Justice and Behavior},
    volume = {43},
    number = {4},
    pages = {437-458},
    year = {2016},
    doi = {10.1177/0093854815628026},
  publisher={Sage Publications Sage CA: Los Angeles, CA}
}

@online{sloan2024racial,
  author={Sloan, CarlyWill},
  year=2024,
  title={Do {P}rosecutor and {D}efendant {R}ace {P}airings {M}atter? {E}vidence from {R}andom {A}ssignment},
  note={Working paper},
  url =    {https://github.com/carlywillsloan/Prosecutors/blob/master/sloan_prosecutor.pdf},
  lastaccessed = {January 6, 2025}
}

@article{rehavi2014racial,
  title={Racial {D}isparity in {F}ederal {C}riminal {S}entences},
  author={Rehavi, M Marit and Starr, Sonja B},
  journal={Journal of Political Economy},
  volume={122},
  number={6},
  pages={1320--1354},
  year={2014},
  publisher={University of Chicago Press Chicago, IL}
}

@article{berdejo2018criminalizing,
  title={Criminalizing {R}ace: {R}acial {D}isparities in {P}lea-{B}argaining},
  author={Berdej{\'o}, Carlos},
  journal={BCL Rev.},
  volume={59},
  pages={1187},
  year={2018},
  publisher={HeinOnline}
}

@report{doj2024guidelines,
  author      = {{California Department of Justice}},
  shortauthor = {{Cal. DOJ}},
  title       = {Race-{B}lind {C}harging {G}uidelines},
  date        = {2024-01-01},
  url         = {https://oag.ca.gov/system/files/media/crim-guidelines-race-blind-charging-2024.pdf}
}

@article{dahl2024large,
    author = {Dahl, Matthew and Magesh, Varun and Suzgun, Mirac and Ho, Daniel E},
    title = {Large {L}egal {F}ictions: {P}rofiling {L}egal {H}allucinations in {L}arge {L}anguage {M}odels},
    journal = {Journal of Legal Analysis},
    volume = {16},
    number = {1},
    pages = {64-93},
    year = {2024},
    month = {06},
    issn = {2161-7201},
    doi = {10.1093/jla/laae003}
}

@online{charlotin2025hallucinations,
  author  = {Damien Charlotin},
  title   = {{AI} {H}allucination {C}ases},
  year    = {2025},
  url     = {https://www.damiencharlotin.com/hallucinations/},
  urldate = {2026-06-06}
}

@article{egami2023using,
  title={Using {I}mperfect {S}urrogates for {D}ownstream {I}nference: {D}esign-based {S}upervised {L}earning for {S}ocial {S}cience {A}pplications of {L}arge {L}anguage {M}odels},
  author={Egami, Naoki and Hinck, Musashi and Stewart, Brandon and Wei, Hanying},
  journal={Advances in Neural Information Processing Systems},
  volume={36},
  pages={68589--68601},
  year={2023}
}

@article{sen2016bundle,
  author = {Maya Sen and Omar Wasow},
  title = {Race as a {\textquoteright}{B}undle of {S}ticks{\textquoteright}: {D}esigns that {E}stimate {E}ffects of {S}eemingly {I}mmutable {C}haracteristics},
  year = {2016},
  journal = {Annual Review of Political Science},
  volume = {19},
  pages = {499-522},
  language = {eng},
}

@software{cpl2026api,
  author  = {{Computational Policy Lab}},
  title   = {blind-charging-api},
  version = {0.13.3},
  date    = {2026},
  url      = {https://github.com/comppolicylab/blind-charging-api}
}

@article{goldin2000orchestrating,
  author = {Goldin, Claudia and Rouse, Cecilia},
  title = {Orchestrating {I}mpartiality: {T}he {I}mpact of ``{B}lind'' {A}uditions on {F}emale {M}usicians},
  journal = {American Economic Review},
  volume = {90},
  number = {4},
  pages = {715--741},
  year = {2000},
  doi = {10.1257/aer.90.4.715}
}

@article{bertrand2004emily,
  author = {Bertrand, Marianne and Mullainathan, Sendhil},
  title = {Are {E}mily and {G}reg {M}ore {E}mployable than {L}akisha and {J}amal? {A} {F}ield {E}xperiment on {L}abor {M}arket {D}iscrimination},
  journal = {American Economic Review},
  volume = {94},
  number = {4},
  pages = {991--1013},
  year = {2004},
  doi = {10.1257/0002828042002561}
}

@article{sah2015blinding,
  author = {Sah, Sunita and Robertson, Christopher T. and Baughman, Shima B.},
  title = {Blinding {P}rosecutors to {D}efendants' {R}ace: {A} {P}olicy {P}roposal to {R}educe {U}nconscious {B}ias in the {C}riminal {J}ustice {S}ystem},
  journal = {Behavioral Science \& Policy},
  volume = {1},
  number = {2},
  pages = {69--76},
  year = {2015},
  url = {https://scholarship.law.bu.edu/faculty_scholarship/923/}
}

@article{robertson2019raceclass,
  author = {Robertson, Christopher and Baughman, Shima Baradaran and Wright, Megan S.},
  title = {Race and {C}lass: {A} {R}andomized {E}xperiment with {P}rosecutors},
  journal = {Journal of Empirical Legal Studies},
  volume = {16},
  number = {4},
  pages = {807--847},
  year = {2019},
  doi = {10.1111/jels.12235}
}

@misc{angwin2016machine,
  author       = {Julia Angwin and Jeff Larson and Surya Mattu and Lauren Kirchner},
  title        = {Machine {B}ias},
  howpublished = {ProPublica},
  year         = {2016},
  month        = may,
  day          = {23},
  url          = {https://www.propublica.org/article/machine-bias-risk-assessments-in-criminal-sentencing},
}

@article{kleinberg2018human,
  author = {Kleinberg, Jon and Lakkaraju, Himabindu and Leskovec, Jure and Ludwig, Jens and Mullainathan, Sendhil},
  title = {Human {D}ecisions and {M}achine {P}redictions},
  journal = {The Quarterly Journal of Economics},
  volume = {133},
  number = {1},
  pages = {237--293},
  year = {2018},
  doi = {10.1093/qje/qjx032}
}

@article{chohlaswood2023designing,
  author  = {Alex Chohlas-Wood and Madison Coots and Sharad Goel and Julian Nyarko},
  title   = {Designing {E}quitable {A}lgorithms},
  journal = {Management Science},
  year    = {2025}
}

@article{corbett2023measure,
  title={The {M}easure and {M}ismeasure of {F}airness},
  author={Corbett-Davies, Sam and Gaebler, Johann D and Nilforoshan, Hamed and Shroff, Ravi and Goel, Sharad},
  journal={Journal of Machine Learning Research},
  volume={24},
  number={312},
  pages={1--117},
  year={2023}
}

@article{dressel2018accuracy,
  title={The accuracy, fairness, and limits of predicting recidivism},
  author={Dressel, Julia and Farid, Hany},
  journal={Science advances},
  volume={4},
  number={1},
  pages={eaao5580},
  year={2018},
  publisher={American Association for the Advancement of Science}
}

@article{lin2020limits,
  title={The limits of human predictions of recidivism},
  author={Lin, Zhiyuan “Jerry” and Jung, Jongbin and Goel, Sharad and Skeem, Jennifer},
  journal={Science advances},
  volume={6},
  number={7},
  pages={eaaz0652},
  year={2020},
  publisher={American Association for the Advancement of Science}
}

@article{koenecke2020racial,
  author = {Koenecke, Allison and Nam, Andrew and Lake, Emily and Nudell, Joe and Quartey, Minnie and Mengesha, Zion and Toups, Connor and Rickford, John R. and Jurafsky, Dan and Goel, Sharad},
  title = {Racial {D}isparities in {A}utomated {S}peech {R}ecognition},
  journal = {Proceedings of the National Academy of Sciences},
  volume = {117},
  number = {14},
  pages = {7684--7689},
  year = {2020},
  doi = {10.1073/pnas.1915768117}
}

@inproceedings{buolamwini2018gendershades,
  title     = {Gender {S}hades: {I}ntersectional {A}ccuracy {D}isparities in {C}ommercial {G}ender {C}lassification},
  author    = {Buolamwini, Joy and Gebru, Timnit},
  booktitle = {Proceedings of the 1st Conference on Fairness, Accountability and Transparency},
  series    = {Proceedings of Machine Learning Research},
  volume    = {81},
  pages     = {77--91},
  year      = {2018},
  editor    = {Friedler, Sorelle A. and Wilson, Christo},
  publisher = {PMLR},
  url       = {https://proceedings.mlr.press/v81/buolamwini18a.html}
}

@techreport{ccj2025principles,
  author       = {{Council on Criminal Justice}},
  title        = {Principles for the {U}se of {A}rtificial {I}ntelligence in {C}riminal {J}ustice},
  institution  = {Council on Criminal Justice},
  year         = {2025},
  month        = oct,
  url          = {https://counciloncj.org/principles-for-the-use-of-ai-in-criminal-justice/}
}

@article{magesh2025hallucination,
  title={Hallucination-free? {A}ssessing the {R}eliability of {L}eading {AI} {L}egal {R}esearch {T}ools},
  author={Magesh, Varun and Surani, Faiz and Dahl, Matthew and Suzgun, Mirac and Manning, Christopher D and Ho, Daniel E},
  journal={Journal of empirical legal studies},
  volume={22},
  number={2},
  pages={216--242},
  year={2025},
  publisher={Wiley Online Library}
}

@inproceedings{raji2020closing,
  author = {Raji, Inioluwa Deborah and Smart, Andrew and White, Rebecca N. and Mitchell, Margaret and Gebru, Timnit and Hutchinson, Ben and Smith-Loud, Jamila and Theron, Daniel and Barnes, Parker},
  title = {Closing the {AI} {A}ccountability {G}ap: {D}efining an {E}nd-to-{E}nd {F}ramework for {I}nternal {A}lgorithmic {A}uditing},
  booktitle = {Proceedings of the 2020 Conference on Fairness, Accountability, and Transparency},
  pages = {33--44},
  year = {2020},
  doi = {10.1145/3351095.3372873}
}

@inproceedings{raji2019actionable,
  title={Actionable {A}uditing: {I}nvestigating the {I}mpact of {P}ublicly {N}aming {B}iased {P}erformance {R}esults of {C}ommercial {AI} {P}roducts},
  author={Raji, Inioluwa Deborah and Buolamwini, Joy},
  booktitle={Proceedings of the 2019 AAAI/ACM Conference on AI, Ethics, and Society},
  pages={429--435},
  year={2019}
}

@article{zheng2023judging,
  title={Judging {LLM}-as-a-judge with mt-bench and chatbot arena},
  author={Zheng, Lianmin and Chiang, Wei-Lin and Sheng, Ying and Zhuang, Siyuan and Wu, Zhanghao and Zhuang, Yonghao and Lin, Zi and Li, Zhuohan and Li, Dacheng and Xing, Eric and others},
  journal={Advances in neural information processing systems},
  volume={36},
  pages={46595--46623},
  year={2023}
}

@article{barrie2026ai,
  title={{AI} and {R}esearch {M}ethods},
  author={Barrie, Christopher and Argyle, Lisa and Bisbee, James and Heseltine, Michael and Lucas, Christopher and Mellon, Jon and Palmer, Alexis and Roberts, Margaret and Spirling, Arthur},
  year={2026}
}

@article{hullman2026human,
  title={This human study did not involve human subjects: {V}alidating {LLM} simulations as behavioral evidence},
  author={Hullman, Jessica and Broska, David and Sun, Huaman and Shaw, Aaron},
  journal={arXiv preprint arXiv:2602.15785},
  year={2026}
}

@article{macdonald2020analysis,
    author = {MacDonald, John and Raphael, Steven},
    title = {Effect of scaling back punishment on racial and ethnic disparities in criminal case outcomes},
    journal = {Criminology \& Public Policy},
    year = 2020,
    doi = {https://doi.org/10.1111/1745-9133.12495},
    url = {https://onlinelibrary.wiley.com/doi/abs/10.1111/1745-9133.12495},
    eprint = {https://onlinelibrary.wiley.com/doi/pdf/10.1111/1745-9133.12495}
}

@book{marquand2026auditing,
  title     = {Auditing {AI}},
    author = {Aidinoff, Marc and
              Armstrong, Lena and
              Bhandari, Esha and
              Biddle, Ellery Roberts and
              Eslami, Motahhare and
              Karahalios, Karrie and
              Matias, J. Nathan and
              Metaxa, Dana{\'e} and
              Nelson, Alondra and
              Sandvig, Christian and
              Vaccaro, Kristen},
    organization = {The Marquand House Collective},  
  year      = {2026},
  publisher = {The MIT Press},
  series    = {The MIT Press Essential Knowledge series},
  address   = {Cambridge, MA},
  pages     = {204},
  isbn      = {9780262051729},
  doi       = {10.7551/mitpress/15997.001.0001}
}

@inproceedings{radford2023robust,
  title = {Robust speech recognition via large-scale weak supervision},
  author = {Radford, Alec and Kim, Jong Wook and Xu, Tao and Brockman, Greg and McLeavey, Christine and Sutskever, Ilya},
  booktitle = {Proceedings of the 40th International Conference on Machine Learning},
  series = {Proceedings of Machine Learning Research},
  volume = {202},
  pages = {28492--28518},
  year = {2023},
  publisher = {PMLR},
  url = {https://proceedings.mlr.press/v202/radford23a.html},
  doi = {10.5555/3618408.3619590}
}

@misc{speechmatics2021breakthrough,
  title = {Speechmatics achieves {AI} breakthrough, beating tech giants in race to reduce bias and improve inclusion in speech recognition},
  author = {{Speechmatics}},
  year = {2021},
  month = oct,
  day = {25},
  howpublished = {\url{https://www.speechmatics.com/company/articles-and-news/breakthrough-ai-bias-inclusion}},
}

@article{kozlowski2025simulating,
  title={Simulating {S}ubjects: {T}he {P}romise and {P}eril of {A}rtificial {I}ntelligence {S}tand-{I}ns for {S}ocial {A}gents and {I}nteractions},
  author={Kozlowski, Austin C and Evans, James},
  journal={Sociological Methods \& Research},
  volume={54},
  number={3},
  pages={1017--1073},
  year={2025},
  publisher={SAGE Publications Sage CA: Los Angeles, CA}
}

@misc{shankar2025docetl,
      title={DocETL: Agentic Query Rewriting and Evaluation for Complex Document Processing}, 
      author={Shreya Shankar and Tristan Chambers and Tarak Shah and Aditya G. Parameswaran and Eugene Wu},
      year={2025},
      eprint={2410.12189},
      archivePrefix={arXiv},
      primaryClass={cs.DB},
      url={https://arxiv.org/abs/2410.12189}, 
}

\clearpage

\appendix
\counterwithin{figure}{section}
\counterwithin{table}{section}
\counterwithin{equation}{section}

\section{Application}
\label{apx:app}

To support deployment in demanding production environments, we developed a separate application, \texttt{blind-charging-api}, that applies \texttt{bc2} within existing prosecutorial workflows. 
The application was designed as a backend service that integrates directly with case management systems (CMSs), allowing prosecutors to continue working within the software they already use rather than requiring a separate user interface. 
The application is deployed on Microsoft Azure because many justice agencies already rely on Azure infrastructure, allowing the application to operate within secure Azure Government Cloud regions. 
Security and privacy considerations were incorporated throughout its design, including the exclusive use of Azure-hosted language models and the automatic deletion of all report data within 4 hours. 
The software is open source~\citep{cpl2026api} and was integrated with multiple prosecutorial CMS platforms, including Karpel Technologies' ProsecutorByKarpel, Journal Technologies’ eProsecutor, HTC's CIBER platform, and custom case management systems such as those used by Los Angeles County. 

\section{California usage statistics, 2025}
\label{apx:usage_stats}

Table~\ref{tab:rbc_solutions}, Table~\ref{tab:usage_stats}, and Table~\ref{tab:rbc_unable_reasons} compile usage information for \texttt{bc2} 
employed across California in 2025,
the first year of the statewide mandate. 
We asked offices to describe how they implemented race-blind charging, including the cases covered by the process, whether redaction was performed manually or through an automated system, the solution used, and any associated validation procedures. 
We also requested case-level usage data, including the outcomes of initial and final charging reviews, explanations for changes between the two reviews, and the cases in which an RBC evaluation could not be completed and the reasons why. 
The tables summarize the redaction solutions adopted by responding offices, the number and outcomes of cases processed using \texttt{bc2}, and the categories of explanations offices provided for unsuccessful or incomplete reviews. 
This information was provided under state law \citet{ab2778}, which requires annual usage information to be collected and furnished upon request to bona fide accredited educational institutions for bona fide research purposes.

\clearpage

\begin{longtable}{@{}ll@{}}
\label{tab:rbc_solutions} \\

\toprule
Office & Solution \\
\midrule
\endfirsthead

\multicolumn{2}{@{}l}{\tablename\ \thetable\ continued} \\
\toprule
Office & Solution \\
\midrule
\endhead

\midrule
\multicolumn{2}{r}{\textit{Continued on next page}} \\
\endfoot

\bottomrule
\noalign{\vskip 6pt}
\caption{
\textit{
Race-blind charging implementation status and software solutions across California prosecutor offices, as identified through public records requests.
}
}
\label{tab:rbc_solutions} \\
\endlastfoot

Alameda
    & Not Participating \\
Alpine
    & Not Participating \\
Amador
    & No response \\
Butte
    & Sicuro \\
Calaveras
    & \texttt{bc2} \\
Colusa
    & No response \\
Contra Costa
    & \texttt{bc2} \\
Del Norte
    & \texttt{bc2} \\
El Dorado
    & \texttt{bc2} \\
Fresno
    & Meadowlark \\
Glenn
    & Not Participating \\
Humboldt
    & \texttt{bc2} \\
Imperial
    & \texttt{bc2} \\
Inyo
    & \texttt{bc2} \\
Kern
    & \texttt{bc2} \\
Kings
    & \texttt{bc2} \\
Lake
    & Meadowlark \\
Lassen
    & \texttt{bc2} \\
Los Angeles
    & \texttt{bc2} \\
Madera
    & \texttt{bc2} \\
Marin
    & Sicuro \\
Mariposa
    & \texttt{bc2} \\
Mendocino
    & \texttt{bc2} \\
Merced
    & Sicuro \\
Modoc
    & \texttt{bc2} \\
Mono
    & \texttt{bc2} \\
Monterey
    & \texttt{bc2} \\
Napa
    & Sicuro \\
Nevada
    & \texttt{bc2} \\
Orange
    & Internally Developed \\
Placer
    & Sicuro \\
Plumas
    & \texttt{bc2} \\
Riverside
    & Internally Developed \\
Sacramento
    & No response \\
San Benito
    & \texttt{bc2} \\
San Bernardino
    & Internally Developed \\
San Diego
    & Sicuro \\
San Francisco
    & Not Participating \\
San Joaquin
    & Sicuro \\
San Luis Obispo
    & \texttt{bc2} \\
San Mateo
    & \texttt{bc2} \\
Santa Barbara
    & Sicuro \\
Santa Clara
    & \texttt{bc2} \\
Santa Cruz
    & \texttt{bc2} \\
Shasta
    & Meadowlark \\
Sierra
    & No response \\
Siskiyou
    & No response \\
Solano
    & Not Participating \\
Sonoma
    & \texttt{bc2} \\
Stanislaus
    & Sicuro \\
Sutter
    & Sicuro \\
Tehama
    & \texttt{bc2} \\
Trinity
    & \texttt{bc2} \\
Tulare
    & No response \\
Tuolumne
    & \texttt{bc2} \\
Ventura
    & Internally Developed \\
Yolo
    & \texttt{bc2} \\
Yuba
    & No response \\

\end{longtable}

\clearpage

\begin{table}
\centering

\begin{tabular}{@{}lllllll@{}}
\toprule
& & & \multicolumn{3}{c}{Failed reviews} \\
\cmidrule(lr){4-6}
Office &
\makecell[l]{RBC\\referrals} &
\makecell[l]{Successful\\reviews} &
\makecell[l]{App\\error} &
\makecell[l]{RBC\\unable} &
\makecell[l]{Case\\ineligible} \\
\midrule
Calaveras
    & 903
    & 433
    & 345
    & 99
    & 26 \\
Contra Costa
    & 17,933
    & 14,967
    & 101
    & 2276
    & 589 \\
El Dorado
    & 2,367
    & 2,140
    & 31
    & 166
    & 30 \\
Inyo
    & 1,115
    & 732
    & 67
    & 221
    & 95 \\
Kings
    & 2,389
    & 2,067
    & 20
    & 293
    & 9 \\
Lassen
    & 943
    & 662
    & 20
    & 90
    & 171 \\
Los Angeles
    & 93,091
    & 58,625
    & ---
    & ---
    & --- \\
Monterey
    & 9,119
    & 7,169
    & 142
    & 1698
    & 110 \\
Nevada
    & 2,732
    & 2,321
    & 55
    & 350
    & 6 \\
Plumas
    & 897
    & 814
    & 4
    & 66
    & 13 \\
San Luis Obispo
    & 10,737
    & 10,734
    & 0
    & 3
    & 0 \\
San Mateo
    & 5,000
    & 3,738
    & 106
    & 980
    & 176 \\
Santa Cruz
    & 9,741
    & 8,568
    & 121
    & 1001
    & 51 \\
Sonoma
    & 3,230
    & 2,801
    & 61
    & 274
    & 94 \\
Tehama
    & 1,917
    & 1,711
    & 27
    & 143
    & 36 \\
Trinity
    & 42
    & 41
    & 1
    & 0
    & 0 \\
Tuolumne
    & 1,902
    & 1,507
    & 26
    & 312
    & 57 \\
\bottomrule
\end{tabular}

\caption{
\textit{
Case counts for offices that provided usage data and use \texttt{bc2}, including cases referred to race-blind charging (RBC) and the outcomes of those referrals, such as successful reviews and various types of failed reviews.
}
}
\label{tab:usage_stats}
\end{table}

\clearpage

\begingroup

\small
\setlength{\LTleft}{\fill}
\setlength{\LTright}{\fill}
\setlength{\LTcapwidth}{0.9\textwidth}
\setlength{\LTpre}{0pt}
\setlength{\LTpost}{0pt}
\renewcommand{\arraystretch}{1.05}

\begin{longtable}{
  @{}
  >{\raggedright\arraybackslash}p{0.22\textwidth}
  >{\raggedright\arraybackslash}p{0.64\textwidth}
  r
  @{}
}

\toprule
\textbf{Office} &
\textbf{Reason RBC review was unable to proceed} &
\textbf{Count} \\
\midrule
\endfirsthead

\toprule
\textbf{Office} &
\textbf{Reason RBC review was unable to proceed} &
\textbf{Count} \\
\midrule
\endhead

\midrule
\multicolumn{3}{r@{}}{\small\itshape Continued on next page} \\
\endfoot

\endlastfoot

  \addlinespace[0.65em]

  \textbf{Calaveras} &
  Prior knowledge / saw identifying information &
  42
  \\*

  &
  Narrative incomplete or missing &
  35
  \\*

  &
  Need more information / evidence &
  1
  \\*

  &
  Multiple suspects / unclear person &
  2
  \\*

  &
  Redaction missed race-related information &
  5
  \\*

  &
  Redacted narrative hard to read &
  12
  \\*

  &
  RBC process / administrative failure &
  1
  \\*

  &
  Other &
  8
  \\

  \addlinespace[0.65em]

  \textbf{Contra Costa} &
  Prior knowledge / saw identifying information &
  1295
  \\*

  &
  Narrative incomplete or missing &
  229
  \\*

  &
  Need more information / evidence &
  18
  \\*

  &
  Multiple suspects / unclear person &
  21
  \\*

  &
  Redaction missed race-related information &
  678
  \\*

  &
  Redacted narrative hard to read &
  41
  \\*

  &
  RBC process / administrative failure &
  43
  \\*

  &
  Other &
  29
  \\

  \addlinespace[0.65em]

  \textbf{El Dorado} &
  Prior knowledge / saw identifying information &
  138
  \\*

  &
  Narrative incomplete or missing &
  8
  \\*

  &
  Need more information / evidence &
  1
  \\*

  &
  Multiple suspects / unclear person &
  2
  \\*

  &
  Redaction missed race-related information &
  9
  \\*

  &
  Redacted narrative hard to read &
  7
  \\*

  &
  RBC process / administrative failure &
  0
  \\*

  &
  Other &
  6
  \\

  \addlinespace[0.65em]

  \textbf{Inyo} &
  Prior knowledge / saw identifying information &
  158
  \\*

  &
  Narrative incomplete or missing &
  40
  \\*

  &
  Need more information / evidence &
  9
  \\*

  &
  Multiple suspects / unclear person &
  0
  \\*

  &
  Redaction missed race-related information &
  4
  \\*

  &
  Redacted narrative hard to read &
  8
  \\*

  &
  RBC process / administrative failure &
  0
  \\*

  &
  Other &
  2
  \\

  \addlinespace[0.65em]

  \textbf{Kings} &
  Prior knowledge / saw identifying information &
  88
  \\*

  &
  Narrative incomplete or missing &
  55
  \\*

  &
  Need more information / evidence &
  31
  \\*

  &
  Multiple suspects / unclear person &
  7
  \\*

  &
  Redaction missed race-related information &
  4
  \\*

  &
  Redacted narrative hard to read &
  54
  \\*

  &
  RBC process / administrative failure &
  0
  \\*

  &
  Other &
  67
  \\

  \addlinespace[0.65em]

  \textbf{Lassen} &
  Prior knowledge / saw identifying information &
  22
  \\*

  &
  Narrative incomplete or missing &
  62
  \\*

  &
  Need more information / evidence &
  1
  \\*

  &
  Multiple suspects / unclear person &
  1
  \\*

  &
  Redaction missed race-related information &
  0
  \\*

  &
  Redacted narrative hard to read &
  2
  \\*

  &
  RBC process / administrative failure &
  0
  \\*

  &
  Other &
  2
  \\

  \addlinespace[0.65em]

  \textbf{Monterey} &
  Prior knowledge / saw identifying information &
  637
  \\*

  &
  Narrative incomplete or missing &
  563
  \\*

  &
  Need more information / evidence &
  0
  \\*

  &
  Multiple suspects / unclear person &
  12
  \\*

  &
  Redaction missed race-related information &
  297
  \\*

  &
  Redacted narrative hard to read &
  418
  \\*

  &
  RBC process / administrative failure &
  2
  \\*

  &
  Other &
  3
  \\

  \addlinespace[0.65em]

  \textbf{Nevada} &
  Prior knowledge / saw identifying information &
  134
  \\*

  &
  Narrative incomplete or missing &
  124
  \\*

  &
  Need more information / evidence &
  58
  \\*

  &
  Multiple suspects / unclear person &
  0
  \\*

  &
  Redaction missed race-related information &
  13
  \\*

  &
  Redacted narrative hard to read &
  36
  \\*

  &
  RBC process / administrative failure &
  0
  \\*

  &
  Other &
  15
  \\

  \addlinespace[0.65em]

  \textbf{Plumas} &
  Prior knowledge / saw identifying information &
  4
  \\*

  &
  Narrative incomplete or missing &
  11
  \\*

  &
  Need more information / evidence &
  33
  \\*

  &
  Multiple suspects / unclear person &
  0
  \\*

  &
  Redaction missed race-related information &
  0
  \\*

  &
  Redacted narrative hard to read &
  0
  \\*

  &
  RBC process / administrative failure &
  0
  \\*

  &
  Other &
  19
  \\

  \addlinespace[0.65em]

  \textbf{San Luis Obispo} &
  Prior knowledge / saw identifying information &
  0
  \\*

  &
  Narrative incomplete or missing &
  1
  \\*

  &
  Need more information / evidence &
  0
  \\*

  &
  Multiple suspects / unclear person &
  0
  \\*

  &
  Redaction missed race-related information &
  2
  \\*

  &
  Redacted narrative hard to read &
  0
  \\*

  &
  RBC process / administrative failure &
  0
  \\*

  &
  Other &
  0
  \\

  \addlinespace[0.65em]

  \textbf{San Mateo} &
  Prior knowledge / saw identifying information &
  454
  \\*

  &
  Narrative incomplete or missing &
  263
  \\*

  &
  Need more information / evidence &
  150
  \\*

  &
  Multiple suspects / unclear person &
  24
  \\*

  &
  Redaction missed race-related information &
  78
  \\*

  &
  Redacted narrative hard to read &
  48
  \\*

  &
  RBC process / administrative failure &
  5
  \\*

  &
  Other &
  27
  \\

  \addlinespace[0.65em]

  \textbf{Santa Cruz} &
  Prior knowledge / saw identifying information &
  323
  \\*

  &
  Narrative incomplete or missing &
  145
  \\*

  &
  Need more information / evidence &
  371
  \\*

  &
  Multiple suspects / unclear person &
  36
  \\*

  &
  Redaction missed race-related information &
  12
  \\*

  &
  Redacted narrative hard to read &
  52
  \\*

  &
  RBC process / administrative failure &
  24
  \\*

  &
  Other &
  103
  \\

  \addlinespace[0.65em]

  \textbf{Sonoma} &
  Prior knowledge / saw identifying information &
  124
  \\*

  &
  Narrative incomplete or missing &
  43
  \\*

  &
  Need more information / evidence &
  49
  \\*

  &
  Multiple suspects / unclear person &
  15
  \\*

  &
  Redaction missed race-related information &
  12
  \\*

  &
  Redacted narrative hard to read &
  35
  \\*

  &
  RBC process / administrative failure &
  0
  \\*

  &
  Other &
  16
  \\

  \addlinespace[0.65em]

  \textbf{Tehama} &
  Prior knowledge / saw identifying information &
  95
  \\*

  &
  Narrative incomplete or missing &
  8
  \\*

  &
  Need more information / evidence &
  5
  \\*

  &
  Multiple suspects / unclear person &
  21
  \\*

  &
  Redaction missed race-related information &
  4
  \\*

  &
  Redacted narrative hard to read &
  10
  \\*

  &
  RBC process / administrative failure &
  1
  \\*

  &
  Other &
  19
  \\

  \addlinespace[0.65em]

  \textbf{Tuolumne} &
  Prior knowledge / saw identifying information &
  166
  \\*

  &
  Narrative incomplete or missing &
  91
  \\*

  &
  Need more information / evidence &
  43
  \\*

  &
  Multiple suspects / unclear person &
  9
  \\*

  &
  Redaction missed race-related information &
  20
  \\*

  &
  Redacted narrative hard to read &
  28
  \\*

  &
  RBC process / administrative failure &
  0
  \\*

  &
  Other &
  7
  \\

\bottomrule
\noalign{\vskip 1.25em}

\caption{
\textit{
Breakdown by office of the RBC unable counts shown in
Table~\ref{tab:usage_stats}.
Categories aggregate the free-text explanations provided by
participating offices.
}
}
\label{tab:rbc_unable_reasons}
\\

\end{longtable}

\endgroup

\section{Records cataloguing}
\label{apx:prr_cataloguing}
Constructing a realistic validation sample through public records requests posed several challenges for our study. 
Jurisdictions varied substantially in their willingness and ability to provide records, partly because state laws differed in the extent to which police reports could be released. 
Even among jurisdictions that fulfilled our requests, the form and content of the records were highly inconsistent. 
Some agencies combined many individual records into a single PDF, while others provided separate files. The materials included also differed across jurisdictions, with some responses containing witness statements, computer-aided dispatch reports, or other supplementary documents in addition to the primary police report. 
Responses could also include administrative or meta-documents describing the public records request itself, which needed to be distinguished from the underlying incident records.

\begin{figure}[htp]
\centering
\includegraphics[width=\linewidth]{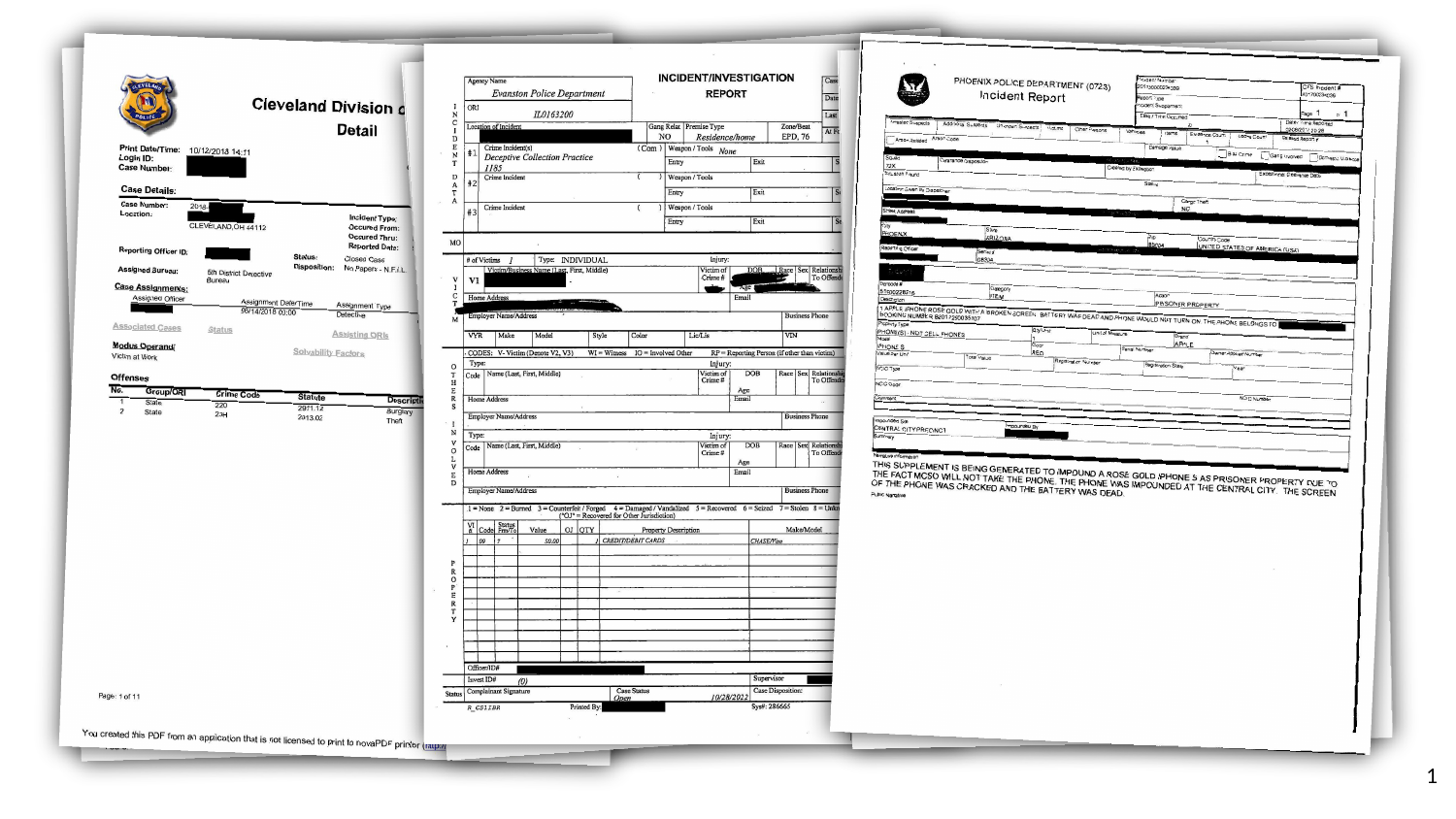}
\caption{
\textit{
Examples of police reports obtained through public records requests, with personally identifying information removed.
}
}
\label{fig:reports}
\end{figure}

These inconsistencies required us to manually construct an inventory of the documents provided in each request. 
The inventory allowed us to identify and extract individual reports embedded within larger PDF files and prepare them for processing with \texttt{bc2}. 
For each document, we recorded information such as the authoring agency, document type, page length, and whether the document contained handwritten material. 
We then used this inventory to construct samples for our downstream analyses.

\subsection{Pre-existing redactions}
\label{apx:preexisting_redactions}

During the course of our analysis,
it became clear that pre-existing redactions---such as black boxes over names, which were imposed to protect identifying information under public records requirements---often interfered with Azure Document Intelligence's ability to recognize a continuous line of text from the underlying PDF,
cascading into further errors downstream.
At the same time, 
these pre-existing redactions also removed information 
that is a target for \texttt{bc2} redaction,
potentially biasing our coverage estimates to appear better 
than they might be in practice.

To address this issue, we developed a page-level classifier to identify documents that had been pre-redacted. 
We rendered each page of every report as an image and submitted the page image to GPT-5.5, which classified whether it contained visible evidence of pre-existing redactions using the prompt reproduced in Appendix~\ref{apx:preredaction_prompt}. 
We then aggregated these page-level classifications to the document level. 
Documents with detected pre-existing redactions on at least one page were excluded from our coverage sample.

We verified the performance of this automatic labeling process by manually examining a small random sample of records.
The classifier achieved a document level recall of 0.925 and precision of 0.649, which was sufficient for excluding records with pre-existing redactions from our coverage sample.

\section{Automated extraction of arrestee race}
\label{apx:arrestee_race}
To assess the accuracy of our automated extraction of arrestee race, we first process each report using Azure Document Intelligence's prebuilt-layout, which extracts the report text and enriches it with structural annotations identifying elements such as paragraphs, tables, and key-value pairs. We then use GPT-5.5 with high reasoning to map this layout analysis onto a structured schema representing common categories of information in police reports, including the names and demographic characteristics of involved individuals. To evaluate the resulting extraction, we compared the automatically identified arrestees and race labels against a hand-labeled sample of 160 reports. After fuzzy-matching names and normalizing race labels, we measured performance by race and obtained a recall of 0.928 and a precision of 0.923.

\section{Narrative extraction performance}
\label{apx:narr_extraction}

To assess the performance of \texttt{bc2}'s narrative extraction step, 
we manually labeled the location and content of free-text narratives across 3,140 records and measured whether our automated process successfully recovered these narratives. 
We define a free-text narrative as unstructured prose describing the incident, including officers' observations, actions, and summaries of statements or events. 
We exclude structured fields, administrative metadata, document headers, and boilerplate material that does not substantively describe the incident.

We find that \texttt{bc2} achieves a recall of
1.000 on the median record, with recall reaching at least 0.899 on 90\% of records. The remaining missed content 
are typically borderline cases for inclusion in a narrative,
e.g., unimportant signatory blocks 
at the end of a narrative.
Additionally, \texttt{bc2} sufficiently ignores non-narrative content across most reports: extracted text has a median precision of 0.954, with precision exceeding 0.800 on 90\% of records.
Most extraneous information is similarly borderline narrative material, such as signature blocks or officer certifications typically included immediately after a narrative.

\section{Design-based supervised learning estimation}
\label{apx:dsl}
To construct the human-validation sample, 
we began with the 4,879 documents for which the LLM critic produced assessments.
We classified documents containing fewer than 250 words as short and documents meeting or exceeding that threshold as long. For each document, we also summarized the LLM critic's binary assessments across four categories---person names, race or ethnicity, physical descriptions, and location information---separately for the original input and the redacted output. 
We then selected documents using independent Bernoulli sampling within the short- and long-document strata.
For feasibility, 
the design targeted 100 documents, 
selecting 80 of the 2081 short documents and 20 of the 2798 long documents for review.
This corresponded to sampling probabilities of approximately 0.0384 and 0.00715, respectively. 
Because documents were selected using independent Bernoulli draws, the realized sample contained 103 documents: 78 short documents and 25 long documents.

A human reviewer assessed each sampled document for all four categories, separately for the input and redacted output, producing eight binary human labels per document.
We combined these human labels with the LLM critic’s assessments using intercept-only logistic DSL models and incorporated the documents’ known sampling probabilities to account for the disproportionate sampling of short documents.

\section{Race-prediction}

\subsection{Embedding-based approach}
\label{apx:embedding_approach}

We calculate the following demographic information for each jurisdiction
to compare to our embedding-based model runs.
First, we calculate the jurisdiction's
Asian, Black, Hispanic, and White residents
as a proportion of the jurisdiction's total residential population,
drawing from the U.S. Census in the 5-year ACS.
We also provide a weighted prior estimate 
of the proportion of records in the training sample from that jurisdiction
where the arrestee is White,
calculated as follows:
\[
\widehat{p}^{\,(-i)}_{j(i)}
=
\frac{W^{(-i)}_{j(i)} + \lambda \bar{W}^{(-i)}}{N^{(-i)}_{j(i)} + \lambda}.
\]
Here \(j(i)\) is the jurisdiction for record \(i\);
\(N^{(-i)}_{j(i)}\) is the number of other model-sample records from that jurisdiction
in the leave-one-out training fold;
\(W^{(-i)}_{j(i)}\) is the number of those records whose arrestee is labeled White;
\(\bar{W}^{(-i)}\) is the overall White arrestee share in the same training fold;
and \(\lambda = 10\) is the shrinkage weight.
If a jurisdiction has no records in the training fold, the prior is set to
\(\bar{W}^{(-i)}\).

To calculate within-jurisdiction AUCs,
we are required to subsample the ``prediction sample'' in two ways.
First, 
we limit to jurisdictions with at least one White and non-White arrestee,
so that we can calculate a valid AUC for each site.
Additionally, 
we limit to jurisdictions with 10 or more records in our sample,
to avoid extreme and noisy AUCs from small sites.

\section{Prompts}

\subsection{\texttt{bc2} narrative extraction prompt}
\label{apx:bc2_extract_prompt}

\begin{flushleft}\footnotesize\ttfamily
\# Your task \\
Review the text from a police report extracted using OCR. Extract ALL full sentences, paragraphs, or blocks of freely-written text. \\
\mbox{} \\
You are looking for any text that is coherent and discusses crime details, related evidence, legal justifications, or testimony from people involved. These may include narratives, synopses, probable cause statements, legal documents, affidavits, formal legal entries, and first-person accounts. In essence, any text providing a descriptive or narrative account in full sentences should be returned, even if it is embedded within an otherwise structured sections. \\
\mbox{} \\
\# What to include \\
- Extract freely-written text marked with labels like "Narrative", "Supplement", "Synopsis", "Statement", "Affadavit", "Transcript", or "Probable Cause Statement." Include labels in the output. \\
- Also be sure to include unlabeled freely-written text if it clearly fits the criteria above. Sometimes the input may start in the middle of a narrative; if so, include that text as well. \\
- Everything from a single standalone (full) sentence to a multi-page narrative should be included. \\
- Stitch freely-written text across pages by removing page headers and page footers. Ensure all freely-written text is extracted in its entirety. \\
- Include ALL freely-written text included in the input, EVEN IF IT IS DUPLICATED IN THE INPUT. NEVER OMIT DUPLICATES for ANY REASON whatsoever. It is CRUCIAL to preserve redundancy from the input IF it exists in the input, even if this might seem unnecessarily repetitive. \\
- You will be punished with a \$1,000 fine if you do not return any freely-written text. And you will be fined an extra \$2,000 if you miss the last narrative in the provided input! \\
\mbox{} \\
\# What to exclude \\
- Exclude structured information (e.g., fields/values like DOBs, addresses, sex, race, dates, etc.). which will constitute most of the report. These are often provided in long unstructured lists with each field or value on their own line. \\
- Exclude field/value pairs that appear at the end of a narrative. \\
- Exclude most "CAD Narratives" except parts that form coherent sentences or paragraphs. \\
- Do not correct grammar or spelling. \\
- Do not create new text. \\
\mbox{} \\
\# Formatting \\
- Return plain-text paragraphs with labels if available. Use dividers (e.g., "------------------------------------------------------") between each distinct body of text. \\
- Examine the entire input for freely-written text. \\
\mbox{} \\
If no freely-written text is found, return "No narratives found." \\
\mbox{} \\
Go slowly and triple-check your work. It is crucial that you faithfully extract any freely-written text contained in the input.
\end{flushleft}

\subsection{\texttt{bc2} race-redaction prompt}
\label{apx:bc2_redact_prompt}

\begin{flushleft}\footnotesize\ttfamily
Your job is to redact all race-related information in the provided text. Race-related information is any word from the following categories: \\
- Explicit mentions of race, ethnicity, nationality, or language (including racial or ethnic slurs, such as the N-word or derivations and any other slurs; and race/gender abbreviations like W/F or B/M) \\
- People\textquotesingle{}s names, nicknames, and social media handles, officer IDs, and signature lines. \\
- Physical descriptions: ONLY hair color, hairstyle, eye color, or skin color \\
- Location information: Addresses, streets, apartment numbers, intersections, zip codes, store names, restaurant names,  major landmarks, neighborhood names, city names, park names, and police department names, beat IDs, mileposts, and abbreviations \\
\mbox{} \\
Do NOT redact ANY other types of information, e.g., do not redact dates, objects, or colors that do not refer to a person\textquotesingle{}s skin tone, or other types of entities (e.g., vehicles) not explicitly listed above. \\
\mbox{} \\
Replace all names specified by the \textasciigrave{}RealName\textasciigrave{} element in the following XML with the pre-specified placeholder given by the \textasciigrave{}ReplacementText\textasciigrave{} element. Do not change this placeholder in any way, use it exactly as it was provided. Consider variants of the \textasciigrave{}RealName\textasciigrave{} if they refer to the same person. \\
\{placeholders\} \\
\mbox{} \\
For other any other person\textquotesingle{}s name or nickname, replace the name or nickname with a placeholder in angular brackets indicating the person\textquotesingle{}s role in the incident, and a number indicating the order the person appeared in the document. For example, for the first mentioned victim, use "[Victim 1]". Then for the second mentioned victim, use "[Victim 2]". If you cannot UNIQUELY identify a person when they are mentioned, DO NOT add a number to the placeholder. E.g., "then victim, victim, and victim reported the crime" should become "then [Victim], [Victim], and [Victim] reported the crime. You should redact the names and nicknames of every single person mentioned. By the time you are done, there should be zero person names and nicknames remaining in the redacted text. \\
\mbox{} \\
When guessing roles in this fashion, be as specific as possible about a person\textquotesingle{}s role (e.g., "Officer Smith and Sergeant Doe" should become "[Officer 1] and [Sergeant 1]"). As a hint, roles are often abbreviated in police reports, e.g., "S1" refers to "Suspect 1", and "V1" refers to "Victim 1". Other common roles are "Witness" (W), "Accused" (A), "Reporting Party" (RP), or "Reporting Victim" (RV). If a person\textquotesingle{}s role in the incident is never mentioned, use a generic "[Man X]" or "[Woman X]" (with X indicating a number when it is appropriate to include). If no gender information is present, you can use "[Person X]" instead. If their role is explicitly mentioned alongside their name, replace everything with a bracketed placeholder, e.g., the phrase "Victim \#9 Jane Doe met (S1) John Doe and went to a party" should become "[Victim 9] met [Suspect 1] and went to a party". \\
\mbox{} \\
Make sure all reporting officer ID, usernames, and signature lines are also redacted. \\
\mbox{} \\
Other listed categories should be redacted in a similar manner, with a generic placeholder like [Street X] or [Neighborhood X] replacing the redacted text. The X represents a counter for that specific redacted concept, NOT any other associated entities. For example, e.g., the phrase "Victim 1 (a black woman) and Victim 2 (an African-American man)" should become "[Victim 1] (a [Race 1] woman) and [Victim 2] (a [Race 1] man)". \\
\mbox{} \\
Be descriptive with the placeholders. For example, the phrase "her Afro hair" should become "her [Hairstyle X]". Make sure to capitalize the placeholder text in title case so it appears professional. \\
\mbox{} \\
You should keep an eye out for nicknames, variations, and social media handles. For example, if "John Doe" appears in the list of individuals, and then "Johnny D." appears in the narrative, use context to decide if "Johnny D." should be redacted with the same replacement as "John Doe." Or if "ANDRE JONES" is mentioned as suspect 1, and then later a witness says, "\textquotesingle{}Yes, it was Andre\textquotesingle{}" then "Andre" should be redacted as "[Suspect 1]". Similarly, if "Safeway" appears in the list of locations with abbreviation [Store 1], "Safeway Deli" should be redacted as "[Store 1] Deli". \\
\mbox{} \\
Race can be represented through abbreviations like "w/f" for White female, "B/M" for Black male, "AM" for Asian male, or "H/F" for Hispanic female. Sometimes gender may be listed before race (e.g., "M/H" for male Hispanic). These abbreviations MUST be treated as explicit race-related information and MUST be redacted. For example, "the suspect was B/M and the victim was described as a Hispanic female" should become "the suspect was [a Race 1 male] and the victim was described as a [Race 2] female." Similarly, "victim: WF" should become "victim: [a Race 1 female]." \\
\mbox{} \\
Make sure you redact race-related information even if it is part of quote. For example, the sentence "The witness said, \textquotesingle{}Yes, I knew them as as Skip or Skipper\textquotesingle{}" should become "The witness said, \textquotesingle{}Yes, I knew the suspect as [Suspect 1] or [Suspect 1].\textquotesingle{}" \\
\mbox{} \\
Foreign languages can also convey race or ethnicity. If you encounter a foreign language, redact it by translating and marking it as translated, for example mask "dos cervesas" as "[Translated: two beers]". \\
\mbox{} \\
For location information, redact the ENTIRE address or place name as a single unit. This includes every component of the address: house/building numbers, street names, apartment/unit/suite numbers, floor numbers. Represent the redacted address using its component parts inside the brackets: "123 Main St." should be redacted as "[Address 1 on Street 1]", and "123 Main St. Apt 2" should become "[Address 1 on Street 1, Apt 1]". NEVER leave any part of an address outside the brackets — e.g., do NOT produce "123 [Street 1] Apt 2" or "[Address 1 on Street 1] Apt 2". When a street is referenced later without a full address, you can reference just the street, e.g., "Then [Suspect 1] escaped on [Street 1]." ALWAYS redact block numbers and apartment numbers: "The 200 block of Richmond Street" should become "The [Block 1] of [Street 1]." \\
\mbox{} \\
Make sure to redact any information that can indicate location even if it is not an explicit address. For example, "downtown" should be replaced with "[Location 1]" and "Holly Park" should be replaced with "[Park 1]." Any names of businesses, neighborhoods, districts, beats, parks, rivers, lakes, and more should be redacted as they indicate location. \\
\mbox{} \\
DO NOT CHANGE ANY characters outside of these brackets. The ONLY changes you make to the narrative should occur entirely within brackets. This means that if you want to to fix a grammatical error, the fix should occur WITHIN brackets. For example, if the original text reads "Ms. Smith black bag went missing", the redacted text should read "[Person 1\textquotesingle{}s] black bag went missing", without an apostrophe. Or if the original text reads "The car went to the house of Ms. Jones" (with a missing period at the end of the sentence), the redacted text should read "The car went to the house of [Person 1.]" DO NOT modify the appearance of text, e.g., by changing capitalization. Also, do not add or subtract spacing between paragraphs. Finally, NEVER truncate the provided text; you should redact the entire block of text that was given to you. \\
\mbox{} \\
Take your time and triple-check your work, quality is better than moving quickly. Make sure to follow these instructions closely. \\
\mbox{} \\
Please provide the redacted text as simple plain-text paragraphs. If you do not find any narratives, or if you are not provided with any text, return "No narratives found." \\
\mbox{} \\
Do not provide any commentary.
\end{flushleft}

\subsection{Pre-redaction detection prompt}
\label{apx:preredaction_prompt}

\begin{flushleft}\footnotesize\ttfamily
You are a helpful assistant in a police  \\
department. Your job is to examine crime \\
reports that are being prepared for public release to see \\
if they have been redacted yet. You will be examining these \\
reports one page at at time. You should examine every page of \\
the report and check whether each page contains redactions. \\
Redactions often appear as: \\
- black boxes on the page covering text \\
- sometimes these black boxes are digitally generated \\
- othertimes black boxes appear to be drawn with a marker \\
- these black-box redactions may also include placeholder text, e.g., the legal statute that justifies \\
the redaction.  \\
- sometimes, instead of black boxes, they just appear as blank white space in the middle of a \\
sentence where a word would be. Blank white space counts as redaction only if it interrupts a  \\
line of text where characters should be (mid-sentence or mid-field), leaving an unnatural gap. \\
\mbox{} \\
Do NOT count whitespace that occurs at: \\
- line breaks \\
- paragraph ends \\
- margins \\
- tables \\
- signature blocks \\
- between sections \\
\mbox{} \\
Do not treat shadows, smudges, staple marks, hole punches, scanner streaks, or page borders as redactions unless they clearly cover text. \\
\mbox{} \\
Respond with \textasciigrave{}is\_redacted=true\textasciigrave{} if the page contains any text obscuration \\
consistent with redaction. \\
Respond with \textasciigrave{}is\_redacted=false\textasciigrave{} if there is no evidence of such redaction. \\
Only mark \textasciigrave{}true\textasciigrave{} if evidence strongly suggests redaction and you\textquotesingle{}re absolutely certain. \\
If there is no clear indication of redactions, mark \textasciigrave{}false\textasciigrave{}. \\
Do not provide any additional commentary.
\end{flushleft}

\subsection{LLM critic prompts}
\label{apx:critic_prompts}

\subsubsection{Critic prompt: Explicit mentions of race}

\begin{flushleft}\footnotesize\ttfamily
Your job is to inspect redacted text and flag any mentions of a person\textquotesingle{}s SPECIFIC race, ethnicity, language, nationality, or racial/ethnic slurs that were not correctly redacted.  \\
\mbox{} \\
Here are some examples of each category: \\
- Race/Ethnicity: \\
  - Black, White, Asian, Hispanic, Native, etc. \\
  - BM, WF, AM, etc. -- when used to describe a person \\
  - B/M, W/F, A/M, etc. \\
- Nationality/Language: \\
  - Greek, Turkish, French, etc. \\
  - Spanish, Hindi, Mandarin, etc. \\
  - Only flag "English" if it clearly refers to a person’s nationality or ethnicity, such as "an English national" or "the suspect is English." \\
- Racial or ethnic slurs: \\
  - N-word \\
  - Other slurs omitted for sensitivity \\
\mbox{} \\
By SPECIFIC, we mean that the text must allow you to clearly and confidently identify the person\textquotesingle{}s actual race or ethnicity. For example, if the text says "the suspect is black", this is a mention of a specific race. However, if the text says "the suspect is a [REDACTED] male", this is NOT a mention of a specific race, since you cannot clearly determine the suspect\textquotesingle{}s race from the text. \\
\mbox{} \\
Your assessment should never flag any of the following as mentions of race: \\
- NEVER flag colors mentioned outside of a racial context. For example, color used to describe the appearance of an object, like "the black car"; describe the appearance of bruise, like "he had a black eye", or even a metaphor, like "a white lie" should all NEVER be considered race-related. \\
- NEVER flag colors that describe non-human animals or inanimate objects. \\
- Be very careful about colors used to describe someone\textquotesingle{}s appearance. For example, if the text says "the suspect is wearing a black hoodie and a black belt", this DOES NOT describe their race! They could be a white person wearing these items, for example. \\
- Do not flag "English" when it refers only to the English language, such as "speaks English," "English-speaking," "English only," "translated into English," or "gave commands in English."  \\
\mbox{} \\
If you find a mention of a specific race, ethnicity, language, nationality, or racial/ethnic slur, indicate the specific race mentioned, one of: \\
- asian \\
- black \\
- hispanic \\
- native \\
- white \\
\mbox{} \\
Next, in a single sentence, explain why this excerpt is a mention of the specific race you indicated. \\
\mbox{} \\
Next, include the excerpt where you think there is a mention of a specific race, ethnicity, nationality, language, or slurs. \\
\mbox{} \\
Next, provide a structured leak classification with exactly these two fields: \\
- person\_role: who the leaked race, ethnicity, language, nationality, or slur refers to. \\
  Strongly prefer one of: suspect, victim, reporting\_party, witness, \\
  law\_enforcement, registered\_owner, driver, passenger, bystander, other\_person, \\
  unknown\_person. \\
- leak\_attribute: the specific attribute leaked. Strongly prefer one of: \\
  race\_label, race\_shorthand, ethnicity\_label, nationality, language, \\
  racial\_slur, ethnic\_slur. \\
If none of the preferred values fit a future leak, use a concise new snake\_case \\
value that is specific but not overly narrow. \\
\mbox{} \\
Finally, return TRUE if you find any such information.  \\
Return FALSE if you do not. \\
\mbox{} \\
For example, given the following excerpt: \\
"The suspect (a B/M, mid-30s, 5\textquotesingle{}10") fled the scene in a white sedan." \\
\mbox{} \\
You should return: \\
\textasciigrave{}\textasciigrave{}\textasciigrave{} \\
\{ \\
  "examples" = [ \\
\qquad{}\qquad{}\{ \\
\qquad{}\qquad{}  "race" = "black", \\
\qquad{}\qquad{}  "explanation" = "The suspect is identified as a black male because of the shorthand \textquotesingle{}B/M\textquotesingle{} used to describe the suspect.", \\
\qquad{}\qquad{}  "excerpt" = "The suspect (a B/M, mid-30s, 5\textquotesingle{}10") fled the scene in a white sedan.", \\
\qquad{}\qquad{}  "person\_role" = "suspect", \\
\qquad{}\qquad{}  "leak\_attribute" = "race\_shorthand" \\
\qquad{}\qquad{}\} \\
  ], \\
  "errors\_detected" = TRUE \\
\} \\
\textasciigrave{}\textasciigrave{}\textasciigrave{} \\
\mbox{} \\
Alternatively, given the following excerpt: \\
"The suspect (a [REDACTED] male, mid-30s, 5\textquotesingle{}10") fled the scene in a white sedan." \\
\mbox{} \\
You should return: \\
\textasciigrave{}\textasciigrave{}\textasciigrave{} \\
\{ \\
  "examples" = [], \\
  "errors\_detected" = FALSE \\
\} \\
\textasciigrave{}\textasciigrave{}\textasciigrave{} \\
\mbox{} \\
Look through the ENTIRE input, from beginning to end. You may be missing something if you don\textquotesingle{}t. It is very important you don\textquotesingle{}t miss even a single example. Go slowly and focus on being comprehensive and thorough, not fast. \\
\mbox{} \\
Do not provide any other commentary.
\end{flushleft}

\subsubsection{Critic prompt: Physical descriptions}
\begin{flushleft}\footnotesize\ttfamily
Your job is to inspect redacted text and flag any mentions of a person\textquotesingle{}s hair color, hair texture, racially/ethnically associated hairstyle, eye color, or skin tone that were not correctly redacted.  \\
\mbox{} \\
Here are some examples of each category: \\
- Hair color: \\
  - Blonde, brunette, black, etc. \\
- Hair texture: \\
  - Curly, straight, kinky, coiled, wavy, etc. \\
- Hairstyle: \\
  - Afro, dreadlocks/locs, cornrows, braids, etc. \\
- Eye color: \\
  - Blue, brown, green, etc. \\
  - Hazel \\
- Skin tone: \\
  - Light, dark, olive, etc. \\
  - White skin, black skin \\
  - Dark complexion, light complexion, etc. \\
  - Fair, tan, etc. \\
  - Vitiglio, albino, etc. \\
\mbox{} \\
Do NOT flag ordinary hair length, grooming, or styling details unless they reveal texture, color, or a clearly racialized hairstyle. For example, do not flag: bald/balding, short hair, long hair, hair up, ponytail, scraggly beard, shorter in front/longer in back, messy hair, shaved head, bangs, or facial hair length. \\
\mbox{} \\
Do NOT flag general race, ethnicity, nationality, language, or slur terms unless they are being used as a physical description. \\
\mbox{} \\
Please be careful to only flag mentions of these physical descriptions when they are used to describe these attributes for a person. For example, "the curly-haired dog" should NOT trigger a flag. \\
\mbox{} \\
Be careful to distinguish between actual descriptions of skin tone vs. other colors used to describe someone\textquotesingle{}s appearance. For example, the phrase "the suspect had a black eye" should NOT be considered a mention of a person\textquotesingle{}s skin tone, since it typically describes a bruised eye, not their normal skin tone. Similarly, descriptions like "flushed" or "red in the face" should NOT be treated as skin tone indicators. \\
\mbox{} \\
If you do not find any mentions of a person\textquotesingle{}s hair color, hairstyle, eye color, or skin tone, return FALSE alongside an empty examples list.  \\
\mbox{} \\
If you do find at least one such example, first return TRUE. Then in the examples list, provide for each example: \\
- The flagged excerpt \\
- In a single sentence, why this qualifies as one of the above categories \\
- A structured leak classification with exactly these two fields: \\
  - person\_role: who the physical description refers to. Strongly prefer one of: \\
\qquad{}\qquad{}suspect, victim, reporting\_party, witness, law\_enforcement, registered\_owner, \\
\qquad{}\qquad{}driver, passenger, bystander, other\_person, unknown\_person. \\
  - leak\_attribute: the specific attribute leaked. Strongly prefer one of: \\
\qquad{}\qquad{}hair\_color, hair\_style, hair\_length, hair\_texture, eye\_color, skin\_tone, \\
\qquad{}\qquad{}complexion, race\_shorthand\_as\_appearance. \\
  If none of the preferred values fit a future leak, use a concise new snake\_case \\
  value that is specific but not overly narrow. \\
\mbox{} \\
For example, if the following excerpt was part of your input: \\
"... The suspect had black hair, styled in dreadlocks. ..." \\
\mbox{} \\
You should return: \\
\textasciigrave{}\textasciigrave{}\textasciigrave{} \\
\{ \\
  "errors\_detected" = TRUE, \\
  "examples" = [ \\
\qquad{}\qquad{}\{ \\
\qquad{}\qquad{}  "excerpt" = "The suspect had black hair", \\
\qquad{}\qquad{}  "explanation" = "The suspect is identified as having black hair, describing the color of their hair.", \\
\qquad{}\qquad{}  "person\_role" = "suspect", \\
\qquad{}\qquad{}  "leak\_attribute" = "hair\_color" \\
\qquad{}\qquad{}\}, \\
\qquad{}\qquad{}\{ \\
\qquad{}\qquad{}  "excerpt" = "styled in dreadlocks", \\
\qquad{}\qquad{}  "explanation" = "The suspect is identified as having dreadlocks, describing the style of their hair.", \\
\qquad{}\qquad{}  "person\_role" = "suspect", \\
\qquad{}\qquad{}  "leak\_attribute" = "hair\_style" \\
\qquad{}\qquad{}\} \\
  ] \\
\} \\
\textasciigrave{}\textasciigrave{}\textasciigrave{} \\
\mbox{} \\
Alternatively, if the following excerpt was part of your input: \\
"... The suspect had [REDACTED] hair, styled in [REDACTED]. ..." \\
\mbox{} \\
You should return: \\
\textasciigrave{}\textasciigrave{}\textasciigrave{} \\
\{ \\
  "errors\_detected" = FALSE, \\
  "examples" = [] \\
\} \\
\textasciigrave{}\textasciigrave{}\textasciigrave{} \\
\mbox{} \\
Look through the ENTIRE input, from beginning to end. You may be missing something if you don\textquotesingle{}t. It is very important you don\textquotesingle{}t miss even a single example. Go slowly and focus on being comprehensive and thorough, not fast. \\
\mbox{} \\
Do not provide any other commentary.
\end{flushleft}

\subsubsection{Critic prompt: Names}
\begin{flushleft}\footnotesize\ttfamily
Your job is to inspect redacted text and flag EVERY SINGLE first name, last name, full name, nickname, or social media handle that was not correctly redacted.  \\
\mbox{} \\
You should keep a particularly close eye out for first names mentioned only once. You must flag first names if you EVER see them. \\
\mbox{} \\
Be particularly careful to identify names that might be in atypical contexts, e.g, quotes, parentheses, or lower case. For example, "Britney", (Christina), and gwen should all be flagged as a person\textquotesingle{}s name. \\
\mbox{} \\
Your assessment should NOT categorize any of the following as person names: \\
- NEVER flag company or business names (e.g., "US Bank) or other proper nouns as person names \\
- NEVER flag standalone job titles like "Officer", "Detective" as person names, where "standalone" here means it is not paired with a person\textquotesingle{}s name. \\
- NEVER flag legal terms that include proper nouns like "Miranda warning" or "Smith Affadavit" or "Terry stop" as person names. \\
- NEVER flag cultural references as person names, including celebrity names like "Taylor Swift" or literature or biblical figures like "Harry Potter" or "Jesus." \\
\mbox{} \\
Return TRUE if you find any person names, and FALSE if you do not. \\
\mbox{} \\
If you find any person names, provide the following information for each name: \\
- A short excerpt of the text surrounding the flagged name for context. This MUST include the name you found. Surround the name in stars (**), like this: "She said to the suspect, *Aaron*, you need to get out of here." \\
- The name of the person you found \\
- An explanation of why you think this is a person\textquotesingle{}s name \\
- IF you found the name of a law enforcement officer specifically: \\
  - Return TRUE for law enforcement officer.  \\
  - Include their rank if it is present (e.g., you should return "Sergeant" for "Sgt. Smith"). \\
(Note you should flag ALL names you find, including both normal people as well as law enforcement officers.) \\
- A structured leak classification with exactly these two fields: \\
  - person\_role: who the leaked name refers to. Strongly prefer one of: \\
\qquad{}\qquad{}suspect, victim, reporting\_party, witness,  \\
\qquad{}\qquad{}registered\_owner, driver, passenger, family\_member, bystander, other\_person, \\
\qquad{}\qquad{}unknown\_person. \\
  - leak\_attribute: the specific kind of name leaked. Strongly prefer one of: \\
\qquad{}\qquad{}full\_name, first\_name, last\_name, nickname, social\_media\_handle, \\
\qquad{}\qquad{}initial\_last\_name, signature\_name. \\
  If none of the preferred values fit a future leak, use a concise new snake\_case \\
  value that is specific but not overly narrow. \\
\mbox{} \\
For example, given the following excerpt: \\
"She said to the suspect, "Aaron, you need to get out of here.", and then I told her (susie) that he needed to leave. Then Sgt. Smith arrived." \\
\mbox{} \\
You should return: \\
\textasciigrave{}\textasciigrave{}\textasciigrave{} \\
\{ \\
  "errors\_detected" = TRUE, \\
  "examples" = [ \\
\qquad{}\qquad{}\{ \\
\qquad{}\qquad{}  "excerpt" = "She said to the suspect, "Aaron, you need to get out of here."", \\
\qquad{}\qquad{}  "person\_name" = "Aaron", \\
\qquad{}\qquad{}  "explanation" = "The suspect\textquotesingle{}s name was clearly identified in a quote.", \\
\qquad{}\qquad{}  "is\_law\_enforcement" = FALSE, \\
\qquad{}\qquad{}  "rank" = null, \\
\qquad{}\qquad{}  "person\_role" = "suspect", \\
\qquad{}\qquad{}  "leak\_attribute" = "first\_name" \\
\qquad{}\qquad{}\}, \\
\qquad{}\qquad{}\{ \\
\qquad{}\qquad{}  "excerpt" = "and then I told him (susie) that he needed to leave.", \\
\qquad{}\qquad{}  "person\_name" = "Susie", \\
\qquad{}\qquad{}  "explanation" = "A person\textquotesingle{}s first name was clearly identified in parentheses", \\
\qquad{}\qquad{}  "is\_law\_enforcement" = FALSE, \\
\qquad{}\qquad{}  "rank" = null, \\
\qquad{}\qquad{}  "person\_role" = "other\_person", \\
\qquad{}\qquad{}  "leak\_attribute" = "first\_name" \\
\qquad{}\qquad{}\}, \\
\qquad{}\qquad{}\{ \\
\qquad{}\qquad{}  "excerpt" = "Then Sgt. Smith arrived.", \\
\qquad{}\qquad{}  "person\_name" = "Smith", \\
\qquad{}\qquad{}  "explanation" = "The law enforcement officer\textquotesingle{}s last name was clearly identified.", \\
\qquad{}\qquad{}  "is\_law\_enforcement" = TRUE, \\
\qquad{}\qquad{}  "rank" = "Sergeant", \\
\qquad{}\qquad{}  "person\_role" = "other\_person", \\
\qquad{}\qquad{}  "leak\_attribute" = "last\_name" \\
\qquad{}\qquad{}\} \\
  ], \\
\} \\
\textasciigrave{}\textasciigrave{}\textasciigrave{} \\
\mbox{} \\
Alternatively, given the following excerpt: \\
"She said to the suspect, "[REDACTED], you need to get out of here.", and then I told him ([REDACTED]) that he needed to leave. Then [REDACTED] arrived." \\
\mbox{} \\
You should return: \\
\textasciigrave{}\textasciigrave{}\textasciigrave{} \\
\{ \\
  "examples" = [], \\
  "errors\_detected" = FALSE \\
\} \\
\textasciigrave{}\textasciigrave{}\textasciigrave{} \\
\mbox{} \\
Look through the ENTIRE input, from beginning to end. You may be missing something if you don\textquotesingle{}t. It is very important you don\textquotesingle{}t miss even a single example. Just because you found one example, does not mean you found them all. Keep looking until you get to the very end of the input! \\
\mbox{} \\
Go slowly and focus on being comprehensive and thorough, not fast. \\
\mbox{} \\
Do not provide any other commentary.
\end{flushleft}

\subsubsection{Critic prompt: Location information}
\begin{flushleft}\footnotesize\ttfamily
Your job is to inspect redacted text and flag any SPECIFIC location information from the below specified categories that was not correctly redacted.  \\
\mbox{} \\
Here are the types of location information that should have been redacted, with illustrative examples of each: \\
- Addresses and street names: \\
  - 123 Main Street. \\
  - Oak \& Elm Avenue \\
  - N. College Avenue \\
  - 1519 [REDACTED] \\
  - 1200 block of Lincoln Blvd. \\
  - I-205 \\
  - 1600 Pennsylvania Ave., Washington, DC 20001 \\
- Named city, neighborhood, and local areas (i.e., NOT county or state): \\
  - New York City, Brooklyn, Manhattan, etc. \\
  - Williamsburg, SoHo, Harlem, etc. \\
  - Named districts or officially recognized local areas, such as uptown, downtown, civic center, waterfront, etc. \\
  - Zipcodes: 10001, 90210, etc. \\
- Specific commerical establishments and landmarks: \\
  - Starbucks, McDonald\textquotesingle{}s, Walmart \\
  - Empire State Building, Statue of Liberty, Central Park \\
  - Nathan Hale High School, St. Mary\textquotesingle{}s Church \\
  - West Allis Memorial Hospital \\
  - Oberlin Municipal Court \\
- Specific city or local law enforcement agencies or patrol areas  (i.e., NOT county or state-level law enforcement): \\
  - New York Police Department \\
  - NYPD \\
  - Western Precinct \\
  - Beat 123 \\
  - University of Michigan Police Department \\
  - Metropolitan Transit Police Department (MTPD) \\
  - Port of Seattle Police Department \\
\mbox{} \\
Your assessment should never consider the following as location information: \\
- NEVER say that any geography at the county level or higher is a location. For example, "the suspect is from California" should NOT be considered a location, nor should "the suspect is from Cook County, Illinois".  \\
- NEVER say that county- or state-associated entity reveals a specific location. For example, none of "King County Jail", "Fulton County Fair", "Salt Lake County DOJ" should be considered a location, since it refers to a county-level facility. \\
- NEVER say that generic locations like "casino" or "corner store" or "hospital" reveal a specific location. \\
- NEVER say that ubiquitous businesses like towing companies, ambulances, or chain stores like Starbucks or McDonald\textquotesingle{}s reveal a specific location. \\
\mbox{} \\
Here are some concrete examples of what NOT to flag: \\
- NEVER flag county names, state names, or associated entities: \\
  - Do NOT flag "The suspect is from Cook County, Illinois." \\
  - Do NOT flag "The suspect is from California." \\
  - Do NOT flag "The suspect is from the Salt Lake County DOJ." \\
  - Do NOT flag "The officer works for the King County Jail." \\
  - Do NOT flag "The suspect was seen at the Fulton County Fair." \\
  - Do NOT flag "Into the Family Justice Center." \\
  - Do NOT flag PO box addresses. \\
  - Do NOT flag apartment or building unit numbers. \\
- NEVER flag generic locations: \\
  - Do NOT flag "The suspect was seen at a casino." \\
  - Do NOT flag "The suspect was seen at a corner store." \\
  - Do NOT flag "The suspect was seen at a hospital." \\
  - Do NOT flag "The suspect lives in an apartment building." \\
  - Do NOT flag "parking lot," "hallway," "lobby," "front entrance," "alleyway," "bridge," "park," etc. even with directions or numbers like "east edge" or "\#1." \\
- NEVER flag businesses that are not specific to a physical location: \\
  - Do NOT flag "NFS towing responded to impound the vehicle." \\
  - Do NOT flag "A bank card from Wells Fargo was found at the scene." \\
- Do NOT flag standalone references. For example, flag "ER Room 6 at West Allis Memorial Hospital," but do NOT flag "ER Room 6" by itself. Flag "south end of the E building at Nathan Hale High School," but do NOT flag "south end of the E building" by itself. \\
\mbox{} \\
When address redactions are incomplete, i.e., 1519 [REDACTED], flag these specific instances as a disclosure of an \_address\_, since the address number is still visible. \\
\mbox{} \\
If you do not find any mentions of location information from one of the above categories, return FALSE alongside an EMPTY examples list.  \\
\mbox{} \\
If you do find any such examples, first return TRUE. Then in the examples list, provide for each example: \\
- The flagged excerpt \\
- In a single sentence, why this qualifies as one of the above categories \\
- A structured leak classification with exactly these two fields: \\
  - person\_role: who or what the location information refers to. Strongly prefer \\
\qquad{}\qquad{}one of: incident\_location, person\_address, reporting\_party, victim, suspect, \\
\qquad{}\qquad{}witness, law\_enforcement, local\_agency, local\_facility, other\_location, \\
\qquad{}\qquad{}unknown\_location. \\
  - leak\_attribute: the specific kind of location leaked. Strongly prefer one of: \\
\qquad{}\qquad{}street\_address, street\_name, intersection, block\_location, city, \\
\qquad{}\qquad{}neighborhood, local\_area, commercial\_establishment, landmark, school, \\
\qquad{}\qquad{}hospital, jail, law\_enforcement\_agency, patrol\_area. \\
  If none of the preferred values fit a future leak, use a concise new snake\_case \\
  value that is specific but not overly narrow. \\
\mbox{} \\
For example, given the following excerpt: \\
"The suspect lived at 100 26th Avenue East." \\
\mbox{} \\
You should return: \\
\textasciigrave{}\textasciigrave{}\textasciigrave{} \\
\{ \\
  "errors\_detected" = TRUE, \\
  "examples" = [ \\
\qquad{}\qquad{}\{ \\
\qquad{}\qquad{}  "excerpt" = "The lived at 100 26th Avenue East.", \\
\qquad{}\qquad{}  "explanation" = "A street address was revealed.", \\
\qquad{}\qquad{}  "person\_role" = "person\_address", \\
\qquad{}\qquad{}  "leak\_attribute" = "street\_address" \\
\qquad{}\qquad{}\} \\
  ] \\
\} \\
\textasciigrave{}\textasciigrave{}\textasciigrave{} \\
\mbox{} \\
Alternatively, given the following excerpt: \\
"The suspect lived at [REDACTED]." \\
\mbox{} \\
You should return: \\
\textasciigrave{}\textasciigrave{}\textasciigrave{} \\
\{ \\
  "errors\_detected" = FALSE, \\
  "examples" = [] \\
\} \\
\textasciigrave{}\textasciigrave{}\textasciigrave{} \\
\mbox{} \\
Look through the ENTIRE input, from beginning to end. You may be missing something if you don\textquotesingle{}t. It is very important you don\textquotesingle{}t miss even a single example. Go slowly and focus on being comprehensive and thorough, not fast. \\
\mbox{} \\
Do not provide any other commentary.
\end{flushleft}

\subsection{Interpretable feature prompts}
\label{apx:interpretable_feature_prompts}
\subsubsection{Interpretable feature prompt: person-name\_\_accused-person\_\_specific}
\paragraph{Scoring prompt}
\begin{flushleft}\footnotesize\ttfamily
Predict the race of accused person based on all names detailed in the provided police report narrative. \\
\mbox{} \\
First, identify everyone who fits the role of accused person using role and event context only. \\
\mbox{} \\
Consider all people in that role who are mentioned in the narrative. \\
\mbox{} \\
Ignore people who clearly belong to a different role. \\
\mbox{} \\
\mbox{} \\
Next, consider all names associated with accused person in the narrative. This includes, e.g., first names, last names, nicknames, aliases, initials, honorifics, or other name-like references. \\
\mbox{} \\
Based ONLY on names like these, and *ANY* others that are present in the narrative, predict whether the selected target person is White or non-White. Provide a one-sentence justification for your binary prediction first, then provide white\_score and non\_white\_score as integers from 0 to 100. \\
\mbox{} \\
The two scores must sum to exactly 100. 50/50 means no directional evidence favors either White or non-White. Scores should reflect the direction and strength of the evidence, not confidence. Indirect cultural, geographic, social, or contextual evidence should move the scores away from 50/50 in proportion to its strength, even when it is not definitive. Do not treat lack of direct demographic identifiers as a reason to automatically return 50/50. \\
\mbox{} \\
You are required to return both scores, even if the names are weak or ambiguous toward whether the accused person is White or non-White.
\end{flushleft}
\paragraph{Relevance prompt}
\begin{flushleft}\footnotesize\ttfamily
Classify if the provided police report narrative contains any mentions of names. \\
\mbox{} \\
Consider all names associated with accused person in the narrative. This includes, e.g., first names, last names, nicknames, aliases, initials, honorifics, or other name-like references. \\
\mbox{} \\
Provide a one-sentence justification for your classification first, then provide a binary classification. \\
\mbox{} \\
Use 1 if the narrative contains any mention of names as described above, even if it\textquotesingle{}s a weak reference. Use 0 if the narrative contains absolutely no matching information.
\end{flushleft}

\subsubsection{Interpretable feature prompt: hair-or-facial-hair\_\_accused-person\_\_specific}
\paragraph{Scoring prompt}
\begin{flushleft}\footnotesize\ttfamily
Predict the race of accused person based on all hair or facial-hair descriptions detailed in the provided police report narrative. \\
\mbox{} \\
First, identify everyone who fits the role of accused person using role and event context only. \\
\mbox{} \\
Consider all people in that role who are mentioned in the narrative. \\
\mbox{} \\
Ignore people who clearly belong to a different role. \\
\mbox{} \\
\mbox{} \\
Next, consider all hair or facial-hair descriptions associated with accused person in the narrative. This includes, e.g., hairstyle, texture, facial hair, and grooming cues such as curly, coily, straight, wavy, tightly curled, coarse, fine, braids, locs, afro, bun, ponytail, fade, buzz cut, shaved head, dyed style, beard, mustache, goatee, stubble, or clean-shaven status. \\
\mbox{} \\
Based ONLY on hair or facial-hair cues like these, and *ANY* others that are present in the narrative, predict whether the selected target person is White or non-White. Provide a one-sentence justification for your binary prediction first, then provide white\_score and non\_white\_score as integers from 0 to 100. \\
\mbox{} \\
The two scores must sum to exactly 100. 50/50 means no directional evidence favors either White or non-White. Scores should reflect the direction and strength of the evidence, not confidence. Indirect cultural, geographic, social, or contextual evidence should move the scores away from 50/50 in proportion to its strength, even when it is not definitive. Do not treat lack of direct demographic identifiers as a reason to automatically return 50/50. \\
\mbox{} \\
You are required to return both scores, even if the hair or facial-hair descriptions are weak or ambiguous toward whether the accused person is White or non-White.
\end{flushleft}
\paragraph{Relevance prompt}
\begin{flushleft}\footnotesize\ttfamily
Classify if the provided police report narrative contains any mentions of hair or facial-hair descriptions. \\
\mbox{} \\
Consider all hair or facial-hair descriptions associated with accused person in the narrative. This includes, e.g., hairstyle, texture, facial hair, and grooming cues such as curly, coily, straight, wavy, tightly curled, coarse, fine, braids, locs, afro, bun, ponytail, fade, buzz cut, shaved head, dyed style, beard, mustache, goatee, stubble, or clean-shaven status. \\
\mbox{} \\
Provide a one-sentence justification for your classification first, then provide a binary classification. \\
\mbox{} \\
Use 1 if the narrative contains any mention of hair or facial-hair descriptions as described above, even if it\textquotesingle{}s a weak reference. Use 0 if the narrative contains absolutely no matching information.
\end{flushleft}

\subsubsection{Interpretable feature prompt: skin-tone\_\_accused-person\_\_specific}
\paragraph{Scoring prompt}
\begin{flushleft}\footnotesize\ttfamily
Predict the race of accused person based on all skin-tone or complexion descriptions detailed in the provided police report narrative. \\
\mbox{} \\
First, identify everyone who fits the role of accused person using role and event context only. \\
\mbox{} \\
Consider all people in that role who are mentioned in the narrative. \\
\mbox{} \\
Ignore people who clearly belong to a different role. \\
\mbox{} \\
\mbox{} \\
Next, consider all skin-tone or complexion descriptions associated with accused person in the narrative. This includes, e.g., light-skinned, dark-skinned, brown-skinned, pale, tan, olive, ruddy, complexion, shade, or similar direct descriptions. \\
\mbox{} \\
Based ONLY on skin-tone or complexion cues like these, and *ANY* others that are present in the narrative, predict whether the selected target person is White or non-White. Provide a one-sentence justification for your binary prediction first, then provide white\_score and non\_white\_score as integers from 0 to 100. \\
\mbox{} \\
The two scores must sum to exactly 100. 50/50 means no directional evidence favors either White or non-White. Scores should reflect the direction and strength of the evidence, not confidence. Indirect cultural, geographic, social, or contextual evidence should move the scores away from 50/50 in proportion to its strength, even when it is not definitive. Do not treat lack of direct demographic identifiers as a reason to automatically return 50/50. \\
\mbox{} \\
You are required to return both scores, even if the skin-tone or complexion descriptions are weak or ambiguous toward whether the accused person is White or non-White.
\end{flushleft}
\paragraph{Relevance prompt}
\begin{flushleft}\footnotesize\ttfamily
Classify if the provided police report narrative contains any mentions of skin-tone or complexion descriptions. \\
\mbox{} \\
Consider all skin-tone or complexion descriptions associated with accused person in the narrative. This includes, e.g., light-skinned, dark-skinned, brown-skinned, pale, tan, olive, ruddy, complexion, shade, or similar direct descriptions. \\
\mbox{} \\
Provide a one-sentence justification for your classification first, then provide a binary classification. \\
\mbox{} \\
Use 1 if the narrative contains any mention of skin-tone or complexion descriptions as described above, even if it\textquotesingle{}s a weak reference. Use 0 if the narrative contains absolutely no matching information.
\end{flushleft}

\subsubsection{Interpretable feature prompt: facial-feature\_\_accused-person\_\_specific}
\paragraph{Scoring prompt}
\begin{flushleft}\footnotesize\ttfamily
Predict the race of accused person based on all facial-feature descriptions detailed in the provided police report narrative. \\
\mbox{} \\
First, identify everyone who fits the role of accused person using role and event context only. \\
\mbox{} \\
Consider all people in that role who are mentioned in the narrative. \\
\mbox{} \\
Ignore people who clearly belong to a different role. \\
\mbox{} \\
\mbox{} \\
Next, consider all facial-feature descriptions associated with accused person in the narrative. This includes, e.g., direct descriptions of facial structure or appearance such as nose shape, eye shape, cheekbones, lips, jawline, or similarly described features. \\
\mbox{} \\
Based ONLY on facial-feature cues like these, and *ANY* others that are present in the narrative, predict whether the selected target person is White or non-White. Provide a one-sentence justification for your binary prediction first, then provide white\_score and non\_white\_score as integers from 0 to 100. \\
\mbox{} \\
The two scores must sum to exactly 100. 50/50 means no directional evidence favors either White or non-White. Scores should reflect the direction and strength of the evidence, not confidence. Indirect cultural, geographic, social, or contextual evidence should move the scores away from 50/50 in proportion to its strength, even when it is not definitive. Do not treat lack of direct demographic identifiers as a reason to automatically return 50/50. \\
\mbox{} \\
You are required to return both scores, even if the facial-feature descriptions are weak or ambiguous toward whether the accused person is White or non-White.
\end{flushleft}
\paragraph{Relevance prompt}
\begin{flushleft}\footnotesize\ttfamily
Classify if the provided police report narrative contains any mentions of facial-feature descriptions. \\
\mbox{} \\
Consider all facial-feature descriptions associated with accused person in the narrative. This includes, e.g., direct descriptions of facial structure or appearance such as nose shape, eye shape, cheekbones, lips, jawline, or similarly described features. \\
\mbox{} \\
Provide a one-sentence justification for your classification first, then provide a binary classification. \\
\mbox{} \\
Use 1 if the narrative contains any mention of facial-feature descriptions as described above, even if it\textquotesingle{}s a weak reference. Use 0 if the narrative contains absolutely no matching information.
\end{flushleft}

\subsubsection{Interpretable feature prompt: eye-color\_\_accused-person\_\_specific}
\paragraph{Scoring prompt}
\begin{flushleft}\footnotesize\ttfamily
Predict the race of accused person based on all eye-color descriptions detailed in the provided police report narrative. \\
\mbox{} \\
First, identify everyone who fits the role of accused person using role and event context only. \\
\mbox{} \\
Consider all people in that role who are mentioned in the narrative. \\
\mbox{} \\
Ignore people who clearly belong to a different role. \\
\mbox{} \\
\mbox{} \\
Next, consider only eye-color descriptions associated with accused person, such as brown, blue, green, gray, hazel, amber, or another stated eye color. Do not use eye shape, facial structure, skin tone, hair, or any other physical description. \\
\mbox{} \\
Based ONLY on eye-color cues like these, and *ANY* others that are present in the narrative, predict whether the selected target person is White or non-White. Provide a one-sentence justification for your binary prediction first, then provide white\_score and non\_white\_score as integers from 0 to 100. \\
\mbox{} \\
The two scores must sum to exactly 100. 50/50 means no directional evidence favors either White or non-White. Scores should reflect the direction and strength of the evidence, not confidence. Indirect cultural, geographic, social, or contextual evidence should move the scores away from 50/50 in proportion to its strength, even when it is not definitive. Do not treat lack of direct demographic identifiers as a reason to automatically return 50/50. \\
\mbox{} \\
You are required to return both scores, even if the eye-color descriptions are weak or ambiguous toward whether the accused person is White or non-White.
\end{flushleft}
\paragraph{Relevance prompt}
\begin{flushleft}\footnotesize\ttfamily
Classify if the provided police report narrative contains any mentions of eye-color descriptions. \\
\mbox{} \\
Consider only eye-color descriptions associated with accused person, such as brown, blue, green, gray, hazel, amber, or another stated eye color. Do not use eye shape, facial structure, skin tone, hair, or any other physical description. \\
\mbox{} \\
Provide a one-sentence justification for your classification first, then provide a binary classification. \\
\mbox{} \\
Use 1 if the narrative contains any mention of eye-color descriptions as described above, even if it\textquotesingle{}s a weak reference. Use 0 if the narrative contains absolutely no matching information.
\end{flushleft}

\subsubsection{Interpretable feature prompt: clothing-or-accessories\_\_accused-person\_\_specific}
\paragraph{Scoring prompt}
\begin{flushleft}\footnotesize\ttfamily
Predict the race of accused person based on all clothing or accessories detailed in the provided police report narrative. \\
\mbox{} \\
First, identify everyone who fits the role of accused person using role and event context only. \\
\mbox{} \\
Consider all people in that role who are mentioned in the narrative. \\
\mbox{} \\
Ignore people who clearly belong to a different role. \\
\mbox{} \\
\mbox{} \\
Next, consider all clothing or accessories associated with accused person in the narrative. This includes, e.g., style or presentation cues such as formalwear, streetwear, workwear, athletic wear, religious attire, ceremonial attire, or culturally distinctive style cues, as well as jewelry, watches, bags, glasses, headwear, or other wearable accessories. \\
\mbox{} \\
Based ONLY on clothing or accessory cues like these, and *ANY* others that are present in the narrative, predict whether the selected target person is White or non-White. Provide a one-sentence justification for your binary prediction first, then provide white\_score and non\_white\_score as integers from 0 to 100. \\
\mbox{} \\
The two scores must sum to exactly 100. 50/50 means no directional evidence favors either White or non-White. Scores should reflect the direction and strength of the evidence, not confidence. Indirect cultural, geographic, social, or contextual evidence should move the scores away from 50/50 in proportion to its strength, even when it is not definitive. Do not treat lack of direct demographic identifiers as a reason to automatically return 50/50. \\
\mbox{} \\
You are required to return both scores, even if the clothing or accessories are weak or ambiguous toward whether the accused person is White or non-White.
\end{flushleft}
\paragraph{Relevance prompt}
\begin{flushleft}\footnotesize\ttfamily
Classify if the provided police report narrative contains any mentions of clothing or accessories. \\
\mbox{} \\
Consider all clothing or accessories associated with accused person in the narrative. This includes, e.g., style or presentation cues such as formalwear, streetwear, workwear, athletic wear, religious attire, ceremonial attire, or culturally distinctive style cues, as well as jewelry, watches, bags, glasses, headwear, or other wearable accessories. \\
\mbox{} \\
Provide a one-sentence justification for your classification first, then provide a binary classification. \\
\mbox{} \\
Use 1 if the narrative contains any mention of clothing or accessories as described above, even if it\textquotesingle{}s a weak reference. Use 0 if the narrative contains absolutely no matching information.
\end{flushleft}

\subsubsection{Interpretable feature prompt: explicit-race-or-ethnicity-mention\_\_accused-person\_\_specific}
\paragraph{Scoring prompt}
\begin{flushleft}\footnotesize\ttfamily
Predict the race of accused person based on all explicit race or ethnicity mentions detailed in the provided police report narrative. \\
\mbox{} \\
First, identify everyone who fits the role of accused person using role and event context only. \\
\mbox{} \\
Consider all people in that role who are mentioned in the narrative. \\
\mbox{} \\
Ignore people who clearly belong to a different role. \\
\mbox{} \\
\mbox{} \\
Next, consider only explicit race or ethnicity labels associated with accused person, such as Black, White, Hispanic, Latino, Asian, Native American, Alaska Native, Pacific Islander, multiracial, tribal affiliation, Indigenous nation or community identity, or another directly stated race or ethnicity. Do not use names, nationality, national origin, ancestry, language, physical descriptions, or indirect cultural cues. \\
\mbox{} \\
Based ONLY on explicit race or ethnicity mentions like these, and *ANY* others that are present in the narrative, predict whether the selected target person is White or non-White. Provide a one-sentence justification for your binary prediction first, then provide white\_score and non\_white\_score as integers from 0 to 100. \\
\mbox{} \\
The two scores must sum to exactly 100. 50/50 means no directional evidence favors either White or non-White. Scores should reflect the direction and strength of the evidence, not confidence. Indirect cultural, geographic, social, or contextual evidence should move the scores away from 50/50 in proportion to its strength, even when it is not definitive. Do not treat lack of direct demographic identifiers as a reason to automatically return 50/50. \\
\mbox{} \\
You are required to return both scores, even if the explicit mentions of race or ethnicity are weak or ambiguous toward whether the accused person is White or non-White.
\end{flushleft}
\paragraph{Relevance prompt}
\begin{flushleft}\footnotesize\ttfamily
Classify if the provided police report narrative contains any explicit mentions of race or ethnicity. \\
\mbox{} \\
Consider only explicit race or ethnicity labels associated with accused person, such as Black, White, Hispanic, Latino, Asian, Native American, Alaska Native, Pacific Islander, multiracial, tribal affiliation, Indigenous nation or community identity, or another directly stated race or ethnicity. Do not use names, nationality, national origin, ancestry, language, physical descriptions, or indirect cultural cues. \\
\mbox{} \\
Provide a one-sentence justification for your classification first, then provide a binary classification. \\
\mbox{} \\
Use 1 if the narrative contains any mention of explicit mentions of race or ethnicity as described above, even if it\textquotesingle{}s a weak reference. Use 0 if the narrative contains absolutely no matching information.
\end{flushleft}

\subsubsection{Interpretable feature prompt: nationality-origin-or-ancestry\_\_accused-person\_\_specific}
\paragraph{Scoring prompt}
\begin{flushleft}\footnotesize\ttfamily
Predict the race of accused person based on all nationality, national-origin, or ancestry information detailed in the provided police report narrative. \\
\mbox{} \\
First, identify everyone who fits the role of accused person using role and event context only. \\
\mbox{} \\
Consider all people in that role who are mentioned in the narrative. \\
\mbox{} \\
Ignore people who clearly belong to a different role. \\
\mbox{} \\
\mbox{} \\
Next, consider only explicit nationality, national-origin, or ancestry information associated with accused person. This includes, e.g., citizenship or nationality labels, country or region of origin, or stated ancestry. Do not use race or ethnicity labels, tribal identity, names, language or accent, current location, religion, cultural practices, or physical descriptions. \\
\mbox{} \\
Based ONLY on nationality, national-origin, or ancestry cues like these, and *ANY* others that are present in the narrative, predict whether the selected target person is White or non-White. Provide a one-sentence justification for your binary prediction first, then provide white\_score and non\_white\_score as integers from 0 to 100. \\
\mbox{} \\
The two scores must sum to exactly 100. 50/50 means no directional evidence favors either White or non-White. Scores should reflect the direction and strength of the evidence, not confidence. Indirect cultural, geographic, social, or contextual evidence should move the scores away from 50/50 in proportion to its strength, even when it is not definitive. Do not treat lack of direct demographic identifiers as a reason to automatically return 50/50. \\
\mbox{} \\
You are required to return both scores, even if the nationality, national-origin, or ancestry information are weak or ambiguous toward whether the accused person is White or non-White.
\end{flushleft}
\paragraph{Relevance prompt}
\begin{flushleft}\footnotesize\ttfamily
Classify if the provided police report narrative contains any mentions of nationality, national origin, or ancestry. \\
\mbox{} \\
Consider only explicit nationality, national-origin, or ancestry information associated with accused person. This includes, e.g., citizenship or nationality labels, country or region of origin, or stated ancestry. Do not use race or ethnicity labels, tribal identity, names, language or accent, current location, religion, cultural practices, or physical descriptions. \\
\mbox{} \\
Provide a one-sentence justification for your classification first, then provide a binary classification. \\
\mbox{} \\
Use 1 if the narrative contains any mention of nationality, national-origin, or ancestry information as described above, even if it\textquotesingle{}s a weak reference. Use 0 if the narrative contains absolutely no matching information.
\end{flushleft}

\subsubsection{Interpretable feature prompt: language-or-accent\_\_accused-person\_\_specific}
\paragraph{Scoring prompt}
\begin{flushleft}\footnotesize\ttfamily
Predict the race of accused person based on all specific-language, interpreter, translation, or accent information detailed in the provided police report narrative. \\
\mbox{} \\
First, identify everyone who fits the role of accused person using role and event context only. \\
\mbox{} \\
Consider all people in that role who are mentioned in the narrative. \\
\mbox{} \\
Ignore people who clearly belong to a different role. \\
\mbox{} \\
\mbox{} \\
Next, consider only language-identity information associated with accused person. This includes, e.g., a specific language the person speaks, understands, prefers, or uses; foreign-language text attributed to the person; use of or need for an interpreter or translation; ability or inability to communicate in a named language; or a directly described foreign or regional accent. Do not use names, race or ethnicity, nationality or origin, cultural practices, slang or word choice, speech clarity, or other communication behavior. \\
\mbox{} \\
Based ONLY on specific-language, interpreter, translation, or accent cues like these, and *ANY* others that are present in the narrative, predict whether the selected target person is White or non-White. Provide a one-sentence justification for your binary prediction first, then provide white\_score and non\_white\_score as integers from 0 to 100. \\
\mbox{} \\
The two scores must sum to exactly 100. 50/50 means no directional evidence favors either White or non-White. Scores should reflect the direction and strength of the evidence, not confidence. Indirect cultural, geographic, social, or contextual evidence should move the scores away from 50/50 in proportion to its strength, even when it is not definitive. Do not treat lack of direct demographic identifiers as a reason to automatically return 50/50. \\
\mbox{} \\
You are required to return both scores, even if the specific-language, interpreter, translation, or accent information are weak or ambiguous toward whether the accused person is White or non-White.
\end{flushleft}
\paragraph{Relevance prompt}
\begin{flushleft}\footnotesize\ttfamily
Classify if the provided police report narrative contains any mentions of a specific language, interpreter or translation use, language proficiency, or accent. \\
\mbox{} \\
Consider only language-identity information associated with accused person. This includes, e.g., a specific language the person speaks, understands, prefers, or uses; foreign-language text attributed to the person; use of or need for an interpreter or translation; ability or inability to communicate in a named language; or a directly described foreign or regional accent. Do not use names, race or ethnicity, nationality or origin, cultural practices, slang or word choice, speech clarity, or other communication behavior. \\
\mbox{} \\
Provide a one-sentence justification for your classification first, then provide a binary classification. \\
\mbox{} \\
Use 1 if the narrative contains any mention of specific-language, interpreter, translation, or accent information as described above, even if it\textquotesingle{}s a weak reference. Use 0 if the narrative contains absolutely no matching information.
\end{flushleft}

\subsubsection{Interpretable feature prompt: communication-presentation\_\_accused-person\_\_specific}
\paragraph{Scoring prompt}
\begin{flushleft}\footnotesize\ttfamily
Predict the race of accused person based on all spoken communication style or presentation detailed in the provided police report narrative. \\
\mbox{} \\
First, identify everyone who fits the role of accused person using role and event context only. \\
\mbox{} \\
Consider all people in that role who are mentioned in the narrative. \\
\mbox{} \\
Ignore people who clearly belong to a different role. \\
\mbox{} \\
\mbox{} \\
Next, consider only communication-style or presentation information associated with accused person. This includes, e.g., slang, colloquial phrasing, dialect, sociolect, register, patterned word choice, profanity, speech tone, clarity, coherence, or descriptions such as slurred, confused, impaired, unusually clear, loud, quiet, or evasive speech. Do not use the identity of a language, interpreter or translation use, accent, names, explicit identity information, or physical descriptions. \\
\mbox{} \\
Based ONLY on spoken communication style or presentation cues like these, and *ANY* others that are present in the narrative, predict whether the selected target person is White or non-White. Provide a one-sentence justification for your binary prediction first, then provide white\_score and non\_white\_score as integers from 0 to 100. \\
\mbox{} \\
The two scores must sum to exactly 100. 50/50 means no directional evidence favors either White or non-White. Scores should reflect the direction and strength of the evidence, not confidence. Indirect cultural, geographic, social, or contextual evidence should move the scores away from 50/50 in proportion to its strength, even when it is not definitive. Do not treat lack of direct demographic identifiers as a reason to automatically return 50/50. \\
\mbox{} \\
You are required to return both scores, even if the spoken communication style or presentation are weak or ambiguous toward whether the accused person is White or non-White.
\end{flushleft}
\paragraph{Relevance prompt}
\begin{flushleft}\footnotesize\ttfamily
Classify if the provided police report narrative contains any mentions of spoken communication style or presentation. \\
\mbox{} \\
Consider only communication-style or presentation information associated with accused person. This includes, e.g., slang, colloquial phrasing, dialect, sociolect, register, patterned word choice, profanity, speech tone, clarity, coherence, or descriptions such as slurred, confused, impaired, unusually clear, loud, quiet, or evasive speech. Do not use the identity of a language, interpreter or translation use, accent, names, explicit identity information, or physical descriptions. \\
\mbox{} \\
Provide a one-sentence justification for your classification first, then provide a binary classification. \\
\mbox{} \\
Use 1 if the narrative contains any mention of spoken communication style or presentation as described above, even if it\textquotesingle{}s a weak reference. Use 0 if the narrative contains absolutely no matching information.
\end{flushleft}

\subsubsection{Interpretable feature prompt: religious-or-cultural-practice\_\_accused-person\_\_specific}
\paragraph{Scoring prompt}
\begin{flushleft}\footnotesize\ttfamily
Predict the race of accused person based on all religious affiliation or cultural-practice information detailed in the provided police report narrative. \\
\mbox{} \\
First, identify everyone who fits the role of accused person using role and event context only. \\
\mbox{} \\
Consider all people in that role who are mentioned in the narrative. \\
\mbox{} \\
Ignore people who clearly belong to a different role. \\
\mbox{} \\
\mbox{} \\
Next, consider only explicit religious affiliation or cultural practices associated with accused person. This includes, e.g., a religion, denomination, sect, religious observance, cultural custom, tradition, holiday, ceremony, or foodway. Do not use names, race or ethnicity, nationality or origin, ancestry or tribal identity, language or accent, location, clothing, social-group affiliation, or physical descriptions. \\
\mbox{} \\
Based ONLY on religious affiliation or cultural-practice cues like these, and *ANY* others that are present in the narrative, predict whether the selected target person is White or non-White. Provide a one-sentence justification for your binary prediction first, then provide white\_score and non\_white\_score as integers from 0 to 100. \\
\mbox{} \\
The two scores must sum to exactly 100. 50/50 means no directional evidence favors either White or non-White. Scores should reflect the direction and strength of the evidence, not confidence. Indirect cultural, geographic, social, or contextual evidence should move the scores away from 50/50 in proportion to its strength, even when it is not definitive. Do not treat lack of direct demographic identifiers as a reason to automatically return 50/50. \\
\mbox{} \\
You are required to return both scores, even if the religious affiliation or cultural-practice information are weak or ambiguous toward whether the accused person is White or non-White.
\end{flushleft}
\paragraph{Relevance prompt}
\begin{flushleft}\footnotesize\ttfamily
Classify if the provided police report narrative contains any mentions of religious affiliation or cultural practices. \\
\mbox{} \\
Consider only explicit religious affiliation or cultural practices associated with accused person. This includes, e.g., a religion, denomination, sect, religious observance, cultural custom, tradition, holiday, ceremony, or foodway. Do not use names, race or ethnicity, nationality or origin, ancestry or tribal identity, language or accent, location, clothing, social-group affiliation, or physical descriptions. \\
\mbox{} \\
Provide a one-sentence justification for your classification first, then provide a binary classification. \\
\mbox{} \\
Use 1 if the narrative contains any mention of religious affiliation or cultural-practice information as described above, even if it\textquotesingle{}s a weak reference. Use 0 if the narrative contains absolutely no matching information.
\end{flushleft}

\subsubsection{Interpretable feature prompt: employment-or-occupation\_\_accused-person\_\_specific}
\paragraph{Scoring prompt}
\begin{flushleft}\footnotesize\ttfamily
Predict the race of accused person based on all employment, work-status, or occupation information detailed in the provided police report narrative. \\
\mbox{} \\
First, identify everyone who fits the role of accused person using role and event context only. \\
\mbox{} \\
Consider all people in that role who are mentioned in the narrative. \\
\mbox{} \\
Ignore people who clearly belong to a different role. \\
\mbox{} \\
\mbox{} \\
Next, consider all employment, work-status, or occupation information associated with accused person in the narrative. This includes, e.g., direct references to employed, unemployed, retired, on duty, off duty, student worker, gig work, job, profession, trade, rank, or occupational role. Do NOT use direct descriptions of the accused person, including explicit race or ethnicity labels or physical descriptions as evidence for this feature. \\
\mbox{} \\
Based ONLY on employment, work-status, or occupation cues like these, and *ANY* others that are present in the narrative, predict whether the selected target person is White or non-White. Provide a one-sentence justification for your binary prediction first, then provide white\_score and non\_white\_score as integers from 0 to 100. \\
\mbox{} \\
The two scores must sum to exactly 100. 50/50 means no directional evidence favors either White or non-White. Scores should reflect the direction and strength of the evidence, not confidence. Indirect cultural, geographic, social, or contextual evidence should move the scores away from 50/50 in proportion to its strength, even when it is not definitive. Do not treat lack of direct demographic identifiers as a reason to automatically return 50/50. \\
\mbox{} \\
You are required to return both scores, even if the employment, work-status, or occupation information are weak or ambiguous toward whether the accused person is White or non-White.
\end{flushleft}
\paragraph{Relevance prompt}
\begin{flushleft}\footnotesize\ttfamily
Classify if the provided police report narrative contains any mentions of employment, work-status, or occupation information. \\
\mbox{} \\
Consider all employment, work-status, or occupation information associated with accused person in the narrative. This includes, e.g., direct references to employed, unemployed, retired, on duty, off duty, student worker, gig work, job, profession, trade, rank, or occupational role. \\
\mbox{} \\
Do NOT use direct descriptions of the accused person, including explicit race or ethnicity labels or physical descriptions as evidence for this feature. \\
\mbox{} \\
Provide a one-sentence justification for your classification first, then provide a binary classification. \\
\mbox{} \\
Use 1 if the narrative contains any mention of employment, work-status, or occupation information as described above, even if it\textquotesingle{}s a weak reference. Use 0 if the narrative contains absolutely no matching information.
\end{flushleft}

\subsubsection{Interpretable feature prompt: vehicle-access-or-type\_\_accused-person\_\_specific}
\paragraph{Scoring prompt}
\begin{flushleft}\footnotesize\ttfamily
Predict the race of accused person based on all vehicle access, vehicle type, or vehicle make/model information detailed in the provided police report narrative. \\
\mbox{} \\
First, identify everyone who fits the role of accused person using role and event context only. \\
\mbox{} \\
Consider all people in that role who are mentioned in the narrative. \\
\mbox{} \\
Ignore people who clearly belong to a different role. \\
\mbox{} \\
\mbox{} \\
Next, consider all vehicle access, vehicle type, or vehicle make and model information associated with accused person in the narrative. This includes, e.g., direct references to private-vehicle ownership or access, specialized vehicles such as work trucks, motorcycles, commercial vehicles, or RVs, and specific vehicle makes, models, trims, or brands associated with the person. Do NOT use direct descriptions of the accused person, including explicit race or ethnicity labels or physical descriptions as evidence for this feature. \\
\mbox{} \\
Based ONLY on vehicle-related cues like these, and *ANY* others that are present in the narrative, predict whether the selected target person is White or non-White. Provide a one-sentence justification for your binary prediction first, then provide white\_score and non\_white\_score as integers from 0 to 100. \\
\mbox{} \\
The two scores must sum to exactly 100. 50/50 means no directional evidence favors either White or non-White. Scores should reflect the direction and strength of the evidence, not confidence. Indirect cultural, geographic, social, or contextual evidence should move the scores away from 50/50 in proportion to its strength, even when it is not definitive. Do not treat lack of direct demographic identifiers as a reason to automatically return 50/50. \\
\mbox{} \\
You are required to return both scores, even if the vehicle access, vehicle type, or vehicle make/model information are weak or ambiguous toward whether the accused person is White or non-White.
\end{flushleft}
\paragraph{Relevance prompt}
\begin{flushleft}\footnotesize\ttfamily
Classify if the provided police report narrative contains any mentions of vehicle access, vehicle type, or vehicle make/model information. \\
\mbox{} \\
Consider all vehicle access, vehicle type, or vehicle make and model information associated with accused person in the narrative. This includes, e.g., direct references to private-vehicle ownership or access, specialized vehicles such as work trucks, motorcycles, commercial vehicles, or RVs, and specific vehicle makes, models, trims, or brands associated with the person. \\
\mbox{} \\
Do NOT use direct descriptions of the accused person, including explicit race or ethnicity labels or physical descriptions as evidence for this feature. \\
\mbox{} \\
Provide a one-sentence justification for your classification first, then provide a binary classification. \\
\mbox{} \\
Use 1 if the narrative contains any mention of vehicle access, vehicle type, or vehicle make/model information as described above, even if it\textquotesingle{}s a weak reference. Use 0 if the narrative contains absolutely no matching information.
\end{flushleft}

\subsubsection{Interpretable feature prompt: weapon\_\_accused-person\_\_specific}
\paragraph{Scoring prompt}
\begin{flushleft}\footnotesize\ttfamily
Predict the race of accused person based on all weapon references detailed in the provided police report narrative. \\
\mbox{} \\
First, identify everyone who fits the role of accused person using role and event context only. \\
\mbox{} \\
Consider all people in that role who are mentioned in the narrative. \\
\mbox{} \\
Ignore people who clearly belong to a different role. \\
\mbox{} \\
\mbox{} \\
Next, consider all weapon references associated with accused person in the narrative. This includes, e.g., direct references to firearms, knives, or other weapons associated with the person. Do NOT use direct descriptions of the accused person, including explicit race or ethnicity labels or physical descriptions as evidence for this feature. \\
\mbox{} \\
Based ONLY on weapon-related cues like these, and *ANY* others that are present in the narrative, predict whether the selected target person is White or non-White. Provide a one-sentence justification for your binary prediction first, then provide white\_score and non\_white\_score as integers from 0 to 100. \\
\mbox{} \\
The two scores must sum to exactly 100. 50/50 means no directional evidence favors either White or non-White. Scores should reflect the direction and strength of the evidence, not confidence. Indirect cultural, geographic, social, or contextual evidence should move the scores away from 50/50 in proportion to its strength, even when it is not definitive. Do not treat lack of direct demographic identifiers as a reason to automatically return 50/50. \\
\mbox{} \\
You are required to return both scores, even if the weapon references are weak or ambiguous toward whether the accused person is White or non-White.
\end{flushleft}
\paragraph{Relevance prompt}
\begin{flushleft}\footnotesize\ttfamily
Classify if the provided police report narrative contains any mentions of weapon references. \\
\mbox{} \\
Consider all weapon references associated with accused person in the narrative. This includes, e.g., direct references to firearms, knives, or other weapons associated with the person. \\
\mbox{} \\
Do NOT use direct descriptions of the accused person, including explicit race or ethnicity labels or physical descriptions as evidence for this feature. \\
\mbox{} \\
Provide a one-sentence justification for your classification first, then provide a binary classification. \\
\mbox{} \\
Use 1 if the narrative contains any mention of weapon references as described above, even if it\textquotesingle{}s a weak reference. Use 0 if the narrative contains absolutely no matching information.
\end{flushleft}

\subsubsection{Interpretable feature prompt: pet-or-animal-reference\_\_accused-person\_\_specific}
\paragraph{Scoring prompt}
\begin{flushleft}\footnotesize\ttfamily
Predict the race of accused person based on all pet or animal references detailed in the provided police report narrative. \\
\mbox{} \\
First, identify everyone who fits the role of accused person using role and event context only. \\
\mbox{} \\
Consider all people in that role who are mentioned in the narrative. \\
\mbox{} \\
Ignore people who clearly belong to a different role. \\
\mbox{} \\
\mbox{} \\
Next, consider all pet or animal references associated with accused person in the narrative. This includes, e.g., direct references to a pet, companion animal, livestock, or another animal associated with the person. Do NOT use direct descriptions of the accused person, including explicit race or ethnicity labels or physical descriptions as evidence for this feature. \\
\mbox{} \\
Based ONLY on animal-related cues like these, and *ANY* others that are present in the narrative, predict whether the selected target person is White or non-White. Provide a one-sentence justification for your binary prediction first, then provide white\_score and non\_white\_score as integers from 0 to 100. \\
\mbox{} \\
The two scores must sum to exactly 100. 50/50 means no directional evidence favors either White or non-White. Scores should reflect the direction and strength of the evidence, not confidence. Indirect cultural, geographic, social, or contextual evidence should move the scores away from 50/50 in proportion to its strength, even when it is not definitive. Do not treat lack of direct demographic identifiers as a reason to automatically return 50/50. \\
\mbox{} \\
You are required to return both scores, even if the pet or animal references are weak or ambiguous toward whether the accused person is White or non-White.
\end{flushleft}
\paragraph{Relevance prompt}
\begin{flushleft}\footnotesize\ttfamily
Classify if the provided police report narrative contains any mentions of pet or animal references. \\
\mbox{} \\
Consider all pet or animal references associated with accused person in the narrative. This includes, e.g., direct references to a pet, companion animal, livestock, or another animal associated with the person. \\
\mbox{} \\
Do NOT use direct descriptions of the accused person, including explicit race or ethnicity labels or physical descriptions as evidence for this feature. \\
\mbox{} \\
Provide a one-sentence justification for your classification first, then provide a binary classification. \\
\mbox{} \\
Use 1 if the narrative contains any mention of pet or animal references as described above, even if it\textquotesingle{}s a weak reference. Use 0 if the narrative contains absolutely no matching information.
\end{flushleft}

\subsubsection{Interpretable feature prompt: consumer-brand-reference\_\_accused-person\_\_specific}
\paragraph{Scoring prompt}
\begin{flushleft}\footnotesize\ttfamily
Predict the race of accused person based on all consumer-brand references detailed in the provided police report narrative. \\
\mbox{} \\
First, identify everyone who fits the role of accused person using role and event context only. \\
\mbox{} \\
Consider all people in that role who are mentioned in the narrative. \\
\mbox{} \\
Ignore people who clearly belong to a different role. \\
\mbox{} \\
\mbox{} \\
Next, consider all consumer-brand references associated with accused person in the narrative. This includes, e.g., direct references to named consumer brands associated with the person\textquotesingle{}s clothing, property, electronics, or possessions. Do NOT use direct descriptions of the accused person, including explicit race or ethnicity labels or physical descriptions as evidence for this feature. \\
\mbox{} \\
Based ONLY on consumer-brand cues like these, and *ANY* others that are present in the narrative, predict whether the selected target person is White or non-White. Provide a one-sentence justification for your binary prediction first, then provide white\_score and non\_white\_score as integers from 0 to 100. \\
\mbox{} \\
The two scores must sum to exactly 100. 50/50 means no directional evidence favors either White or non-White. Scores should reflect the direction and strength of the evidence, not confidence. Indirect cultural, geographic, social, or contextual evidence should move the scores away from 50/50 in proportion to its strength, even when it is not definitive. Do not treat lack of direct demographic identifiers as a reason to automatically return 50/50. \\
\mbox{} \\
You are required to return both scores, even if the consumer-brand references are weak or ambiguous toward whether the accused person is White or non-White.
\end{flushleft}
\paragraph{Relevance prompt}
\begin{flushleft}\footnotesize\ttfamily
Classify if the provided police report narrative contains any mentions of consumer-brand references. \\
\mbox{} \\
Consider all consumer-brand references associated with accused person in the narrative. This includes, e.g., direct references to named consumer brands associated with the person\textquotesingle{}s clothing, property, electronics, or possessions. \\
\mbox{} \\
Do NOT use direct descriptions of the accused person, including explicit race or ethnicity labels or physical descriptions as evidence for this feature. \\
\mbox{} \\
Provide a one-sentence justification for your classification first, then provide a binary classification. \\
\mbox{} \\
Use 1 if the narrative contains any mention of consumer-brand references as described above, even if it\textquotesingle{}s a weak reference. Use 0 if the narrative contains absolutely no matching information.
\end{flushleft}

\subsubsection{Interpretable feature prompt: social-group-affiliation\_\_accused-person\_\_specific}
\paragraph{Scoring prompt}
\begin{flushleft}\footnotesize\ttfamily
Predict the race of accused person based on all social-group affiliations detailed in the provided police report narrative. \\
\mbox{} \\
First, identify everyone who fits the role of accused person using role and event context only. \\
\mbox{} \\
Consider all people in that role who are mentioned in the narrative. \\
\mbox{} \\
Ignore people who clearly belong to a different role. \\
\mbox{} \\
\mbox{} \\
Next, consider all social-group affiliations associated with accused person in the narrative. This includes, e.g., direct references to gangs, cliques, crews, clubs, teams, fraternities, social groups, or recurring organized affiliations tied to the person. Do NOT use direct descriptions of the accused person, including explicit race or ethnicity labels or physical descriptions as evidence for this feature. \\
\mbox{} \\
Based ONLY on social-group-affiliation cues like these, and *ANY* others that are present in the narrative, predict whether the selected target person is White or non-White. Provide a one-sentence justification for your binary prediction first, then provide white\_score and non\_white\_score as integers from 0 to 100. \\
\mbox{} \\
The two scores must sum to exactly 100. 50/50 means no directional evidence favors either White or non-White. Scores should reflect the direction and strength of the evidence, not confidence. Indirect cultural, geographic, social, or contextual evidence should move the scores away from 50/50 in proportion to its strength, even when it is not definitive. Do not treat lack of direct demographic identifiers as a reason to automatically return 50/50. \\
\mbox{} \\
You are required to return both scores, even if the social-group affiliations are weak or ambiguous toward whether the accused person is White or non-White.
\end{flushleft}
\paragraph{Relevance prompt}
\begin{flushleft}\footnotesize\ttfamily
Classify if the provided police report narrative contains any mentions of social-group affiliations. \\
\mbox{} \\
Consider all social-group affiliations associated with accused person in the narrative. This includes, e.g., direct references to gangs, cliques, crews, clubs, teams, fraternities, social groups, or recurring organized affiliations tied to the person. \\
\mbox{} \\
Do NOT use direct descriptions of the accused person, including explicit race or ethnicity labels or physical descriptions as evidence for this feature. \\
\mbox{} \\
Provide a one-sentence justification for your classification first, then provide a binary classification. \\
\mbox{} \\
Use 1 if the narrative contains any mention of social-group affiliations as described above, even if it\textquotesingle{}s a weak reference. Use 0 if the narrative contains absolutely no matching information.
\end{flushleft}

\subsubsection{Interpretable feature prompt: neighborhood-name\_\_accused-person\_\_specific}
\paragraph{Scoring prompt}
\begin{flushleft}\footnotesize\ttfamily
Predict the race of accused person based on all neighborhoods, districts, subdivisions, or named local areas detailed in the provided police report narrative. \\
\mbox{} \\
First, identify everyone who fits the role of accused person using role and event context only. \\
\mbox{} \\
Consider all people in that role who are mentioned in the narrative. \\
\mbox{} \\
Ignore people who clearly belong to a different role. \\
\mbox{} \\
\mbox{} \\
Next, consider all listed neighborhoods, districts, subdivisions, or named local areas central to the incident. This includes, e.g., neighborhood names, district names, subdivision names, and other named local areas tied to the incident. Do NOT use direct descriptions of the accused person, including explicit race or ethnicity labels or physical descriptions as evidence for this feature. \\
\mbox{} \\
Based ONLY on neighborhoods or named local areas like these, and *ANY* others that are present in the narrative, predict whether the selected target person is White or non-White. You can consider higher-level geographies, like city, county, and state, in combination with neighborhood or named local area when making your prediction. Do not consider more fine-grained geographic information, like street names, intersections, or address numbers. Provide a one-sentence justification for your binary prediction first, then provide white\_score and non\_white\_score as integers from 0 to 100. \\
\mbox{} \\
The two scores must sum to exactly 100. 50/50 means no directional evidence favors either White or non-White. Scores should reflect the direction and strength of the evidence, not confidence. Indirect cultural, geographic, social, or contextual evidence should move the scores away from 50/50 in proportion to its strength, even when it is not definitive. Do not treat lack of direct demographic identifiers as a reason to automatically return 50/50. \\
\mbox{} \\
You are required to return both scores, even if the neighborhoods, districts, subdivisions, or named local areas are weak or ambiguous toward whether the accused person is White or non-White.
\end{flushleft}
\paragraph{Relevance prompt}
\begin{flushleft}\footnotesize\ttfamily
Classify if the provided police report narrative contains any mentions of neighborhoods, districts, subdivisions, or named local areas. \\
\mbox{} \\
Consider all listed neighborhoods, districts, subdivisions, or named local areas central to the incident. This includes, e.g., neighborhood names, district names, subdivision names, and other named local areas tied to the incident. \\
\mbox{} \\
Do NOT use direct descriptions of the accused person, including explicit race or ethnicity labels or physical descriptions as evidence for this feature. \\
\mbox{} \\
Provide a one-sentence justification for your classification first, then provide a binary classification. \\
\mbox{} \\
Use 1 if the narrative contains any mention of neighborhoods, districts, subdivisions, or named local areas as described above, even if it\textquotesingle{}s a weak reference. Use 0 if the narrative contains absolutely no matching information.
\end{flushleft}

\subsubsection{Interpretable feature prompt: city-name\_\_accused-person\_\_specific}
\paragraph{Scoring prompt}
\begin{flushleft}\footnotesize\ttfamily
Predict the race of accused person based on all cities detailed in the provided police report narrative. \\
\mbox{} \\
First, identify everyone who fits the role of accused person using role and event context only. \\
\mbox{} \\
Consider all people in that role who are mentioned in the narrative. \\
\mbox{} \\
Ignore people who clearly belong to a different role. \\
\mbox{} \\
\mbox{} \\
Next, consider all listed cities central to the incident. This includes, e.g., city names, municipality names, or equivalent named cities tied to the incident. Do NOT use direct descriptions of the accused person, including explicit race or ethnicity labels or physical descriptions as evidence for this feature. \\
\mbox{} \\
Based ONLY on cities like these, and *ANY* others that are present in the narrative, predict whether the selected target person is White or non-White. You can consider higher-level geographies, like county and state, in combination with city when making your prediction. Do not consider more fine-grained geographic information, like neighborhood, street, intersection, or address number. Provide a one-sentence justification for your binary prediction first, then provide white\_score and non\_white\_score as integers from 0 to 100. \\
\mbox{} \\
The two scores must sum to exactly 100. 50/50 means no directional evidence favors either White or non-White. Scores should reflect the direction and strength of the evidence, not confidence. Indirect cultural, geographic, social, or contextual evidence should move the scores away from 50/50 in proportion to its strength, even when it is not definitive. Do not treat lack of direct demographic identifiers as a reason to automatically return 50/50. \\
\mbox{} \\
You are required to return both scores, even if the cities are weak or ambiguous toward whether the accused person is White or non-White.
\end{flushleft}
\paragraph{Relevance prompt}
\begin{flushleft}\footnotesize\ttfamily
Classify if the provided police report narrative contains any mentions of cities. \\
\mbox{} \\
Consider all listed cities central to the incident. This includes, e.g., city names, municipality names, or equivalent named cities tied to the incident. \\
\mbox{} \\
Do NOT use direct descriptions of the accused person, including explicit race or ethnicity labels or physical descriptions as evidence for this feature. \\
\mbox{} \\
Provide a one-sentence justification for your classification first, then provide a binary classification. \\
\mbox{} \\
Use 1 if the narrative contains any mention of cities as described above, even if it\textquotesingle{}s a weak reference. Use 0 if the narrative contains absolutely no matching information.
\end{flushleft}

\subsubsection{Interpretable feature prompt: county-name\_\_accused-person\_\_specific}
\paragraph{Scoring prompt}
\begin{flushleft}\footnotesize\ttfamily
Predict the race of accused person based on all counties detailed in the provided police report narrative. \\
\mbox{} \\
First, identify everyone who fits the role of accused person using role and event context only. \\
\mbox{} \\
Consider all people in that role who are mentioned in the narrative. \\
\mbox{} \\
Ignore people who clearly belong to a different role. \\
\mbox{} \\
\mbox{} \\
Next, consider all listed counties central to the incident. This includes, e.g., county names, parish names, borough names, or equivalent county-level areas tied to the incident. Do NOT use direct descriptions of the accused person, including explicit race or ethnicity labels or physical descriptions as evidence for this feature. \\
\mbox{} \\
Based ONLY on counties like these, and *ANY* others that are present in the narrative, predict whether the selected target person is White or non-White. You can consider higher-level geographies, like state, in combination with county when making your prediction. Do not consider more fine-grained geographic information, like city, neighborhood, street, intersection, or address number. Provide a one-sentence justification for your binary prediction first, then provide white\_score and non\_white\_score as integers from 0 to 100. \\
\mbox{} \\
The two scores must sum to exactly 100. 50/50 means no directional evidence favors either White or non-White. Scores should reflect the direction and strength of the evidence, not confidence. Indirect cultural, geographic, social, or contextual evidence should move the scores away from 50/50 in proportion to its strength, even when it is not definitive. Do not treat lack of direct demographic identifiers as a reason to automatically return 50/50. \\
\mbox{} \\
You are required to return both scores, even if the counties are weak or ambiguous toward whether the accused person is White or non-White.
\end{flushleft}
\paragraph{Relevance prompt}
\begin{flushleft}\footnotesize\ttfamily
Classify if the provided police report narrative contains any mentions of counties. \\
\mbox{} \\
Consider all listed counties central to the incident. This includes, e.g., county names, parish names, borough names, or equivalent county-level areas tied to the incident. \\
\mbox{} \\
Do NOT use direct descriptions of the accused person, including explicit race or ethnicity labels or physical descriptions as evidence for this feature. \\
\mbox{} \\
Provide a one-sentence justification for your classification first, then provide a binary classification. \\
\mbox{} \\
Use 1 if the narrative contains any mention of counties as described above, even if it\textquotesingle{}s a weak reference. Use 0 if the narrative contains absolutely no matching information.
\end{flushleft}

\subsubsection{Interpretable feature prompt: state-name\_\_accused-person\_\_specific}
\paragraph{Scoring prompt}
\begin{flushleft}\footnotesize\ttfamily
Predict the race of accused person based on all states detailed in the provided police report narrative. \\
\mbox{} \\
First, identify everyone who fits the role of accused person using role and event context only. \\
\mbox{} \\
Consider all people in that role who are mentioned in the narrative. \\
\mbox{} \\
Ignore people who clearly belong to a different role. \\
\mbox{} \\
\mbox{} \\
Next, consider all listed states central to the incident. This includes, e.g., full state names, state abbreviations, or equivalent state-level references tied to the incident. Do NOT use direct descriptions of the accused person, including explicit race or ethnicity labels or physical descriptions as evidence for this feature. \\
\mbox{} \\
Based ONLY on states like these, and *ANY* others that are present in the narrative, predict whether the selected target person is White or non-White. Provide a one-sentence justification for your binary prediction first, then provide white\_score and non\_white\_score as integers from 0 to 100. \\
\mbox{} \\
The two scores must sum to exactly 100. 50/50 means no directional evidence favors either White or non-White. Scores should reflect the direction and strength of the evidence, not confidence. Indirect cultural, geographic, social, or contextual evidence should move the scores away from 50/50 in proportion to its strength, even when it is not definitive. Do not treat lack of direct demographic identifiers as a reason to automatically return 50/50. \\
\mbox{} \\
You are required to return both scores, even if the states are weak or ambiguous toward whether the accused person is White or non-White.
\end{flushleft}
\paragraph{Relevance prompt}
\begin{flushleft}\footnotesize\ttfamily
Classify if the provided police report narrative contains any mentions of states. \\
\mbox{} \\
Consider all listed states central to the incident. This includes, e.g., full state names, state abbreviations, or equivalent state-level references tied to the incident. \\
\mbox{} \\
Do NOT use direct descriptions of the accused person, including explicit race or ethnicity labels or physical descriptions as evidence for this feature. \\
\mbox{} \\
Provide a one-sentence justification for your classification first, then provide a binary classification. \\
\mbox{} \\
Use 1 if the narrative contains any mention of states as described above, even if it\textquotesingle{}s a weak reference. Use 0 if the narrative contains absolutely no matching information.
\end{flushleft}

\subsubsection{Interpretable feature prompt: street-name\_\_accused-person\_\_specific}
\paragraph{Scoring prompt}
\begin{flushleft}\footnotesize\ttfamily
Predict the race of accused person based on all streets or roads detailed in the provided police report narrative. \\
\mbox{} \\
First, identify everyone who fits the role of accused person using role and event context only. \\
\mbox{} \\
Consider all people in that role who are mentioned in the narrative. \\
\mbox{} \\
Ignore people who clearly belong to a different role. \\
\mbox{} \\
\mbox{} \\
Next, consider all listed streets or roads central to the incident. This includes, e.g., street names, road names, avenue names, boulevard names, highway names, or similar named roads tied to the incident. Do NOT use direct descriptions of the accused person, including explicit race or ethnicity labels or physical descriptions as evidence for this feature. \\
\mbox{} \\
Based ONLY on streets or roads like these, and *ANY* others that are present in the narrative, predict whether the selected target person is White or non-White. You can consider higher-level geographies, like city, county, and state, in combination with street or road when making your prediction. Do not consider more fine-grained geographic information, like address number, unit number, or apartment number. Provide a one-sentence justification for your binary prediction first, then provide white\_score and non\_white\_score as integers from 0 to 100. \\
\mbox{} \\
The two scores must sum to exactly 100. 50/50 means no directional evidence favors either White or non-White. Scores should reflect the direction and strength of the evidence, not confidence. Indirect cultural, geographic, social, or contextual evidence should move the scores away from 50/50 in proportion to its strength, even when it is not definitive. Do not treat lack of direct demographic identifiers as a reason to automatically return 50/50. \\
\mbox{} \\
You are required to return both scores, even if the streets or roads are weak or ambiguous toward whether the accused person is White or non-White.
\end{flushleft}
\paragraph{Relevance prompt}
\begin{flushleft}\footnotesize\ttfamily
Classify if the provided police report narrative contains any mentions of streets or roads. \\
\mbox{} \\
Consider all listed streets or roads central to the incident. This includes, e.g., street names, road names, avenue names, boulevard names, highway names, or similar named roads tied to the incident. \\
\mbox{} \\
Do NOT use direct descriptions of the accused person, including explicit race or ethnicity labels or physical descriptions as evidence for this feature. \\
\mbox{} \\
Provide a one-sentence justification for your classification first, then provide a binary classification. \\
\mbox{} \\
Use 1 if the narrative contains any mention of streets or roads as described above, even if it\textquotesingle{}s a weak reference. Use 0 if the narrative contains absolutely no matching information.
\end{flushleft}

\subsubsection{Interpretable feature prompt: intersection-name\_\_accused-person\_\_specific}
\paragraph{Scoring prompt}
\begin{flushleft}\footnotesize\ttfamily
Predict the race of accused person based on all intersections detailed in the provided police report narrative. \\
\mbox{} \\
First, identify everyone who fits the role of accused person using role and event context only. \\
\mbox{} \\
Consider all people in that role who are mentioned in the narrative. \\
\mbox{} \\
Ignore people who clearly belong to a different role. \\
\mbox{} \\
\mbox{} \\
Next, consider all listed intersections central to the incident. This includes, e.g., named cross-streets, intersections, junctions, or similar intersection references tied to the incident. Do NOT use direct descriptions of the accused person, including explicit race or ethnicity labels or physical descriptions as evidence for this feature. \\
\mbox{} \\
Based ONLY on intersections like these, and *ANY* others that are present in the narrative, predict whether the selected target person is White or non-White. You can consider higher-level geographies, like city, county, and state, in combination with intersection when making your prediction. Do not consider more fine-grained geographic information, like address number, unit number, or apartment number. Provide a one-sentence justification for your binary prediction first, then provide white\_score and non\_white\_score as integers from 0 to 100. \\
\mbox{} \\
The two scores must sum to exactly 100. 50/50 means no directional evidence favors either White or non-White. Scores should reflect the direction and strength of the evidence, not confidence. Indirect cultural, geographic, social, or contextual evidence should move the scores away from 50/50 in proportion to its strength, even when it is not definitive. Do not treat lack of direct demographic identifiers as a reason to automatically return 50/50. \\
\mbox{} \\
You are required to return both scores, even if the intersections are weak or ambiguous toward whether the accused person is White or non-White.
\end{flushleft}
\paragraph{Relevance prompt}
\begin{flushleft}\footnotesize\ttfamily
Classify if the provided police report narrative contains any mentions of intersections. \\
\mbox{} \\
Consider all listed intersections central to the incident. This includes, e.g., named cross-streets, intersections, junctions, or similar intersection references tied to the incident. \\
\mbox{} \\
Do NOT use direct descriptions of the accused person, including explicit race or ethnicity labels or physical descriptions as evidence for this feature. \\
\mbox{} \\
Provide a one-sentence justification for your classification first, then provide a binary classification. \\
\mbox{} \\
Use 1 if the narrative contains any mention of intersections as described above, even if it\textquotesingle{}s a weak reference. Use 0 if the narrative contains absolutely no matching information.
\end{flushleft}

\subsubsection{Interpretable feature prompt: address-name\_\_accused-person\_\_specific}
\paragraph{Scoring prompt}
\begin{flushleft}\footnotesize\ttfamily
Predict the race of accused person based on all street addresses detailed in the provided police report narrative. \\
\mbox{} \\
First, identify everyone who fits the role of accused person using role and event context only. \\
\mbox{} \\
Consider all people in that role who are mentioned in the narrative. \\
\mbox{} \\
Ignore people who clearly belong to a different role. \\
\mbox{} \\
\mbox{} \\
Next, consider all listed street addresses central to the incident. This includes, e.g., numbered street addresses, block addresses, or similar address references tied to the incident. Do NOT use direct descriptions of the accused person, including explicit race or ethnicity labels or physical descriptions as evidence for this feature. \\
\mbox{} \\
Based ONLY on addresses like these, and *ANY* others that are present in the narrative, predict whether the selected target person is White or non-White. You can consider higher-level geographies, like neighborhood, city, county, and state, in combination with the address when making your prediction. Provide a one-sentence justification for your binary prediction first, then provide white\_score and non\_white\_score as integers from 0 to 100. \\
\mbox{} \\
The two scores must sum to exactly 100. 50/50 means no directional evidence favors either White or non-White. Scores should reflect the direction and strength of the evidence, not confidence. Indirect cultural, geographic, social, or contextual evidence should move the scores away from 50/50 in proportion to its strength, even when it is not definitive. Do not treat lack of direct demographic identifiers as a reason to automatically return 50/50. \\
\mbox{} \\
You are required to return both scores, even if the street addresses are weak or ambiguous toward whether the accused person is White or non-White.
\end{flushleft}
\paragraph{Relevance prompt}
\begin{flushleft}\footnotesize\ttfamily
Classify if the provided police report narrative contains any mentions of street addresses. \\
\mbox{} \\
Consider all listed street addresses central to the incident. This includes, e.g., numbered street addresses, block addresses, or similar address references tied to the incident. \\
\mbox{} \\
Do NOT use direct descriptions of the accused person, including explicit race or ethnicity labels or physical descriptions as evidence for this feature. \\
\mbox{} \\
Provide a one-sentence justification for your classification first, then provide a binary classification. \\
\mbox{} \\
Use 1 if the narrative contains any mention of street addresses as described above, even if it\textquotesingle{}s a weak reference. Use 0 if the narrative contains absolutely no matching information.
\end{flushleft}

\subsubsection{Interpretable feature prompt: named-location\_\_accused-person\_\_specific}
\paragraph{Scoring prompt}
\begin{flushleft}\footnotesize\ttfamily
Predict the race of accused person based on all named locations detailed in the provided police report narrative. \\
\mbox{} \\
First, identify everyone who fits the role of accused person using role and event context only. \\
\mbox{} \\
Consider all people in that role who are mentioned in the narrative. \\
\mbox{} \\
Ignore people who clearly belong to a different role. \\
\mbox{} \\
\mbox{} \\
Next, consider all named locations central to the incident. This includes, e.g., named businesses such as restaurants, grocery stores, bars, gas stations, gyms, commercial establishments, and private firms; commercial-establishment types such as convenience stores, luxury retailers, pawn shops, bars, restaurants, or hotels; landmarks or named local places such as parks, plazas, monuments, named gathering spots, or other named local places; institutions such as hospitals, clinics, jails, schools, courthouses, shelters, religious institutions, social-service offices, and government offices; and transit locations such as stations, stops, terminals, platforms, transit centers, or other named transit locations. Do NOT use direct descriptions of the accused person, including explicit race or ethnicity labels or physical descriptions as evidence for this feature. \\
\mbox{} \\
Based ONLY on named locations like these, and *ANY* others that are present in the narrative, predict whether the selected target person is White or non-White. You can consider higher-level geographies, like city, county, and state, in combination with the named location when making your prediction. Provide a one-sentence justification for your binary prediction first, then provide white\_score and non\_white\_score as integers from 0 to 100. \\
\mbox{} \\
The two scores must sum to exactly 100. 50/50 means no directional evidence favors either White or non-White. Scores should reflect the direction and strength of the evidence, not confidence. Indirect cultural, geographic, social, or contextual evidence should move the scores away from 50/50 in proportion to its strength, even when it is not definitive. Do not treat lack of direct demographic identifiers as a reason to automatically return 50/50. \\
\mbox{} \\
You are required to return both scores, even if the named locations are weak or ambiguous toward whether the accused person is White or non-White.
\end{flushleft}
\paragraph{Relevance prompt}
\begin{flushleft}\footnotesize\ttfamily
Classify if the provided police report narrative contains any mentions of named locations. \\
\mbox{} \\
Consider all named locations central to the incident. This includes, e.g., named businesses such as restaurants, grocery stores, bars, gas stations, gyms, commercial establishments, and private firms; commercial-establishment types such as convenience stores, luxury retailers, pawn shops, bars, restaurants, or hotels; landmarks or named local places such as parks, plazas, monuments, named gathering spots, or other named local places; institutions such as hospitals, clinics, jails, schools, courthouses, shelters, religious institutions, social-service offices, and government offices; and transit locations such as stations, stops, terminals, platforms, transit centers, or other named transit locations. \\
\mbox{} \\
Do NOT use direct descriptions of the accused person, including explicit race or ethnicity labels or physical descriptions as evidence for this feature. \\
\mbox{} \\
Provide a one-sentence justification for your classification first, then provide a binary classification. \\
\mbox{} \\
Use 1 if the narrative contains any mention of named locations as described above, even if it\textquotesingle{}s a weak reference. Use 0 if the narrative contains absolutely no matching information.
\end{flushleft}

\subsubsection{Interpretable feature prompt: housing-context\_\_accused-person\_\_specific}
\paragraph{Scoring prompt}
\begin{flushleft}\footnotesize\ttfamily
Predict the race of accused person based on all housing-context information detailed in the provided police report narrative. \\
\mbox{} \\
First, identify everyone who fits the role of accused person using role and event context only. \\
\mbox{} \\
Consider all people in that role who are mentioned in the narrative. \\
\mbox{} \\
Ignore people who clearly belong to a different role. \\
\mbox{} \\
\mbox{} \\
Next, consider all housing-context information associated with accused person or central to the incident. This includes, e.g., housing-status cues such as homelessness, unstable housing, shelter use, encampment residence, apartment living, single-family homes, large homes, luxury residences, or similar housing-status cues, and housing-type cues such as apartment complex, single-family home, trailer, motel, encampment, assisted living facility, or similar residential types. Do NOT use direct descriptions of the accused person, including explicit race or ethnicity labels or physical descriptions as evidence for this feature. \\
\mbox{} \\
Based ONLY on housing-context cues like these, and *ANY* others that are present in the narrative, predict whether the selected target person is White or non-White. Provide a one-sentence justification for your binary prediction first, then provide white\_score and non\_white\_score as integers from 0 to 100. \\
\mbox{} \\
The two scores must sum to exactly 100. 50/50 means no directional evidence favors either White or non-White. Scores should reflect the direction and strength of the evidence, not confidence. Indirect cultural, geographic, social, or contextual evidence should move the scores away from 50/50 in proportion to its strength, even when it is not definitive. Do not treat lack of direct demographic identifiers as a reason to automatically return 50/50. \\
\mbox{} \\
You are required to return both scores, even if the housing-context information are weak or ambiguous toward whether the accused person is White or non-White.
\end{flushleft}
\paragraph{Relevance prompt}
\begin{flushleft}\footnotesize\ttfamily
Classify if the provided police report narrative contains any mentions of housing-context information. \\
\mbox{} \\
Consider all housing-context information associated with accused person or central to the incident. This includes, e.g., housing-status cues such as homelessness, unstable housing, shelter use, encampment residence, apartment living, single-family homes, large homes, luxury residences, or similar housing-status cues, and housing-type cues such as apartment complex, single-family home, trailer, motel, encampment, assisted living facility, or similar residential types. \\
\mbox{} \\
Do NOT use direct descriptions of the accused person, including explicit race or ethnicity labels or physical descriptions as evidence for this feature. \\
\mbox{} \\
Provide a one-sentence justification for your classification first, then provide a binary classification. \\
\mbox{} \\
Use 1 if the narrative contains any mention of housing-context information as described above, even if it\textquotesingle{}s a weak reference. Use 0 if the narrative contains absolutely no matching information.
\end{flushleft}

\subsubsection{Interpretable feature prompt: built-environment\_\_accused-person\_\_specific}
\paragraph{Scoring prompt}
\begin{flushleft}\footnotesize\ttfamily
Predict the race of accused person based on all built-environment cues detailed in the provided police report narrative. \\
\mbox{} \\
First, identify everyone who fits the role of accused person using role and event context only. \\
\mbox{} \\
Consider all people in that role who are mentioned in the narrative. \\
\mbox{} \\
Ignore people who clearly belong to a different role. \\
\mbox{} \\
\mbox{} \\
Next, consider all built-environment cues central to the incident. This includes, e.g., abandoned property, boarded windows, upscale finishings, visible disrepair, gated entry, security barriers, industrial buildout, or similar built-environment signals. Do NOT use direct descriptions of the accused person, including explicit race or ethnicity labels or physical descriptions as evidence for this feature. \\
\mbox{} \\
Based ONLY on built-environment cues like these, and *ANY* others that are present in the narrative, predict whether the selected target person is White or non-White. Provide a one-sentence justification for your binary prediction first, then provide white\_score and non\_white\_score as integers from 0 to 100. \\
\mbox{} \\
The two scores must sum to exactly 100. 50/50 means no directional evidence favors either White or non-White. Scores should reflect the direction and strength of the evidence, not confidence. Indirect cultural, geographic, social, or contextual evidence should move the scores away from 50/50 in proportion to its strength, even when it is not definitive. Do not treat lack of direct demographic identifiers as a reason to automatically return 50/50. \\
\mbox{} \\
You are required to return both scores, even if the built-environment cues are weak or ambiguous toward whether the accused person is White or non-White.
\end{flushleft}
\paragraph{Relevance prompt}
\begin{flushleft}\footnotesize\ttfamily
Classify if the provided police report narrative contains any mentions of built-environment cues. \\
\mbox{} \\
Consider all built-environment cues central to the incident. This includes, e.g., abandoned property, boarded windows, upscale finishings, visible disrepair, gated entry, security barriers, industrial buildout, or similar built-environment signals. \\
\mbox{} \\
Do NOT use direct descriptions of the accused person, including explicit race or ethnicity labels or physical descriptions as evidence for this feature. \\
\mbox{} \\
Provide a one-sentence justification for your classification first, then provide a binary classification. \\
\mbox{} \\
Use 1 if the narrative contains any mention of built-environment cues as described above, even if it\textquotesingle{}s a weak reference. Use 0 if the narrative contains absolutely no matching information.
\end{flushleft}

\subsubsection{Interpretable feature prompt: behavior\_\_accused-person\_\_specific}
\paragraph{Scoring prompt}
\begin{flushleft}\footnotesize\ttfamily
Predict the race of accused person based on all behavior detailed in the provided police report narrative. \\
\mbox{} \\
First, identify everyone who fits the role of accused person using role and event context only. \\
\mbox{} \\
Consider all people in that role who are mentioned in the narrative. \\
\mbox{} \\
Ignore people who clearly belong to a different role. \\
\mbox{} \\
\mbox{} \\
Next, consider all behavior associated with accused person in the narrative. This includes, e.g., cooperation, refusal, noncooperation, flight, evasion, aggression, confrontational behavior, intoxication, apparent impairment, or similar observed behavior. Do NOT use direct descriptions of the accused person, including explicit race or ethnicity labels or physical descriptions as evidence for this feature. \\
\mbox{} \\
Based ONLY on behavioral cues like these, and *ANY* others that are present in the narrative, predict whether the selected target person is White or non-White. Provide a one-sentence justification for your binary prediction first, then provide white\_score and non\_white\_score as integers from 0 to 100. \\
\mbox{} \\
The two scores must sum to exactly 100. 50/50 means no directional evidence favors either White or non-White. Scores should reflect the direction and strength of the evidence, not confidence. Indirect cultural, geographic, social, or contextual evidence should move the scores away from 50/50 in proportion to its strength, even when it is not definitive. Do not treat lack of direct demographic identifiers as a reason to automatically return 50/50. \\
\mbox{} \\
You are required to return both scores, even if the behavior are weak or ambiguous toward whether the accused person is White or non-White.
\end{flushleft}
\paragraph{Relevance prompt}
\begin{flushleft}\footnotesize\ttfamily
Classify if the provided police report narrative contains any mentions of behavior. \\
\mbox{} \\
Consider all behavior associated with accused person in the narrative. This includes, e.g., cooperation, refusal, noncooperation, flight, evasion, aggression, confrontational behavior, intoxication, apparent impairment, or similar observed behavior. \\
\mbox{} \\
Do NOT use direct descriptions of the accused person, including explicit race or ethnicity labels or physical descriptions as evidence for this feature. \\
\mbox{} \\
Provide a one-sentence justification for your classification first, then provide a binary classification. \\
\mbox{} \\
Use 1 if the narrative contains any mention of behavior as described above, even if it\textquotesingle{}s a weak reference. Use 0 if the narrative contains absolutely no matching information.
\end{flushleft}

\subsubsection{Interpretable feature prompt: substance-use\_\_accused-person\_\_specific}
\paragraph{Scoring prompt}
\begin{flushleft}\footnotesize\ttfamily
Predict the race of accused person based on all drug, tobacco, or other substance-use information detailed in the provided police report narrative. \\
\mbox{} \\
First, identify everyone who fits the role of accused person using role and event context only. \\
\mbox{} \\
Consider all people in that role who are mentioned in the narrative. \\
\mbox{} \\
Ignore people who clearly belong to a different role. \\
\mbox{} \\
\mbox{} \\
Next, consider all drug, tobacco, or other substance-use information associated with accused person in the narrative. This includes, e.g., direct or indirect references to marijuana, cannabis, weed, cocaine, crack cocaine, heroin, methamphetamine, fentanyl, PCP, pills, prescription drug misuse, drug paraphernalia, intoxication from drugs, cigarettes, menthol cigarettes, tobacco, vaping, or similar substance-use cues. Do NOT use direct descriptions of the accused person, including explicit race or ethnicity labels or physical descriptions as evidence for this feature. \\
\mbox{} \\
Based ONLY on substance-use cues like these, and *ANY* others that are present in the narrative, predict whether the selected target person is White or non-White. Provide a one-sentence justification for your binary prediction first, then provide white\_score and non\_white\_score as integers from 0 to 100. \\
\mbox{} \\
The two scores must sum to exactly 100. 50/50 means no directional evidence favors either White or non-White. Scores should reflect the direction and strength of the evidence, not confidence. Indirect cultural, geographic, social, or contextual evidence should move the scores away from 50/50 in proportion to its strength, even when it is not definitive. Do not treat lack of direct demographic identifiers as a reason to automatically return 50/50. \\
\mbox{} \\
You are required to return both scores, even if the drug, tobacco, or other substance-use information are weak or ambiguous toward whether the accused person is White or non-White.
\end{flushleft}
\paragraph{Relevance prompt}
\begin{flushleft}\footnotesize\ttfamily
Classify if the provided police report narrative contains any mentions of drug, tobacco, or other substance-use information. \\
\mbox{} \\
Consider all drug, tobacco, or other substance-use information associated with accused person in the narrative. This includes, e.g., direct or indirect references to marijuana, cannabis, weed, cocaine, crack cocaine, heroin, methamphetamine, fentanyl, PCP, pills, prescription drug misuse, drug paraphernalia, intoxication from drugs, cigarettes, menthol cigarettes, tobacco, vaping, or similar substance-use cues. \\
\mbox{} \\
Do NOT use direct descriptions of the accused person, including explicit race or ethnicity labels or physical descriptions as evidence for this feature. \\
\mbox{} \\
Provide a one-sentence justification for your classification first, then provide a binary classification. \\
\mbox{} \\
Use 1 if the narrative contains any mention of drug, tobacco, or other substance-use information as described above, even if it\textquotesingle{}s a weak reference. Use 0 if the narrative contains absolutely no matching information.
\end{flushleft}

\subsubsection{Interpretable feature prompt: education-attainment\_\_accused-person\_\_specific}
\paragraph{Scoring prompt}
\begin{flushleft}\footnotesize\ttfamily
Predict the race of accused person based on all education-attainment information detailed in the provided police report narrative. \\
\mbox{} \\
First, identify everyone who fits the role of accused person using role and event context only. \\
\mbox{} \\
Consider all people in that role who are mentioned in the narrative. \\
\mbox{} \\
Ignore people who clearly belong to a different role. \\
\mbox{} \\
\mbox{} \\
Next, consider all explicit education-attainment information associated with accused person in the narrative. This includes, e.g., direct references to school completion, degrees, credentials, enrollment, or educational background. Do NOT use direct descriptions of the accused person, including explicit race or ethnicity labels or physical descriptions as evidence for this feature. \\
\mbox{} \\
Based ONLY on education-attainment cues like these, and *ANY* others that are present in the narrative, predict whether the selected target person is White or non-White. Provide a one-sentence justification for your binary prediction first, then provide white\_score and non\_white\_score as integers from 0 to 100. \\
\mbox{} \\
The two scores must sum to exactly 100. 50/50 means no directional evidence favors either White or non-White. Scores should reflect the direction and strength of the evidence, not confidence. Indirect cultural, geographic, social, or contextual evidence should move the scores away from 50/50 in proportion to its strength, even when it is not definitive. Do not treat lack of direct demographic identifiers as a reason to automatically return 50/50. \\
\mbox{} \\
You are required to return both scores, even if the education-attainment information are weak or ambiguous toward whether the accused person is White or non-White.
\end{flushleft}
\paragraph{Relevance prompt}
\begin{flushleft}\footnotesize\ttfamily
Classify if the provided police report narrative contains any mentions of education-attainment information. \\
\mbox{} \\
Consider all explicit education-attainment information associated with accused person in the narrative. This includes, e.g., direct references to school completion, degrees, credentials, enrollment, or educational background. \\
\mbox{} \\
Do NOT use direct descriptions of the accused person, including explicit race or ethnicity labels or physical descriptions as evidence for this feature. \\
\mbox{} \\
Provide a one-sentence justification for your classification first, then provide a binary classification. \\
\mbox{} \\
Use 1 if the narrative contains any mention of education-attainment information as described above, even if it\textquotesingle{}s a weak reference. Use 0 if the narrative contains absolutely no matching information.
\end{flushleft}

\subsubsection{Interpretable feature prompt: justice-history-or-supervision-status\_\_accused-person\_\_specific}
\paragraph{Scoring prompt}
\begin{flushleft}\footnotesize\ttfamily
Predict the race of accused person based on all prior incarceration, custody-history, warrant, or active justice-supervision information detailed in the provided police report narrative. \\
\mbox{} \\
First, identify everyone who fits the role of accused person using role and event context only. \\
\mbox{} \\
Consider all people in that role who are mentioned in the narrative. \\
\mbox{} \\
Ignore people who clearly belong to a different role. \\
\mbox{} \\
\mbox{} \\
Next, consider all prior incarceration, custody-history, warrant, or active justice-supervision information associated with accused person in the narrative. This includes, e.g., direct references to prior incarceration, prior jail or prison history, recent release, warrants, probation, parole, supervision, court monitoring, or similar custody-history or justice-supervision status. Do NOT use direct descriptions of the accused person, including explicit race or ethnicity labels or physical descriptions as evidence for this feature. \\
\mbox{} \\
Based ONLY on custody-history or justice-supervision cues like these, and *ANY* others that are present in the narrative, predict whether the selected target person is White or non-White. Provide a one-sentence justification for your binary prediction first, then provide white\_score and non\_white\_score as integers from 0 to 100. \\
\mbox{} \\
The two scores must sum to exactly 100. 50/50 means no directional evidence favors either White or non-White. Scores should reflect the direction and strength of the evidence, not confidence. Indirect cultural, geographic, social, or contextual evidence should move the scores away from 50/50 in proportion to its strength, even when it is not definitive. Do not treat lack of direct demographic identifiers as a reason to automatically return 50/50. \\
\mbox{} \\
You are required to return both scores, even if the prior incarceration, custody-history, warrant, or active justice-supervision information are weak or ambiguous toward whether the accused person is White or non-White.
\end{flushleft}
\paragraph{Relevance prompt}
\begin{flushleft}\footnotesize\ttfamily
Classify if the provided police report narrative contains any mentions of prior incarceration, custody-history, warrant, or active justice-supervision information. \\
\mbox{} \\
Consider all prior incarceration, custody-history, warrant, or active justice-supervision information associated with accused person in the narrative. This includes, e.g., direct references to prior incarceration, prior jail or prison history, recent release, warrants, probation, parole, supervision, court monitoring, or similar custody-history or justice-supervision status. \\
\mbox{} \\
Do NOT use direct descriptions of the accused person, including explicit race or ethnicity labels or physical descriptions as evidence for this feature. \\
\mbox{} \\
Provide a one-sentence justification for your classification first, then provide a binary classification. \\
\mbox{} \\
Use 1 if the narrative contains any mention of prior incarceration, custody-history, warrant, or active justice-supervision information as described above, even if it\textquotesingle{}s a weak reference. Use 0 if the narrative contains absolutely no matching information.
\end{flushleft}

\subsubsection{Interpretable feature prompt: number-of-children-count\_\_score}
\paragraph{Scoring prompt}
\begin{flushleft}\footnotesize\ttfamily
Count children referenced in the accused person\textquotesingle{}s incident or household context in the provided police report narrative. \\
\mbox{} \\
Use only the specific feature named in the prompt. \\
\mbox{} \\
Consider all relevant evidence for that feature across the narrative. \\
\mbox{} \\
Ignore all non-feature information unless it is necessary to understand the requested feature. \\
\mbox{} \\
\mbox{} \\
Consider all evidence in the narrative about children referenced in the accused person\textquotesingle{}s incident or household context. This includes, e.g., references to children being present, children living in the household, children mentioned by relation, or children otherwise tied to the incident context. \\
\mbox{} \\
Based on evidence like this, how many children are explicitly referenced in relation to the accused person\textquotesingle{}s incident or household context? Count only children, not adults. \\
\mbox{} \\
Provide a one-sentence justification for your count first, then provide one non-negative integer. \\
\mbox{} \\
Count only instances matching the requested feature as described above. \\
\mbox{} \\
You are required to return a value whenever the narrative contains information relevant to the requested count. Use 0 when the relevant information indicates that there are no matching instances.
\end{flushleft}

\subsubsection{Interpretable feature prompt: household-composition-or-structure-indicator\_\_score}
\paragraph{Scoring prompt}
\begin{flushleft}\footnotesize\ttfamily
Classify whether the provided police report narrative indicates a notable household composition or structure context. \\
\mbox{} \\
Use only the specific feature named in the prompt. \\
\mbox{} \\
Consider all relevant evidence for that feature across the narrative. \\
\mbox{} \\
Ignore all non-feature information unless it is necessary to understand the requested feature. \\
\mbox{} \\
\mbox{} \\
Consider all evidence in the narrative about household composition or structure involving the accused person. This includes, e.g., a stepfamily or blended-family context, a multigenerational household, an extended-family or broader kinship arrangement, a separated or estranged household, split-custody, an absent-member household, or another non-nuclear or incomplete household context. \\
\mbox{} \\
Based on evidence like this, does the narrative indicate a notable household composition or structure context involving the accused person? \\
\mbox{} \\
Provide a one-sentence justification for your classification first, then provide a binary classification. \\
\mbox{} \\
Use 1 if the narrative indicates the requested feature is present, even if it is a weak or ambiguous reference. Use 0 if the narrative indicates the requested feature is absent. \\
\mbox{} \\
You are required to return a value whenever the narrative contains information relevant to the requested feature.
\end{flushleft}

\subsubsection{Interpretable feature prompt: public-transit-dependence-indicator\_\_score}
\paragraph{Scoring prompt}
\begin{flushleft}\footnotesize\ttfamily
Classify whether the provided police report narrative indicates public-transit dependence or use. \\
\mbox{} \\
Use only the specific feature named in the prompt. \\
\mbox{} \\
Consider all relevant evidence for that feature across the narrative. \\
\mbox{} \\
Ignore all non-feature information unless it is necessary to understand the requested feature. \\
\mbox{} \\
\mbox{} \\
Consider all evidence in the narrative about transportation used by or available to the accused person in the incident context. This includes, e.g., references to buses, trains, transit stops, transit passes, waiting for transit, or relying on public transit to travel. \\
\mbox{} \\
Based on evidence like this, does the narrative indicate that the accused person depends on or is using public transit as part of the incident context? \\
\mbox{} \\
Provide a one-sentence justification for your classification first, then provide a binary classification. \\
\mbox{} \\
Use 1 if the narrative indicates the requested feature is present, even if it is a weak or ambiguous reference. Use 0 if the narrative indicates the requested feature is absent. \\
\mbox{} \\
You are required to return a value whenever the narrative contains information relevant to the requested feature.
\end{flushleft}

\subsubsection{Interpretable feature prompt: timing-context-indicator\_\_score}
\paragraph{Scoring prompt}
\begin{flushleft}\footnotesize\ttfamily
Classify whether the provided police report narrative provides a meaningful timing context for the incident. \\
\mbox{} \\
Use only the specific feature named in the prompt. \\
\mbox{} \\
Consider all relevant evidence for that feature across the narrative. \\
\mbox{} \\
Ignore all non-feature information unless it is necessary to understand the requested feature. \\
\mbox{} \\
\mbox{} \\
Consider all evidence in the narrative about when the incident occurred. This includes, e.g., references to time of day, day of week, overnight timing, workday timing, commute timing, or other meaningful temporal context. \\
\mbox{} \\
Based on evidence like this, does the narrative provide a meaningful timing context for the incident, such as late night, overnight, ordinary workday hours, or a commute-related window? \\
\mbox{} \\
Provide a one-sentence justification for your classification first, then provide a binary classification. \\
\mbox{} \\
Use 1 if the narrative indicates the requested feature is present, even if it is a weak or ambiguous reference. Use 0 if the narrative indicates the requested feature is absent. \\
\mbox{} \\
You are required to return a value whenever the narrative contains information relevant to the requested feature.
\end{flushleft}

\subsubsection{Interpretable feature prompt: explicit-race-mentions-count\_\_score}
\paragraph{Scoring prompt}
\begin{flushleft}\footnotesize\ttfamily
Count explicit mentions of race, ethnicity, nationality, tribal identity, or comparable ethnoracial labels in the provided police report narrative. \\
\mbox{} \\
Use only the specific feature named in the prompt. \\
\mbox{} \\
Consider all relevant evidence for that feature across the narrative. \\
\mbox{} \\
Ignore all non-feature information unless it is necessary to understand the requested feature. \\
\mbox{} \\
\mbox{} \\
Consider all explicit mentions in the narrative of race, ethnicity, nationality, tribal identity, or comparable ethnoracial labels. This includes, e.g., direct mentions of race labels, ethnicity labels, nationality labels, tribal identity labels, or similar ethnoracial descriptors. \\
\mbox{} \\
Based on evidence like this, how many explicit mentions of race, ethnicity, nationality, tribal identity, or comparable ethnoracial labels appear anywhere in the narrative? \\
\mbox{} \\
Provide a one-sentence justification for your count first, then provide one non-negative integer. \\
\mbox{} \\
Count only instances matching the requested feature as described above. \\
\mbox{} \\
You are required to return a value whenever the narrative contains information relevant to the requested count. Use 0 when the relevant information indicates that there are no matching instances.
\end{flushleft}

\subsubsection{Interpretable feature prompt: incident-consistent-with-crime-type\_\_arson\_\_score}
\paragraph{Scoring prompt}
\begin{flushleft}\footnotesize\ttfamily
Classify how consistent the main incident described in the provided police report narrative is with arson. \\
\mbox{} \\
Use only the specific feature named in the prompt. \\
\mbox{} \\
Consider all relevant evidence for that feature across the narrative. \\
\mbox{} \\
Ignore all non-feature information unless it is necessary to understand the requested feature. \\
\mbox{} \\
\mbox{} \\
Consider all incident facts described in the narrative that are relevant to whether the main incident is consistent with arson. This includes, e.g., actions taken, harms described, property involved, threats, injuries, methods, intent cues, or other incident facts bearing on whether the incident matches arson. \\
\mbox{} \\
Based on incident facts like these, how likely is it that the main incident described in the narrative is consistent with arson? Use only the narrative facts about the incident itself, not any downstream charging decisions. \\
\mbox{} \\
Provide a one-sentence justification for your decision first, then provide one integer between 0 and 100, where 0 represents that the main incident described in the narrative is extremely unlikely to be consistent with arson, and 100 represents that the main incident described in the narrative is extremely likely to be consistent with arson. 50 means no relevant evidence favors either endpoint; indirect cultural, geographic, or contextual associations should move the score away from 50 in proportion to their strength, even when they are not definitive. \\
\mbox{} \\
Use only the narrative facts about the incident itself, not any downstream charging decisions. \\
\mbox{} \\
You are required to return a value whenever the narrative contains incident facts relevant to consistency with arson, even if the evidence is weak, indirect, subtle, or ambiguous.
\end{flushleft}

\subsubsection{Interpretable feature prompt: incident-consistent-with-crime-type\_\_assault\_\_score}
\paragraph{Scoring prompt}
\begin{flushleft}\footnotesize\ttfamily
Classify how consistent the main incident described in the provided police report narrative is with assault. \\
\mbox{} \\
Use only the specific feature named in the prompt. \\
\mbox{} \\
Consider all relevant evidence for that feature across the narrative. \\
\mbox{} \\
Ignore all non-feature information unless it is necessary to understand the requested feature. \\
\mbox{} \\
\mbox{} \\
Consider all incident facts described in the narrative that are relevant to whether the main incident is consistent with assault. This includes, e.g., actions taken, harms described, property involved, threats, injuries, methods, intent cues, or other incident facts bearing on whether the incident matches assault. \\
\mbox{} \\
Based on incident facts like these, how likely is it that the main incident described in the narrative is consistent with assault? Use only the narrative facts about the incident itself, not any downstream charging decisions. \\
\mbox{} \\
Provide a one-sentence justification for your decision first, then provide one integer between 0 and 100, where 0 represents that the main incident described in the narrative is extremely unlikely to be consistent with assault, and 100 represents that the main incident described in the narrative is extremely likely to be consistent with assault. 50 means no relevant evidence favors either endpoint; indirect cultural, geographic, or contextual associations should move the score away from 50 in proportion to their strength, even when they are not definitive. \\
\mbox{} \\
Use only the narrative facts about the incident itself, not any downstream charging decisions. \\
\mbox{} \\
You are required to return a value whenever the narrative contains incident facts relevant to consistency with assault, even if the evidence is weak, indirect, subtle, or ambiguous.
\end{flushleft}

\subsubsection{Interpretable feature prompt: incident-consistent-with-crime-type\_\_bribery\_\_score}
\paragraph{Scoring prompt}
\begin{flushleft}\footnotesize\ttfamily
Classify how consistent the main incident described in the provided police report narrative is with bribery. \\
\mbox{} \\
Use only the specific feature named in the prompt. \\
\mbox{} \\
Consider all relevant evidence for that feature across the narrative. \\
\mbox{} \\
Ignore all non-feature information unless it is necessary to understand the requested feature. \\
\mbox{} \\
\mbox{} \\
Consider all incident facts described in the narrative that are relevant to whether the main incident is consistent with bribery. This includes, e.g., actions taken, harms described, property involved, threats, injuries, methods, intent cues, or other incident facts bearing on whether the incident matches bribery. \\
\mbox{} \\
Based on incident facts like these, how likely is it that the main incident described in the narrative is consistent with bribery? Use only the narrative facts about the incident itself, not any downstream charging decisions. \\
\mbox{} \\
Provide a one-sentence justification for your decision first, then provide one integer between 0 and 100, where 0 represents that the main incident described in the narrative is extremely unlikely to be consistent with bribery, and 100 represents that the main incident described in the narrative is extremely likely to be consistent with bribery. 50 means no relevant evidence favors either endpoint; indirect cultural, geographic, or contextual associations should move the score away from 50 in proportion to their strength, even when they are not definitive. \\
\mbox{} \\
Use only the narrative facts about the incident itself, not any downstream charging decisions. \\
\mbox{} \\
You are required to return a value whenever the narrative contains incident facts relevant to consistency with bribery, even if the evidence is weak, indirect, subtle, or ambiguous.
\end{flushleft}

\subsubsection{Interpretable feature prompt: incident-consistent-with-crime-type\_\_burglary\_\_score}
\paragraph{Scoring prompt}
\begin{flushleft}\footnotesize\ttfamily
Classify how consistent the main incident described in the provided police report narrative is with burglary. \\
\mbox{} \\
Use only the specific feature named in the prompt. \\
\mbox{} \\
Consider all relevant evidence for that feature across the narrative. \\
\mbox{} \\
Ignore all non-feature information unless it is necessary to understand the requested feature. \\
\mbox{} \\
\mbox{} \\
Consider all incident facts described in the narrative that are relevant to whether the main incident is consistent with burglary. This includes, e.g., actions taken, harms described, property involved, threats, injuries, methods, intent cues, or other incident facts bearing on whether the incident matches burglary. \\
\mbox{} \\
Based on incident facts like these, how likely is it that the main incident described in the narrative is consistent with burglary? Use only the narrative facts about the incident itself, not any downstream charging decisions. \\
\mbox{} \\
Provide a one-sentence justification for your decision first, then provide one integer between 0 and 100, where 0 represents that the main incident described in the narrative is extremely unlikely to be consistent with burglary, and 100 represents that the main incident described in the narrative is extremely likely to be consistent with burglary. 50 means no relevant evidence favors either endpoint; indirect cultural, geographic, or contextual associations should move the score away from 50 in proportion to their strength, even when they are not definitive. \\
\mbox{} \\
Use only the narrative facts about the incident itself, not any downstream charging decisions. \\
\mbox{} \\
You are required to return a value whenever the narrative contains incident facts relevant to consistency with burglary, even if the evidence is weak, indirect, subtle, or ambiguous.
\end{flushleft}

\subsubsection{Interpretable feature prompt: incident-consistent-with-crime-type\_\_destruction-of-property\_\_score}
\paragraph{Scoring prompt}
\begin{flushleft}\footnotesize\ttfamily
Classify how consistent the main incident described in the provided police report narrative is with destruction of property. \\
\mbox{} \\
Use only the specific feature named in the prompt. \\
\mbox{} \\
Consider all relevant evidence for that feature across the narrative. \\
\mbox{} \\
Ignore all non-feature information unless it is necessary to understand the requested feature. \\
\mbox{} \\
\mbox{} \\
Consider all incident facts described in the narrative that are relevant to whether the main incident is consistent with destruction of property. This includes, e.g., actions taken, harms described, property involved, threats, injuries, methods, intent cues, or other incident facts bearing on whether the incident matches destruction of property. \\
\mbox{} \\
Based on incident facts like these, how likely is it that the main incident described in the narrative is consistent with destruction of property? Use only the narrative facts about the incident itself, not any downstream charging decisions. \\
\mbox{} \\
Provide a one-sentence justification for your decision first, then provide one integer between 0 and 100, where 0 represents that the main incident described in the narrative is extremely unlikely to be consistent with destruction of property, and 100 represents that the main incident described in the narrative is extremely likely to be consistent with destruction of property. 50 means no relevant evidence favors either endpoint; indirect cultural, geographic, or contextual associations should move the score away from 50 in proportion to their strength, even when they are not definitive. \\
\mbox{} \\
Use only the narrative facts about the incident itself, not any downstream charging decisions. \\
\mbox{} \\
You are required to return a value whenever the narrative contains incident facts relevant to consistency with destruction of property, even if the evidence is weak, indirect, subtle, or ambiguous.
\end{flushleft}

\subsubsection{Interpretable feature prompt: incident-consistent-with-crime-type\_\_drug-offenses\_\_score}
\paragraph{Scoring prompt}
\begin{flushleft}\footnotesize\ttfamily
Classify how consistent the main incident described in the provided police report narrative is with drug offenses. \\
\mbox{} \\
Use only the specific feature named in the prompt. \\
\mbox{} \\
Consider all relevant evidence for that feature across the narrative. \\
\mbox{} \\
Ignore all non-feature information unless it is necessary to understand the requested feature. \\
\mbox{} \\
\mbox{} \\
Consider all incident facts described in the narrative that are relevant to whether the main incident is consistent with drug offenses. This includes, e.g., actions taken, harms described, property involved, threats, injuries, methods, intent cues, or other incident facts bearing on whether the incident matches drug offenses. \\
\mbox{} \\
Based on incident facts like these, how likely is it that the main incident described in the narrative is consistent with drug offenses? Use only the narrative facts about the incident itself, not any downstream charging decisions. \\
\mbox{} \\
Provide a one-sentence justification for your decision first, then provide one integer between 0 and 100, where 0 represents that the main incident described in the narrative is extremely unlikely to be consistent with drug offenses, and 100 represents that the main incident described in the narrative is extremely likely to be consistent with drug offenses. 50 means no relevant evidence favors either endpoint; indirect cultural, geographic, or contextual associations should move the score away from 50 in proportion to their strength, even when they are not definitive. \\
\mbox{} \\
Use only the narrative facts about the incident itself, not any downstream charging decisions. \\
\mbox{} \\
You are required to return a value whenever the narrative contains incident facts relevant to consistency with drug offenses, even if the evidence is weak, indirect, subtle, or ambiguous.
\end{flushleft}

\subsubsection{Interpretable feature prompt: incident-consistent-with-crime-type\_\_fraud-forgery\_\_score}
\paragraph{Scoring prompt}
\begin{flushleft}\footnotesize\ttfamily
Classify how consistent the main incident described in the provided police report narrative is with fraud/forgery. \\
\mbox{} \\
Use only the specific feature named in the prompt. \\
\mbox{} \\
Consider all relevant evidence for that feature across the narrative. \\
\mbox{} \\
Ignore all non-feature information unless it is necessary to understand the requested feature. \\
\mbox{} \\
\mbox{} \\
Consider all incident facts described in the narrative that are relevant to whether the main incident is consistent with fraud/forgery. This includes, e.g., actions taken, harms described, property involved, threats, injuries, methods, intent cues, or other incident facts bearing on whether the incident matches fraud/forgery. \\
\mbox{} \\
Based on incident facts like these, how likely is it that the main incident described in the narrative is consistent with fraud/forgery? Use only the narrative facts about the incident itself, not any downstream charging decisions. \\
\mbox{} \\
Provide a one-sentence justification for your decision first, then provide one integer between 0 and 100, where 0 represents that the main incident described in the narrative is extremely unlikely to be consistent with fraud/forgery, and 100 represents that the main incident described in the narrative is extremely likely to be consistent with fraud/forgery. 50 means no relevant evidence favors either endpoint; indirect cultural, geographic, or contextual associations should move the score away from 50 in proportion to their strength, even when they are not definitive. \\
\mbox{} \\
Use only the narrative facts about the incident itself, not any downstream charging decisions. \\
\mbox{} \\
You are required to return a value whenever the narrative contains incident facts relevant to consistency with fraud/forgery, even if the evidence is weak, indirect, subtle, or ambiguous.
\end{flushleft}

\subsubsection{Interpretable feature prompt: incident-consistent-with-crime-type\_\_homicide\_\_score}
\paragraph{Scoring prompt}
\begin{flushleft}\footnotesize\ttfamily
Classify how consistent the main incident described in the provided police report narrative is with homicide. \\
\mbox{} \\
Use only the specific feature named in the prompt. \\
\mbox{} \\
Consider all relevant evidence for that feature across the narrative. \\
\mbox{} \\
Ignore all non-feature information unless it is necessary to understand the requested feature. \\
\mbox{} \\
\mbox{} \\
Consider all incident facts described in the narrative that are relevant to whether the main incident is consistent with homicide. This includes, e.g., actions taken, harms described, property involved, threats, injuries, methods, intent cues, or other incident facts bearing on whether the incident matches homicide. \\
\mbox{} \\
Based on incident facts like these, how likely is it that the main incident described in the narrative is consistent with homicide? Use only the narrative facts about the incident itself, not any downstream charging decisions. \\
\mbox{} \\
Provide a one-sentence justification for your decision first, then provide one integer between 0 and 100, where 0 represents that the main incident described in the narrative is extremely unlikely to be consistent with homicide, and 100 represents that the main incident described in the narrative is extremely likely to be consistent with homicide. 50 means no relevant evidence favors either endpoint; indirect cultural, geographic, or contextual associations should move the score away from 50 in proportion to their strength, even when they are not definitive. \\
\mbox{} \\
Use only the narrative facts about the incident itself, not any downstream charging decisions. \\
\mbox{} \\
You are required to return a value whenever the narrative contains incident facts relevant to consistency with homicide, even if the evidence is weak, indirect, subtle, or ambiguous.
\end{flushleft}

\subsubsection{Interpretable feature prompt: incident-consistent-with-crime-type\_\_kidnapping-abduction\_\_score}
\paragraph{Scoring prompt}
\begin{flushleft}\footnotesize\ttfamily
Classify how consistent the main incident described in the provided police report narrative is with kidnapping/abduction. \\
\mbox{} \\
Use only the specific feature named in the prompt. \\
\mbox{} \\
Consider all relevant evidence for that feature across the narrative. \\
\mbox{} \\
Ignore all non-feature information unless it is necessary to understand the requested feature. \\
\mbox{} \\
\mbox{} \\
Consider all incident facts described in the narrative that are relevant to whether the main incident is consistent with kidnapping/abduction. This includes, e.g., actions taken, harms described, property involved, threats, injuries, methods, intent cues, or other incident facts bearing on whether the incident matches kidnapping/abduction. \\
\mbox{} \\
Based on incident facts like these, how likely is it that the main incident described in the narrative is consistent with kidnapping/abduction? Use only the narrative facts about the incident itself, not any downstream charging decisions. \\
\mbox{} \\
Provide a one-sentence justification for your decision first, then provide one integer between 0 and 100, where 0 represents that the main incident described in the narrative is extremely unlikely to be consistent with kidnapping/abduction, and 100 represents that the main incident described in the narrative is extremely likely to be consistent with kidnapping/abduction. 50 means no relevant evidence favors either endpoint; indirect cultural, geographic, or contextual associations should move the score away from 50 in proportion to their strength, even when they are not definitive. \\
\mbox{} \\
Use only the narrative facts about the incident itself, not any downstream charging decisions. \\
\mbox{} \\
You are required to return a value whenever the narrative contains incident facts relevant to consistency with kidnapping/abduction, even if the evidence is weak, indirect, subtle, or ambiguous.
\end{flushleft}

\subsubsection{Interpretable feature prompt: incident-consistent-with-crime-type\_\_larceny-theft\_\_score}
\paragraph{Scoring prompt}
\begin{flushleft}\footnotesize\ttfamily
Classify how consistent the main incident described in the provided police report narrative is with larceny/theft. \\
\mbox{} \\
Use only the specific feature named in the prompt. \\
\mbox{} \\
Consider all relevant evidence for that feature across the narrative. \\
\mbox{} \\
Ignore all non-feature information unless it is necessary to understand the requested feature. \\
\mbox{} \\
\mbox{} \\
Consider all incident facts described in the narrative that are relevant to whether the main incident is consistent with larceny/theft. This includes, e.g., actions taken, harms described, property involved, threats, injuries, methods, intent cues, or other incident facts bearing on whether the incident matches larceny/theft. \\
\mbox{} \\
Based on incident facts like these, how likely is it that the main incident described in the narrative is consistent with larceny/theft? Use only the narrative facts about the incident itself, not any downstream charging decisions. \\
\mbox{} \\
Provide a one-sentence justification for your decision first, then provide one integer between 0 and 100, where 0 represents that the main incident described in the narrative is extremely unlikely to be consistent with larceny/theft, and 100 represents that the main incident described in the narrative is extremely likely to be consistent with larceny/theft. 50 means no relevant evidence favors either endpoint; indirect cultural, geographic, or contextual associations should move the score away from 50 in proportion to their strength, even when they are not definitive. \\
\mbox{} \\
Use only the narrative facts about the incident itself, not any downstream charging decisions. \\
\mbox{} \\
You are required to return a value whenever the narrative contains incident facts relevant to consistency with larceny/theft, even if the evidence is weak, indirect, subtle, or ambiguous.
\end{flushleft}

\subsubsection{Interpretable feature prompt: incident-consistent-with-crime-type\_\_motor-vehicle-theft\_\_score}
\paragraph{Scoring prompt}
\begin{flushleft}\footnotesize\ttfamily
Classify how consistent the main incident described in the provided police report narrative is with motor vehicle theft. \\
\mbox{} \\
Use only the specific feature named in the prompt. \\
\mbox{} \\
Consider all relevant evidence for that feature across the narrative. \\
\mbox{} \\
Ignore all non-feature information unless it is necessary to understand the requested feature. \\
\mbox{} \\
\mbox{} \\
Consider all incident facts described in the narrative that are relevant to whether the main incident is consistent with motor vehicle theft. This includes, e.g., actions taken, harms described, property involved, threats, injuries, methods, intent cues, or other incident facts bearing on whether the incident matches motor vehicle theft. \\
\mbox{} \\
Based on incident facts like these, how likely is it that the main incident described in the narrative is consistent with motor vehicle theft? Use only the narrative facts about the incident itself, not any downstream charging decisions. \\
\mbox{} \\
Provide a one-sentence justification for your decision first, then provide one integer between 0 and 100, where 0 represents that the main incident described in the narrative is extremely unlikely to be consistent with motor vehicle theft, and 100 represents that the main incident described in the narrative is extremely likely to be consistent with motor vehicle theft. 50 means no relevant evidence favors either endpoint; indirect cultural, geographic, or contextual associations should move the score away from 50 in proportion to their strength, even when they are not definitive. \\
\mbox{} \\
Use only the narrative facts about the incident itself, not any downstream charging decisions. \\
\mbox{} \\
You are required to return a value whenever the narrative contains incident facts relevant to consistency with motor vehicle theft, even if the evidence is weak, indirect, subtle, or ambiguous.
\end{flushleft}

\subsubsection{Interpretable feature prompt: incident-consistent-with-crime-type\_\_robbery\_\_score}
\paragraph{Scoring prompt}
\begin{flushleft}\footnotesize\ttfamily
Classify how consistent the main incident described in the provided police report narrative is with robbery. \\
\mbox{} \\
Use only the specific feature named in the prompt. \\
\mbox{} \\
Consider all relevant evidence for that feature across the narrative. \\
\mbox{} \\
Ignore all non-feature information unless it is necessary to understand the requested feature. \\
\mbox{} \\
\mbox{} \\
Consider all incident facts described in the narrative that are relevant to whether the main incident is consistent with robbery. This includes, e.g., actions taken, harms described, property involved, threats, injuries, methods, intent cues, or other incident facts bearing on whether the incident matches robbery. \\
\mbox{} \\
Based on incident facts like these, how likely is it that the main incident described in the narrative is consistent with robbery? Use only the narrative facts about the incident itself, not any downstream charging decisions. \\
\mbox{} \\
Provide a one-sentence justification for your decision first, then provide one integer between 0 and 100, where 0 represents that the main incident described in the narrative is extremely unlikely to be consistent with robbery, and 100 represents that the main incident described in the narrative is extremely likely to be consistent with robbery. 50 means no relevant evidence favors either endpoint; indirect cultural, geographic, or contextual associations should move the score away from 50 in proportion to their strength, even when they are not definitive. \\
\mbox{} \\
Use only the narrative facts about the incident itself, not any downstream charging decisions. \\
\mbox{} \\
You are required to return a value whenever the narrative contains incident facts relevant to consistency with robbery, even if the evidence is weak, indirect, subtle, or ambiguous.
\end{flushleft}

\subsubsection{Interpretable feature prompt: incident-consistent-with-crime-type\_\_sex-offenses\_\_score}
\paragraph{Scoring prompt}
\begin{flushleft}\footnotesize\ttfamily
Classify how consistent the main incident described in the provided police report narrative is with sex offenses. \\
\mbox{} \\
Use only the specific feature named in the prompt. \\
\mbox{} \\
Consider all relevant evidence for that feature across the narrative. \\
\mbox{} \\
Ignore all non-feature information unless it is necessary to understand the requested feature. \\
\mbox{} \\
\mbox{} \\
Consider all incident facts described in the narrative that are relevant to whether the main incident is consistent with sex offenses. This includes, e.g., actions taken, harms described, property involved, threats, injuries, methods, intent cues, or other incident facts bearing on whether the incident matches sex offenses. \\
\mbox{} \\
Based on incident facts like these, how likely is it that the main incident described in the narrative is consistent with sex offenses? Use only the narrative facts about the incident itself, not any downstream charging decisions. \\
\mbox{} \\
Provide a one-sentence justification for your decision first, then provide one integer between 0 and 100, where 0 represents that the main incident described in the narrative is extremely unlikely to be consistent with sex offenses, and 100 represents that the main incident described in the narrative is extremely likely to be consistent with sex offenses. 50 means no relevant evidence favors either endpoint; indirect cultural, geographic, or contextual associations should move the score away from 50 in proportion to their strength, even when they are not definitive. \\
\mbox{} \\
Use only the narrative facts about the incident itself, not any downstream charging decisions. \\
\mbox{} \\
You are required to return a value whenever the narrative contains incident facts relevant to consistency with sex offenses, even if the evidence is weak, indirect, subtle, or ambiguous.
\end{flushleft}

\end{document}